\documentclass[aps,prd,amsmath,amssymb,eqsecnum,nofootinbib]{revtex4}

\usepackage{color}
\usepackage{graphicx}       
\usepackage{graphics}

\newcommand{\beq}{\begin{equation}}
\newcommand{\eeq}{\end{equation}}

\newcommand{\fld}{\Phi}
\newcommand{\Gret}{G_{\text{ret}}}
\newcommand{\Grad}{\tilde{g}}
\newcommand{\fldtail}{\fld_\mu^{\text{tail}}}
\newcommand{\mass}{m}

\newcommand{\alp}{\alpha}
\newcommand{\bet}{\beta}
\newcommand{\gam}{\gamma}
\newcommand{\eps}{\epsilon}
\newcommand{\lam}{\lambda}
\newcommand{\sig}{\sigma}

\newcommand{\rstar}{r_\ast}

\newcommand{\phif}{\Delta \phi}

\newcommand{\Bef}{\mathcal{B}}
\newcommand{\unorm}{\tilde{u}}

\newcommand{\Aout}{A^{\text{(out)}}}
\newcommand{\Ain}{A^{\text{(in)}}}

\newcommand{\uin}{u^{\text{(in)}}_{l \omega}}
\newcommand{\uup}{u^{\text{(up)}}_{l \omega}}

\newcommand{\sech}{\text{sech}}
\newcommand{\nn}{\nonumber}

\newcommand{\lmax}{l_{\text{cut}}}
\newcommand{\nmax}{n_{\text{max}}}
\newcommand{\Ztt}{{\sigma^t}_t}
\newcommand{\Zyy}{{\sigma^y}_y}

\newcommand{\Phipartial}{\Phi_{\text{partial}}}
\newcommand{\GQNM}{\Gret^{\text{QNM}}}
\newcommand{\II}{\mathcal{I}}
\newcommand{\Agam}{\mathcal{A}(\gam)}

\newcommand{\R}{\rho}
\newcommand{\Rs}{\rho_\ast}

\begin{document}

\title{Self-Force Calculations with Matched Expansions and Quasinormal Mode Sums}

\author{Marc Casals}
\email{marc.casals@dcu.ie}
\affiliation{CENTRA, Instituto Superior T\'{e}cnico, Lisbon, Portugal \\ School of Mathematical Sciences, Dublin City University, Glasnevin, Dublin 9, Ireland}

\author{Sam Dolan}
\email{sam.dolan@ucd.ie}
\author{Adrian C. Ottewill}
\email{adrian.ottewill@ucd.ie}
\author{Barry Wardell}
\email{barry.wardell@ucd.ie}
\affiliation{Complex and Adaptive Systems Laboratory and School of Mathematical Sciences, University College Dublin, Belfield, Dublin 4, Ireland}

\date{\today}

\begin{abstract}
Accurate modelling of gravitational wave emission by extreme mass ratio inspirals is essential for their detection by the LISA mission. A leading perturbative approach involves the calculation of the self-force acting upon the smaller orbital body. In this work, we present the first application of the Poisson-Wiseman-Anderson method of `matched expansions' to compute the self-force acting on a point particle moving in a curved spacetime. The method employs two expansions for the Green function which are respectively valid in the `quasilocal' and `distant past' regimes, and which may be matched together within the normal neighbourhood.  We perform our calculation in a static region of the spherically symmetric Nariai spacetime ($dS_2 \times  \mathbb{S}^2$), in which scalar field perturbations are governed by a radial equation with a P\"{o}schl-Teller potential (frequently used as an approximation to the Schwarzschild radial potential) whose solutions are known in closed form. 

The key new ingredients in our study are: (i) very high order quasilocal expansions, and (ii) expansion of the `distant past' Green function in quasinormal modes. In combination, these tools enable a detailed study of the properties of the scalar-field Green function. We demonstrate that the Green function is singular whenever $x$ and $x^\prime$ are connected by a null geodesic and apply asymptotic methods to determine the structure of the Green function near the null wavefront. We show that the singular part of the Green function undergoes a transition each time the null wavefront passes through a caustic point, following a repeating four-fold sequence $\delta(\sig)$, $1/\pi \sig$, $-\delta(\sig)$, $-1/\pi \sig, $ etc., where $\sig$ is Synge's world function. 

The matched expansion method provides insight into the non-local properties of the self-force. We show that the self-force generated by the segment of worldline lying outside the normal neighbourhood is not negligible. We apply the matched expansion method to compute the scalar self-force acting on a static particle on the Nariai spacetime, and validate against an alternative method, obtaining agreement to six decimal places. 

We conclude with a discussion of the implications for wave propagation and self-force calculations. On black hole spacetimes, any expansion of the Green function in quasinormal modes must be augmented by a branch cut integral. Nevertheless we expect the Green function in Schwarzschild spacetime to inherit certain key features, such as a four-fold singular structure linked to the asymptotic behaviour of quasinormal modes. In this way, the Nariai spacetime provides a fertile testing ground for developing insight into the non-local part of the self-force on black hole spacetimes.
\end{abstract}
\maketitle

\section{Introduction}\label{sec:intro}

The last decade has seen a surge of interest in the nascent field of gravitational wave astronomy.  Gravitational waves -- propagating ripples in spacetime -- are generated by some of the most violent processes in the known universe, such as supernovae, black hole mergers and galaxy collisions. These powerful processes are hidden from the view of `traditional' electromagnetic-wave telescopes behind shrouds of dust and radiation. On the other hand, gravitational waves are not strongly absorbed or scattered by intervening matter, and carry information about the dynamics at the heart of such processes. The prospects seem good for direct detection of gravitational waves in the near future. A number of ground-based detectors (such as LIGO \cite{LIGO}, VIRGO \cite{VIRGO} and GEO600 \cite{GEO}) are now in the data collection phase. 

Gravitational wave astronomy will enter a new era with the launch of the first space-based observatory: the Laser Interferometer Space Antenna (LISA) \cite{LISA}. It is hoped that this joint NASA/ESA mission, presently in the design and planning phase, will be launched within a decade. It will be preceded by a pathfinder mission, due for launch at the end of this year \cite{Bell:2008}.

Black hole binary systems are a key target for gravitational wave (GW) observatories worldwide. Data analysis methods such as matched filtering may be applied to separate a weak GW signal from a noisy background \cite{Vallisneri:2009}. 
An essential prerequisite for detection via matched filtering is accurate templates for the gravitational wave emission from black hole binaries. Breakthroughs in numerical relativity in the last five years have led to a rapid advance in the modelling of comparable-mass binaries, where the partners are of similar mass. Progress in numerical relativity continues apace.

A key target for the LISA mission are the so-called \emph{Extreme Mass Ratio Inspirals} (EMRIs): compact binaries in which one partner (mass $M$) is significantly more massive than the other (mass $m$). Mass ratios of $\mu \equiv m/M \gtrsim 10^{-8}$ are possible, for example for a solar-mass black hole orbiting a supermassive black hole \cite{Hughes:Drasco:2005}. Mass ratios of up to $m/M\sim 1/10$ have been studied by numerical relativists \cite{Gonzalez-Sperhake-Bruegman-2008}; smaller ratios are presently beyond the scope of numerical relativity due to the existence of two distinct and dissimilar length scales in the system. Perturbative approaches seem more likely to succeed in the extreme-mass regime.

The smaller compact mass $m$ distorts the curvature of the spacetime in which it is moving. Hence, rather than following a geodesic of the background spacetime generated by the larger mass $M$, the smaller mass follows a geodesic of the \emph{total} spacetime \cite{Detweiler-Whiting-2003}. However, if the mass ratio is extreme, the deviation of the smaller body's motion from the background geodesic will be (locally) small. The deviation may be interpreted as arising from a \emph{self-force}, created by the smaller mass $m$ interacting with its own gravitational field. To leading order, the self-force acceleration is proportional to $m$. With knowledge of the leading term in the self-force, one may model the evolution of the orbit and subsequent inspiral of the smaller mass, and compute the gravitational wave emission to high accuracy. However, finding the instantaneous self-force in a curved spacetime is not at all straightforward; it turns out to depend on the \emph{entire past history} of the
smaller mass, $m$.

The idea of a self-force has a long history in physics. In the late 19th century it was well-known that a charge undergoing an acceleration 
in flat spacetime
will generate electromagnetic radiation, and will feel a corresponding \emph{radiation reaction}. The self-acceleration of a charged point particle in flat spacetime is given by the well-known Abraham-Lorentz-Dirac formula \cite{Dirac-1938}. Radiation reaction implies that the `classical' model of the atom (a point-particle electron orbiting a compact nucleus) is unstable. The observed stability of the atom remained a puzzle for many years, and provided a key motivation for the development of quantum mechanics. 
In the 1960s, DeWitt and Brehme \cite{DeWitt:1960} derived a formula for the self-force acting on an electrically-charged point particle in a curved background, and a correction was later provided by Hobbs \cite{Hobbs:1968a}. The gravitational self-force acting on a point mass was found in 1997 by two groups working concurrently and independently: Mino, Sasaki and Tanaka \cite{Mino:Sasaki:Tanaka:1996} and Quinn and Wald \cite{Quinn:Wald:1997}. Shortly after, Quinn derived the self-force acting on a minimally-coupled scalar charge \cite{Quinn:2000}. These developments are summarized in 2004/05 reviews by Poisson \cite{Poisson:2003} and Detweiler \cite{Detweiler:2005}. In the subsequent period, a range of complementary approaches to the self-force problem have been developed \cite{Galley:Hu:Lin:2006, Harte:2008, Gralla:Wald:2008, Futamase:Hogan:Itoh:2008, Flanagan:Hinderer:2008}. 

The self-force expressions for scalar, electromagnetic and gravitational cases take similar form \cite{Poisson:2003}. In this paper, we restrict our attention to the simplest case: a point-like scalar charge $q$ of mass $\mass$ coupled to a massless scalar field $\fld(x)$ moving on a curved background geometry. The scalar field $\fld(x)$ satisfies the field equation
\beq
 \left( \square - \xi R \right) \fld(x)  = - 4 \pi \rho(x)  \label{scalar-field-eq1}
\eeq
where $\square$ is the d'Alembertian on the curved background created by the larger mass $M$, $R$ is the Ricci scalar, and $\xi$ is the curvature coupling constant. The charge density, $\rho$, of the point particle is 
\beq
 \rho (x) = \int_{\gamma} q \, \frac{\delta^4 (x^\mu - z^\mu(\tau) )}{\sqrt{-g}} \, d \tau
\eeq
where $z(\tau)$ describes the worldline $\gamma$ of the particle with proper time $\tau$, 
$g_{\mu \nu}$ is the background metric, $g = \text{det}( g_{\mu \nu} )$, and $\delta^4(\cdot)$ is the four-dimensional Dirac distribution. The field 
exerts a radiation reaction on the particle, creating a self-force \cite{Quinn:2000}
\beq
f_\mu^{\text{self}} = q \nabla_\mu \fld_R
\eeq
which leads to the equations of motion for the scalar particle
\beq
\label{eq:ma}
ma^\mu = (g^{\mu \nu} + u^\mu u^\nu ) f_\nu^{\text{self}} = q (g^{\mu \nu} + u^\mu u^\nu ) \nabla_\nu \fld_R
\eeq
where $u^\mu$ is the particle's four-velocity and  $\fld_R$ is the {\it radiative} part of the field. Identifying the correct radiative field (which is regular at the particle's position) is the essential step in the derivation of the self-force \cite{Poisson:2003}. Note that the projection operator $g^{\mu \nu} + u^\mu u^\nu$ has been applied here to ensure that $u_{\mu} f^\mu_{\text{self}} = 0$.
The mass $m$ appearing in (\ref{eq:ma}) is the `dynamical' (and renormalized) particle's mass, which
in the scalar case is not necessarily a constant of motion \cite{Quinn:2000}. Rather, it evolves according to 
\beq
\label{eq:dmdtau}
\frac{d \mass}{d \tau} = - q u^\mu \nabla_\mu \fld_R .
\eeq
In other words, a spinless particle may radiate away its mass through the emission of monopolar waves.

A leading method for computing the derivative of the radiative field, $\nabla_\nu \fld_R$, and hence the self-force, is based on \emph{mode sum regularization} (MSR). The MSR approach was developed by Barack, Ori and collaborators \cite{Barack:Ori:2000, Barack:2001, Barack:Mino:Nakano:2002, Barack:Sago:2007} and Detweiler and coworkers \cite{Detweiler:Messaritaki:Whiting:2002, Detweiler:2005, Vega:Detweiler:2008}. The method has been applied to the Schwarzschild spacetime to compute, for example, the gravitational self-force for circular orbits \cite{Barack:Sago:2007} and the scalar self-force for eccentric orbits \cite{Haas:2007}. The application to Kerr is in progress \cite{Barack:Golbourn:Sago:2007, Barack:Ori:Sago:2008}. It was recently shown \cite{Sago:Barack:Detweiler:2008} that the gravitational self-force computed in the Lorenz gauge is in agreement with that found in the Regge-Wheeler gauge \cite{Detweiler:Messaritaki:Whiting:2002, Detweiler:2005}. Further gauge-invariant comparisons, and comparison with the predictions of Post-Newtonian theory \cite{Blanchet:Grishchuk:Schaefer:2009, Damour:Nagar:2009} are presently under consideration \cite{Detweiler:2008}.

One drawback of the MSR method is that it gives relatively little geometric insight into the physical origin of the self-force. An alternative approach, based on \emph{matched expansions}, was suggested by Poisson and Wiseman in 1998 \cite{Poisson:Wiseman:1998}. Their idea was to compute the self-force by matching together two independent expansions for the Green function, valid in `quasilocal' and `distant past' regimes. This suggestion was analysed by Anderson and Wiseman \cite{Anderson:Wiseman:2005}, who concluded in 2005 that ``this approach remains, in our opinion, in the category of `promising but possessing some technical challenges'.'' The present paper represents the first practical attempt to implement this method. 

In the following sections we demonstrate that accurate self-force calculations via matched expansions are indeed feasible. We apply the method to compute the self-force for a scalar charge at fixed position on the product spacetime $dS_2 \times  \mathbb{S}^2$ (i.e. the product of a two-sphere and a two-dimensional de Sitter spacetime) introduced long ago by Nariai \cite{Nariai:1950,Nariai:1951}.  We introduce a method for calculating the `distant past` Green function using an expansion in quasinormal modes. The effect of \emph{caustics} upon wave propagation is examined. This work is intended to lay a foundation for future studies of self-force in black hole spacetimes through matched expansions. The prospects for extending the calculation to the Schwarzschild spacetime appear good, although the work remains to be conducted.

The remainder of this paper is organised as follows. In Sec.~\ref{sec:matched-expansions} we define the self-force and outline the Poisson-Anderson-Wiseman method of matched expansions. In Sec.~\ref{sec:Nariai} we consider wave propagation on the Schwarzschild spacetime. A radial equation of standard form is obtained via the well-known `trick' of replacing the Schwarzschild potential with a so-called `P\"{o}schl-Teller' potential. We show that a P\"{o}schl-Teller potential arises more naturally if we consider wave propagation on the `Nariai' spacetime, whose properties are described in detail.

Section \ref{sec:scalarGF} is concerned with the scalar Green function on the Nariai spacetime. We begin in Sec.~\ref{subsec:FG} by expressing the Green function as a mode sum over angular modes and integral over frequency. We show in Sec.~\ref{subsec:DP} that performing the integral over frequency leaves a sum of residues: a so-called `quasinormal mode sum', which may be matched onto a `quasilocal' Green function, briefly described in Sec.~\ref{subsec:QL}.

In Sec.~\ref{sec:singularities} we consider the singular structure of the Green function. In Sec.~\ref{subsec:large-l} we demonstrate that the Green function is singular on the null surface, even beyond the boundary of the normal neighbourhood and through caustics. We show that the singular behaviour arises from the large-$l$ asymptotics of the quasinormal mode sums. To investigate further, we employ two closely-related methods for converting sums into integrals, namely, the Watson transform and Poisson sum (Sec.~\ref{subsec:watson-and-poisson}). The form of the Green function close to the null cone is studied in detail in Secs.~\ref{subsec:Poisson} and \ref{subsec:Hadamard}, and asymptotic expressions are derived. 

Section \ref{sec:sf} describes the calculation of the self-force for the specific case of the static particle. For the Schwarzschild spacetime, the static case has been well-studied. We show in Sec.~\ref{subsec:static full Green} that the massive-field approach of Rosenthal \cite{Rosenthal:2004} may be adapted to the Nariai spacetime. This provides an independent check on the matched expansion calculation which is described in Sec.~\ref{subsec:matched-static}. Relevant numerical methods are outlined in Sec.~\ref{subsec:nummeth}.

In Sec.~\ref{sec:results} we present a selection of significant numerical results. We start in Sec.~\ref{subsec:results:GF} by examining the properties of the quasinormal mode Green function. In Sec.~\ref{subsec:results-asymptotics} we test the asymptotic expressions describing the singularity structure. In Sec.~\ref{subsec:results:matched} we show that the `quasilocal' and `distant past' Green functions match in an appropriate regime. In Sec.~\ref{subsec:results:SF} we present results for the self-force on a static particle.

We conclude in Sec.~\ref{sec:conclusions} with a discussion of the implications of this study. 
Throughout the paper, we employ geometrized units $G = c = 1$, and the metric sign convention $\{-+++\}$.

\section{The Method of Matched Expansions}\label{sec:matched-expansions}
Here we briefly outline the Poisson-Wiseman-Anderson method of `matched expansions' \cite{Poisson:Wiseman:1998, Anderson:Wiseman:2005}. We start with an expression for the covariant derivative of the radiative scalar field \cite{Poisson:2003,Quinn:2000},
\beq
\label{eq:rad-field-deriv}
\nabla_\mu \fld_R \left( z(\tau) \right) = - \frac{1}{12}(1 - 6\xi) q R u_\mu + q ( g_{\mu \nu} + u_{\mu} u_{\nu} ) \left( \frac{1}{3} \dot{a}^\nu + \frac{1}{6} {R^\nu}_\lam u^\lam  \right) + \fldtail\left( z(\tau) \right)
\eeq
Here, ${R^\nu}_\lam$ is the Ricci tensor of the background metric and $\dot{a}^\nu$ is the derivative with respect to proper time of the four-acceleration $a^\nu = \tfrac{d u^{\nu}}{d \tau}$. The first two sets of terms are evaluated locally \cite{Quinn:Wald:1997, Quinn:2000, Poisson:2003}. 
The final term $\fldtail$ is non-local; it is the so-called \emph{tail integral},
\beq
\fldtail \left( z(\tau) \right)= q \lim_{\eps \rightarrow 0^+} \int_{-\infty}^{\tau-\eps} \nabla_\mu \Gret (z(\tau), z(\tau^\prime)) d \tau^\prime  \label{tail-integral}
\eeq
where $\Gret(x, x^\prime)$ is the retarded Green function, defined by 
\beq
 \square_x \Gret(x, x^{\prime}) = -4 \pi \frac{\delta^4(x^{\mu} - x^{\mu \prime})}{\sqrt{-g}}
 \label{eq:gf-waveeq}
\eeq
together with appropriate causality conditions (which we describe in Sec.\ref{subsec:FG}).

\begin{figure}
 \begin{center}
  \includegraphics[width=5cm]{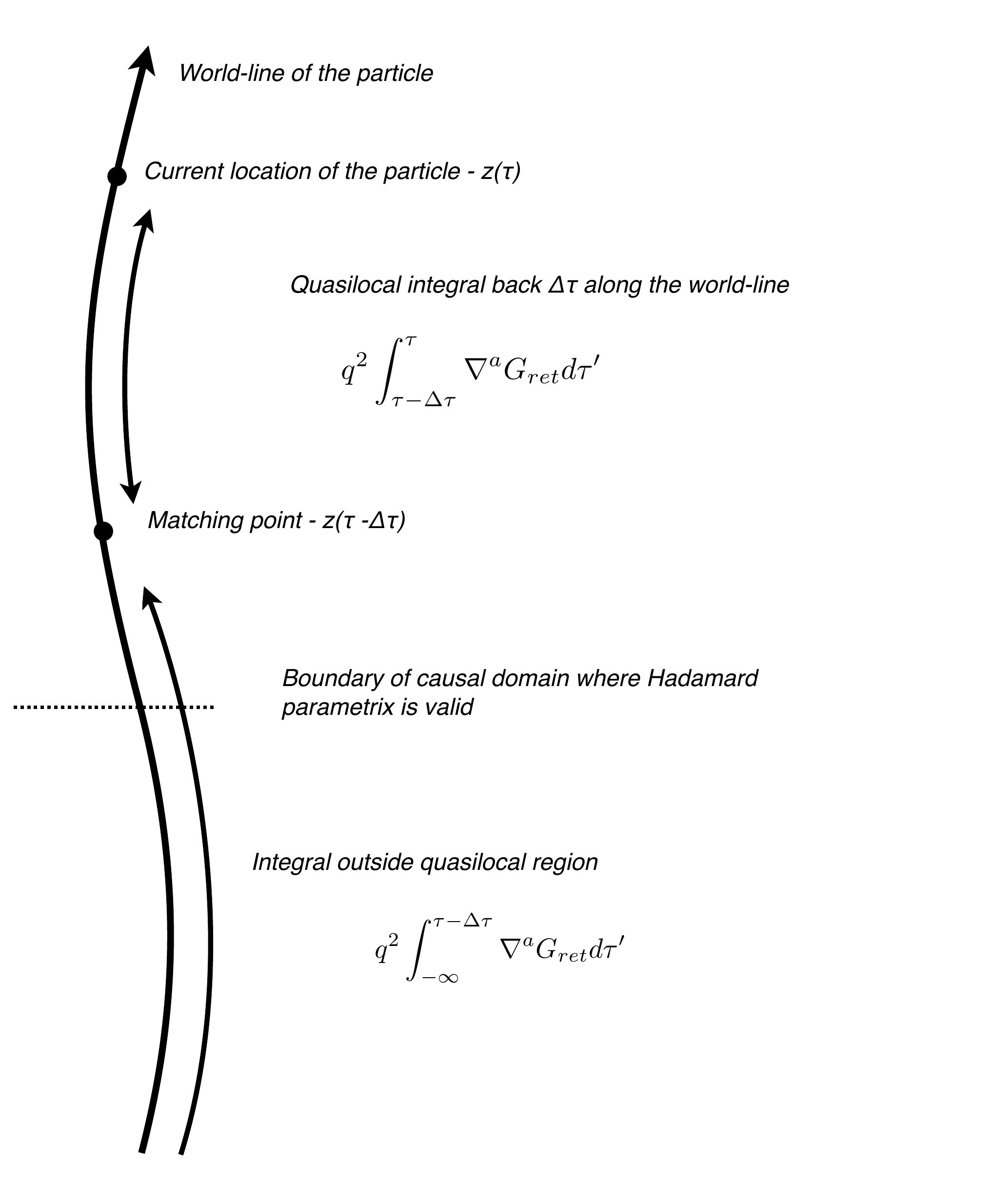}
 \end{center}
 \caption{\emph{In the method of matched expansions, the tail integral is split into quasilocal (QL) and distant past (DP) parts. }}
 \label{fig:matchedexpansion}
\end{figure}

Note that the tail integral depends on the entire past history of the particle's motion. Its evaluation is the main obstacle to progress. 
The tail integral (\ref{tail-integral}) may be split into so-called \emph{quasilocal} (QL) and \emph{distant past} (DP) parts, as shown in Fig.~\ref{fig:matchedexpansion}. That is,
\begin{eqnarray}
 \fldtail \left( z(\tau) \right) &=&  \fld^{\text{(QL)}}_\mu \left( z(\tau) \right) +  \fld^{\text{(DP)}}_\mu  \left( z(\tau) \right) \label{eq:fld-QL-DP} \nonumber \\
&=& q \lim_{\eps \rightarrow 0^+} \int_{\tau - \Delta \tau}^{\tau - \eps} \nabla_\mu \Gret(z(\tau), z(\tau^\prime)) d\tau^\prime + q \int^{\tau - \Delta \tau}_{-\infty} \nabla_\mu \Gret(z(\tau), z(\tau^\prime)) d\tau^\prime
\end{eqnarray}
where $\tau - \Delta \tau$ is the \emph{matching time}, with $\Delta \tau$ being a free parameter in the method (see Fig.~\ref{fig:matchedexpansion}).

The QL and DP parts may be evaluated separately using independent methods. In particular, if we choose $\Delta \tau$  to be sufficiently small that $z(\tau)$ and $z(\tau - \Delta \tau)$ are within a \emph{convex normal neighbourhood} \cite{Friedlander}, then the QL part may be evaluated by expressing the Green function in the Hadamard parametrix \cite{Hadamard}. In other words, if $z(\tau)$ and $z(\tau - \Delta \tau)$ are connected by a unique timelike geodesic, then the QL integral is simply
\beq
q^{-1} \fld^{\text{(QL)}}_\mu  \left( z(\tau) \right) = - \lim_{\eps \rightarrow 0^+} \int_{\tau - \Delta \tau}^{\tau - \eps} \nabla_\mu V( z(\tau), z(\tau^\prime)) d\tau^\prime
\eeq
where $V(x,x^\prime)$ is the smooth symmetric biscalar describing the propagation of radiation within the light cone (see Sec. \ref{subsec:QL} for full details).  
The approach ultimately yields a series expansion for the QL self force in the coordinate separation of the points $x$ and $x^\prime$. The Hadamard-expansion method is now well advanced for several spacetimes of physical relevance, such as Schwarzschild and Kerr \cite{Anderson:2003, Anderson:2003:err1, Anderson:2003:err2, Anderson:Flanagan:Ottewill:2004, Ottewill:Wardell:2008, Ottewill:Wardell:2009}. In Sec. \ref{subsec:QL} we apply this method to determine the quasilocal Green function and self-force in the Nariai spacetime.

Evaluating the contribution to the Green function from the `distant past' is a greater challenge, and is the main focus of this work. One possibility is to decompose the Green function into a sum over \emph{angular modes} and an integral over frequency. 
In a spherically symmetric spacetime the Green function may be defined in terms of an integral transform and mode decomposition as follows,
\beq
\Gret (x, x^{\prime}) = \frac{1}{2\pi} \int_{-\infty+ic}^{+\infty+ic} d \omega e^{-i \omega (t-t^\prime)} \sum_{l=0}^\infty (2l+1) P_l(\cos \gam) \Grad_{l \omega}(r, r^\prime)
\label{modesum1}
\eeq
Here $c$ is a positive constant, $t$ and $r$ are appropriate time and radial coordinates, and $\cos \gam = \cos \theta \cos \theta^\prime + \sin \theta \sin \theta^\prime \cos( \phi - \phi^\prime )$, where $\gam$ is the angle between the spacetime points $x$ and $x^\prime$. The radial Green function $\Grad_{l \omega}(r, r^\prime)$ may be constructed from two linearly-independent solutions of a radial equation. Since the DP Green function does not need to be extended to coincidence ($\tau^\prime \rightarrow \tau$), the mode sum does not require regularization (though it may still be regularized if desired). However, Anderson and Wiseman \cite{Anderson:Wiseman:2005} found the convergence of the mode sum to be poor, noting that going from 10 modes to 100 increased the accuracy by only a factor of three. 

In this paper we explore a new method for evaluating the `\emph{distant past}' contribution, based on an expansion in so-called \emph{quasinormal modes}. 
The integral over frequency in equation (\ref{modesum1}) may be evaluated by deforming the contour in the complex plane \cite{Leaver:1986, Andersson:1997}. This is shown in Fig. \ref{fig:contours}. In the Schwarzschild case there arise three distinct contributions to the Green function, from the three sections of the frequency integral in (\ref{modesum1}):
 \begin{enumerate}
  \item A prompt response, arising from the integral along high-frequency arcs.
  \item A `quasinormal mode sum', arising from the residues of poles in the lower half-plane of complex frequency $\omega$.
  \item Power-law tail, arising from an integral along a branch cut.
 \end{enumerate}
The three parts (1--3) are commonly supposed to dominate the scattered signal at early, intermediate and late times, respectively \cite{Leaver:1986,Andersson:1997}. (This may be slightly misleading, however; Leaver \cite{Leaver:1986} notes that, in addition, the branch cut integral (part 3) ``contributes heavily to the initial burst of radiation''). In this work, we investigate an alternative spacetime, introduced by Nariai in 1950~\cite{Nariai:1950,Nariai:1951}, in which the power-law tail (part 3) is absent. We demonstrate that, on the Nariai spacetime, at suitably `late times', the distant past Green function may be written as a sum over quasinormal modes (defined in Sec. \ref{subsec:DP}). We use the sum to compute the Green function, the radiative field and the self-force for a static particle.

The key question addressed in this work is the following: how much of the self-force arises from the quasilocal region, and how much from the distant past? If the Green function falls off fast enough then only the QL integral would be needed, and, since the QL integral is restricted to the normal neighbourhood, only the Hadamard parametrix is required. Unfortunately this is not necessarily the case; Anderson and Wiseman \cite{Anderson:Wiseman:2005} note that there are simple situations in which the DP integral in (\ref{modesum1}) gives the dominant contribution to the self-force.

Using the methods presented in this paper we are able to compute the retarded Green function and the integrand of Eq.~(\ref{tail-integral}) as a function of time along the past worldline. We show that the DP contribution cannot be neglected. In particular, we find that the Green function and the integrand of Eq.~(\ref{tail-integral}) is singular whenever the two points $z^{\mu}(\tau)$ and $z^\mu(\tau^\prime)$ are connected by a null geodesic. We show that the singular form of the Green function changes every time a null geodesic passes through a \emph{caustic}. On a spherically-symmetric spacetime, caustics occur at the antipodal points.

\begin{figure}
 \begin{center}
  \includegraphics[width=10cm]{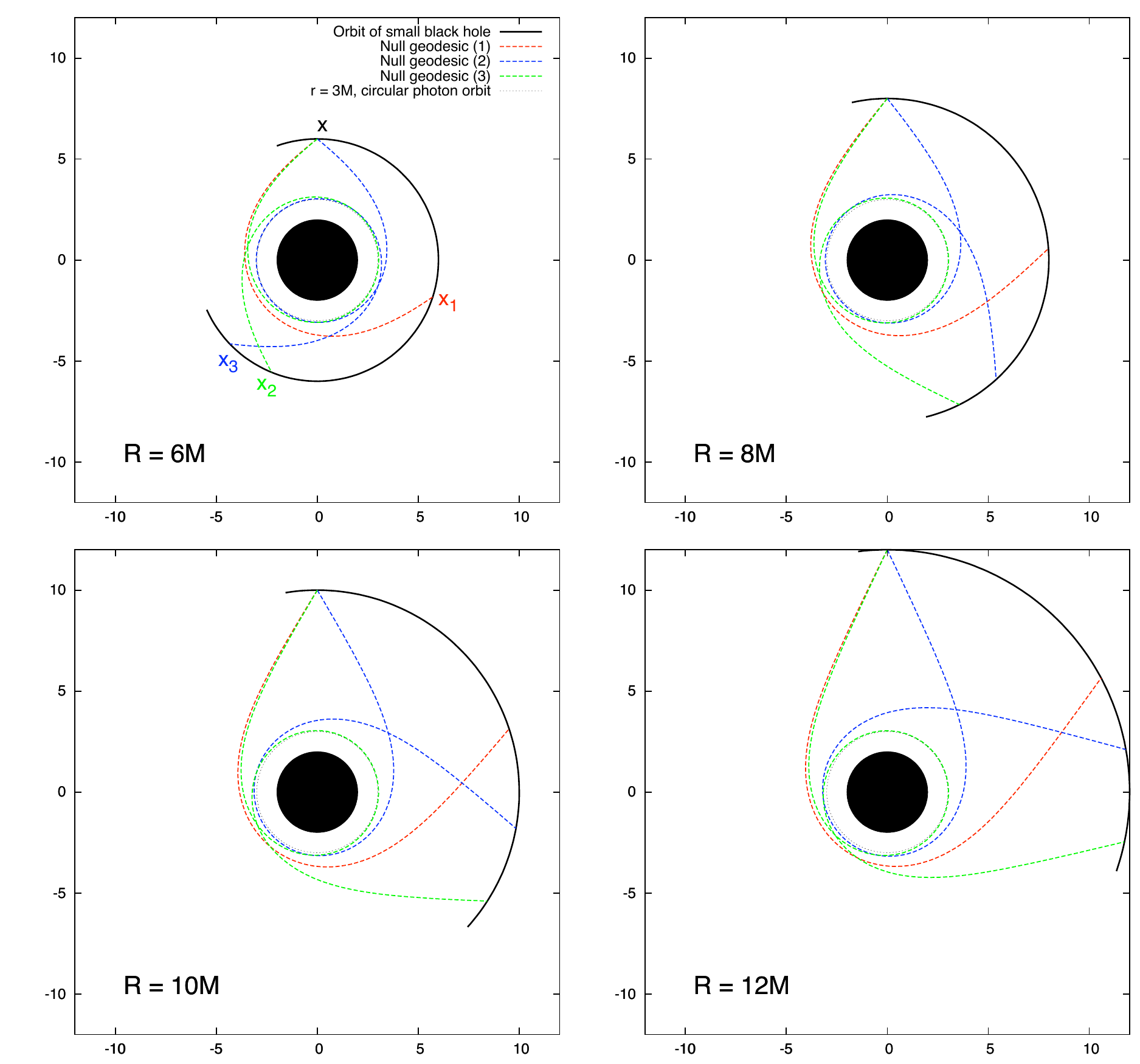}
 \end{center}
 \caption{\emph{Orbiting null geodesics on the Schwarzschild spacetime that intersect timelike circular orbits of various radii $R = 6M$, $8M$, $10M$ and $12M$}. The null geodesics are shown as coloured dotted lines, and the timelike circular orbit is shown as a black line. The spacetime point $x$ is connected to $x_1, x_2, $ etc.~by null geodesics, as well as by the timelike circular geodesic. The Green function is singular when $x^\prime = x_1, x_2, $ etc. Note that between $R=6M$ and $R=8M$ the ordering of the points $x_2$ and $x_3$ becomes reversed.}
 \label{fig:circ_orbits}
\end{figure}

On Schwarzschild spacetime, the presence of an unstable photon orbit at $r=3M$ implies that a null geodesic originating on a timelike worldline may later re-intersect the timelike worldline, by orbiting around the black hole. Hence the effect of caustics may be significant. For example, Fig. \ref{fig:circ_orbits} shows orbiting null geodesics on the Schwarzschild spacetime which intersect timelike circular orbits of various radii. We believe that understanding the singular behaviour of the integrand of Eq.~(\ref{tail-integral}) is a crucial step in understanding the origin of the non-local part of the self-force. As we shall see, the Nariai spacetime proves a fertile testing ground. 

\section{Schwarzschild and Nariai Spacetimes}\label{sec:Nariai}

To evaluate the retarded Green function (\ref{modesum1}) we require solutions to the homogeneous scalar field equation on the appropriate curved background. In the absence of sources, the scalar field equation (\ref{scalar-field-eq1}) is  
\beq
 \frac{1}{\sqrt{-g}} \, \partial_\mu \left( \sqrt{-g} g^{\mu \nu} \partial_{\nu} \fld \right) - \xi R \Phi = 0. \label{scalar-field-eq2}
\eeq
For the Schwarzschild spacetime, the line element is
\beq
\label{eq:schwle}
ds^2 = -f_S(r) dt_S^2 + f_S^{-1}(r) dr^2 + r^2 d \Omega^2_2, \quad \quad \quad d\Omega_2^2=d\theta^2+\sin^2\theta d\phi^2,
\eeq
where $f_S(r) = 1 - 2M/r$ and the label `S' denotes `Schwarzschild'. Decomposing the field in the usual way, 
\beq
\fld(x) = \int^{\infty}_{-\infty} d\omega_S \sum_{l=0}^{+\infty}\sum_{m=-l}^{+l} c_{l m \omega_S} \Phi_{l m \omega_S}(x)
\quad \quad \text{where} \quad  \Phi_{l m \omega_S} (x)=
 \frac{u_{l\omega_S}^{(S)}(r)}{r} Y_{lm}(\theta, \phi) e^{-i \omega_S t_S},
\eeq
where $Y_{lm}(\theta, \phi)$ are the spherical harmonics, $c_{l m \omega_S}$ are the coefficients in the mode decomposition, and the radial function $u_{l \omega_S}^{(S)}(r)$ satisfies the radial equation
\beq
\left[ \frac{d^2}{d \rstar^2} + \omega_S^2 - V_l^{(S)}(r) \right] u_{l\omega_S}^{(S)}(r) = 0  \label{rad-eq-schw}
\eeq
with an effective potential 
\beq
V_l^{(S)}(r) = f_S(r) \left( \frac{l(l+1)}{r^2} + \frac{f_S^\prime(r)}{r} \right) = \left(1 - \frac{2M}{r} \right) \left( \frac{l(l+1)}{r^2} + \frac{2M}{r^3} \right)  . \label{Veff-Schw}
\eeq
Here $\rstar$ is a tortoise (Regge-Wheeler) coordinate, defined by 
\beq
\frac{d\rstar}{dr} = f^{-1}_S(r)
\quad \quad \Rightarrow \quad 
\rstar = r + 2M \ln( r/2M - 1 )  -  (3 M - 2M \ln 2 ) .  \label{rstar-defn}
\eeq
The outer region $r\in (2M,+\infty)$ of the Schwarzschild black hole is now covered by $\rstar\in(-\infty,+\infty)$.
Note that we have chosen the integration constant for our convenience so that, in the high-$l$ limit, the peak of the potential barrier (at $r=3M$) coincides with $\rstar = 0$.

\subsection{P\"{o}schl-Teller Potential and Nariai Spacetime}
\begin{figure}
 \begin{center}
  \includegraphics[width=8cm]{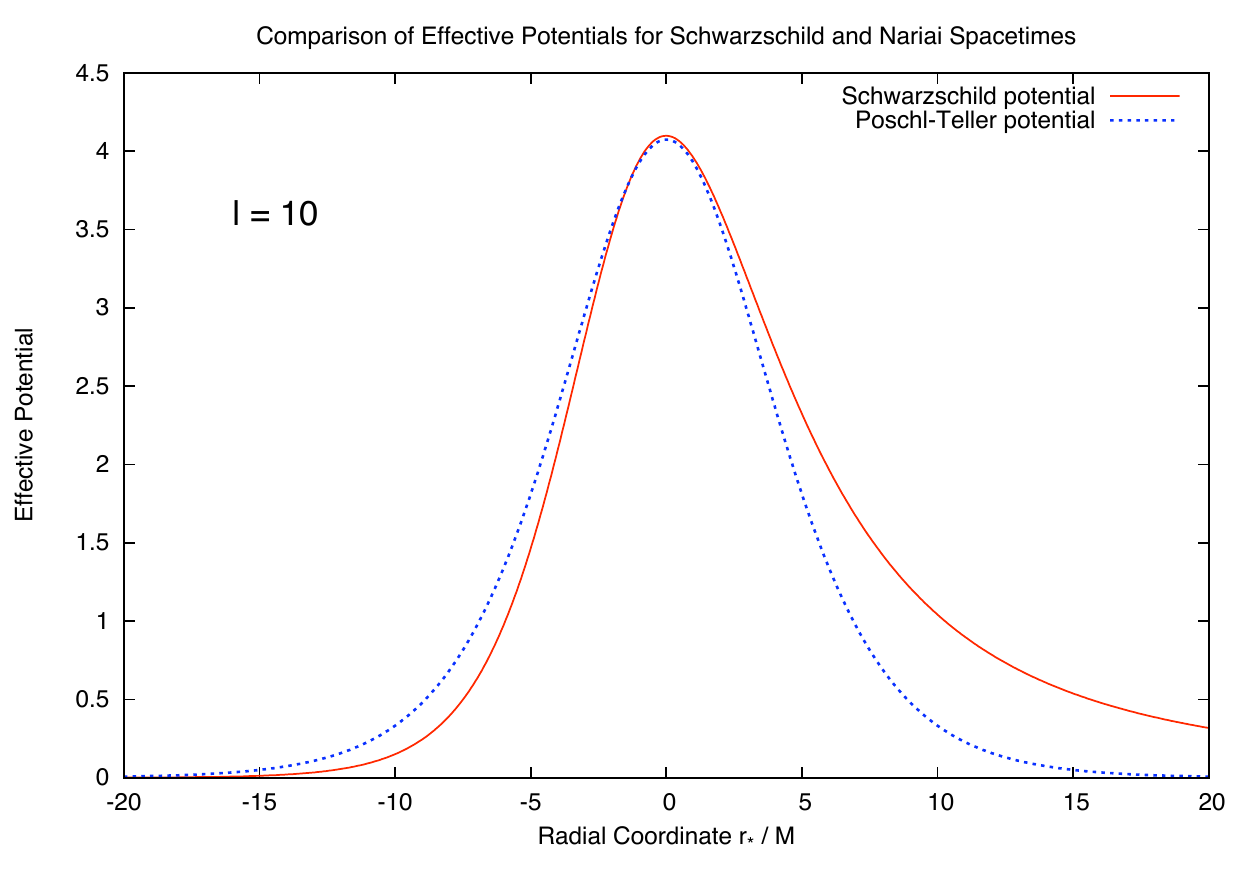}
 \end{center}
 \caption{\emph{Effective Potentials for Schwarzschild (\ref{Veff-Schw}) and P\"{o}schl-Teller (\ref{VPT}) radial wave equations}. Note that here the Schwarzschild tortoise coordinate is defined in (\ref{rstar-defn}) so that the peak is near $\rstar = 0$. The constants in (\ref{VPT}) are $V_0 = l(l+1)$, $\rstar^{(0)} = 0$ and $\alp = 1/(\sqrt{27}M)$.}
 \label{fig:Veff}
\end{figure}

Unfortunately, to the best of our knowledge, closed-form solutions to (\ref{rad-eq-schw}) with potential (\ref{Veff-Schw}) are not known. However, there is a closely-related potential for which exact solutions are available: the so-called P\"{o}schl-Teller potential~\cite{Poschl:Teller:1933},
\beq
V_l^{(PT)}(\rstar) =  \frac{\alp^2 V_0}{\cosh^2(\alp (\rstar-\rstar^{(0)}))}  \label{VPT}
\eeq
where $\alp$, $V_0$ and $\rstar^{(0)}$ are constants ($V_0$ may depend on $l$). Unlike the Schwarzschild potential, the P\"oschl-Teller potential is symmetric about $\rstar^{(0)}$, and decays exponentially in the limit $\rstar \rightarrow \infty$. Yet, like the Schwarzschild potential it has single peak, and with appropriate choice of constants, the P\"{o}schl-Teller potential can be made to fit the Schwarzschild potential in the vicinity of this peak (see Fig. \ref{fig:Veff}). 
In the Schwarzschild spacetime, the peak of the potential barrier is associated with the unstable photon orbit at $r = 3M$. As mentioned in the previous section (see Fig.~\ref{fig:circ_orbits}), the photon orbit may lead to singularities in the `distant past' Green function, and in the integrand of (\ref{tail-integral}). Hence by building a toy model which includes an unstable null orbit, we hope to capture the essential features of the distant past Green function. Authors have found that the P\"{o}schl-Teller potential is a useful model for exploring (some of the) properties of the Schwarzschild solution, for example the quasinormal mode frequency spectrum \cite{Ferrari:Mashhoon:1984, Berti:Cardoso:2006}. In this work, we hope to gain some insight into the `Distant Past' integral on the Schwarzschild spacetime by using the exact wavefunctions for the P\"{o}schl-Teller potential, given later in Sec.~\ref{subsec:radial-solns}.


An obvious question follows: is there a spacetime on which the scalar field equation reduces to a radial wave equation with a P\"oschl-Teller potential? The answer turns out to be: yes \cite{Cardoso:Lemos:2003, Zerbini:Vanzo:2004}! The relevant spacetime was first introduced by Nariai in 1950 \cite{Nariai:1950, Nariai:1951}.

To show the correspondence explicitly, let us define the line element
\beq 
 ds^2 = -f(\R) dt_N^2 + f^{-1}(\R) d\R^2 + d\Omega^2_2,   \label{Nariai-le}
\eeq
where $f(\R) = 1 - \R^2$ and $\R \in (-1,+1)$. Line element (\ref{Nariai-le}) describes the central diamond of the Penrose diagram of the Nariai spacetime (Fig. \ref{fig:penrose}), which is described more fully in Sec.~\ref{subsec:nariai}.  Consider the wave equation (\ref{scalar-field-eq2}) on this spacetime.
We seek separable solutions of the form $\fld (x)= u^{(N)}_{l\omega_N}(\R) Y_{lm}(\theta, \phi) e^{-i \omega_N t_N}$, where the label `N' denotes `Nariai'. The radial function satisfies the equation
\beq
f(\R) \frac{d}{d \R} \left( f(\R) \frac{d u^{(N)}_{l\omega_N} }{d \R} \right)   +   \left( \omega_N^2 - f(\R) [ l (l+1) + \xi R] \right) u^{(N)}_{l\omega_N} (\R) = 0  \label{Nar-rad-eq1}
\eeq
where $\xi$ is the curvature coupling constant and $R = 4$ is the Ricci scalar. Now let us define a new tortoise coordinate in the usual way,
\beq
\frac{d \Rs}{d \R} = f^{-1}(\R) \quad \quad \Rightarrow \quad \Rs = \tanh^{-1} \R .   \label{eq:rhostar}
\eeq
Note that $f(\R) = \sech^2(\Rs)$ and the tortoise coordinate is in the range $\Rs \in (-\infty,+\infty)$. Hence radial equation (\ref{Nar-rad-eq1}) may be rewritten in P\"oschl-Teller form,
\beq
\left(   \frac{d^2 }{d \Rs^2} + \omega_N^2 - \frac{U_0}{\cosh^2 \Rs}  \right) u^{(N)}_{l\omega_N}(\Rs) =  0
\label{rad-eq-nar}
\eeq
where
$
U_0 =  l(l+1) + 4\xi   \label{U0-def}.
$
We take the point of view that, as well as being of interest in its own right, the Nariai spacetime can provide insight into the propagation of waves on the Schwarzschild spacetime. The closest analogy between the two spacetimes is found by making the associations
\beq
 \Rs   \rightleftharpoons  \alp \rstar  ,   \quad \quad t_N   \rightleftharpoons    \alp t_S   , \quad \quad \omega_N   \rightleftharpoons \omega_S / \alp \quad \quad \text{where} \quad \alp = 1/(\sqrt{27}M) .
\eeq
Fig.~\ref{fig:Veff} shows the corresponding match between the potential barriers $V_l^{(S)}(\rstar)$ and $V_l^{(PT)}(\Rs)$. In the following sections, we drop the label `N', so that $t \equiv t_N$ and $\omega = \omega_N$. 

The solutions of Eq. (\ref{rad-eq-nar}) are presented in Sec.~\ref{subsec:radial-solns}. First, though, we consider the Nariai spacetime in more detail.

\subsection{Nariai spacetime}\label{subsec:nariai}
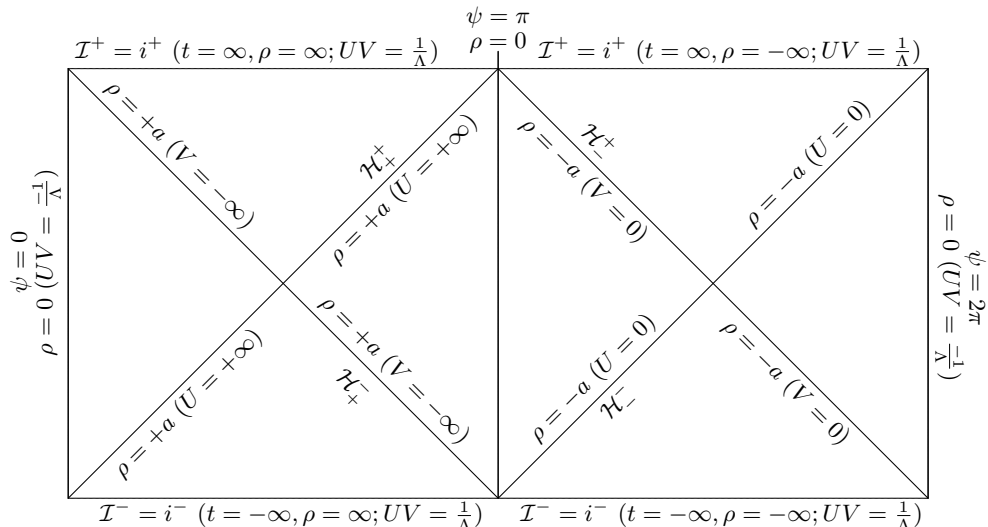
\begin{figure}[htb]
\setlength{\unitlength}{2.3pt}
\begin{center}
\begin{picture}(150,100)
\put(71,142){
\rotatebox{45}{\makebox(0,0){
\begin{picture}(250,250)
\put(25,25){\line(1,0){100}}
\put(-25,75){\line(1,0){100}}
\put(25,25){\line(0,1){100}}
\put(75,-25){\line(0,1){100}}
\qbezier[1000](75,-25)(100,0)(125,25) 
\qbezier[1000](-25,75)(0,100)(25,125) 
\qbezier[1000](25,25)(51,51)(77,77) 
\qbezier[1000](-25,75)(25,25)(75,-25) 
\qbezier[1000](25,125)(75,75)(125,25) 
\put(50,22){\makebox(0,0){$\mathcal{H}^-_-$}}
\put(50,28){\makebox(0,0){$\R=-a\ (U=0)$}}
\put(100,28){\makebox(0,0){$\R=-a\ (U=0)$}}
\put(50,78){\makebox(0,0){$\mathcal{H}^+_+$}}
\put(50,72){\makebox(0,0){$\R=+a\ (U=+\infty)$}}
\put(0,72){\makebox(0,0){$\R=+a\ (U=+\infty)$}}
\put(78,55){\rotatebox{-90}{\makebox(0,0){$\mathcal{H}^+_-$}}}
\put(72,52){\rotatebox{-90}{\makebox(0,0){$\R=-a\ (V=0)$}}}
\put(21,55){\rotatebox{-90}{\makebox(0,0){$\mathcal{H}^-_+$}}}
\put(28,52){\rotatebox{-90}{\makebox(0,0){$\R=+a\ (V=-\infty)$}}}
\put(72,5){\rotatebox{-90}{\makebox(0,0){$\R=-a\ (V=0)$}}}
\put(28,102){\rotatebox{-90}{\makebox(0,0){$\R=+a\ (V=-\infty)$}}}
\put(78,78){\rotatebox{-45}{\makebox(0,0){$\R=0$}}}
\put(81,81){\rotatebox{-45}{\makebox(0,0){$\psi=\pi$}}}
\put(0,105){\rotatebox{+45}{\makebox(0,0){$\R=0\ (UV=\frac{-1}{\Lambda})$}}}
\put(-3,108){\rotatebox{+45}{\makebox(0,0){$\psi=0$}}}
\put(102,-3){\rotatebox{-135}{\makebox(0,0){$\R=0\ (UV=\frac{-1}{\Lambda})$}}}
\put(105,-6){\rotatebox{-135}{\makebox(0,0){$\psi=2\pi$}}}
\put(49,105){\rotatebox{-45}{\makebox(0,0){$\mathcal{I}^+=i^+\ (t=\infty, \R=\infty; UV=\frac{1}{\Lambda})$}}}
\put(104,50){\rotatebox{-45}{\makebox(0,0){$\mathcal{I}^+=i^+\ (t=\infty, \R=-\infty; UV=\frac{1}{\Lambda})$}}}
\put(49,-3){\rotatebox{-45}{\makebox(0,0){$\mathcal{I}^-=i^-\ (t=-\infty, \R=-\infty; UV=\frac{1}{\Lambda})$}}}
\put(-1,47){\rotatebox{-45}{\makebox(0,0){$\mathcal{I}^-=i^-\ (t=-\infty, \R=\infty; UV=\frac{1}{\Lambda})$}}}
\end{picture}}}}
\end{picture}
\caption{\emph{Penrose diagram for the Nariai spacetime} in coordinates $(\psi,\zeta)$. 
The hypersurfaces $\psi=0$ and $\psi=2\pi$ are identified. 
Past/future timelike infinity $i^{-/+}$ coincides with past/future null infinity 
$\mathcal{I}^{-/+}$, and they are all spacelike hypersurfaces.
Thus, there exist observer-dependent past and future cosmological event-horizons, here marked as $\mathcal{H}^{\pm}_{\pm}$
for an observer along $\psi=\pi$.
}
\label{fig:penrose}
\end{center}
\end{figure}

The Nariai spacetime~\cite{Nariai:1950, Nariai:1951} may be constructed from an embedding in a 6-dimensional Minkowski space
\beq
ds^2=-dZ_0^2+\sum_{i=1}^5dZ_i^2
\eeq
of a 4-D surface determined by the two constraints,
\beq
-Z_0^2+Z_1^2+Z_2^2=a^2, \qquad Z_3^2+Z_4^2+Z_5^2=a^2, \qquad \text{where\ } a>0,
\eeq
corresponding to a hyperboloid and a sphere, respectively.
The entire manifold is covered by the coordinates $\{\mathcal{T},\psi,\theta,\phi\}$ defined via
\begin{align}
&Z_0=a\sinh\left(\frac{\mathcal{T}}{a}\right),\quad &Z_1&=a\cosh\left(\frac{\mathcal{T}}{a}\right)\cos\psi,\quad  &Z_2&=a\cosh\left(\frac{\mathcal{T}}{a}\right)\sin\psi,\quad \\
&Z_3=a\sin\theta\cos\phi,\quad  &Z_4&=a\sin\theta\sin\phi,\quad  & Z_5&=a\cos\theta,
\end{align}
with $\mathcal{T}\in(-\infty,+\infty), \psi\in [0,2\pi), \theta\in [0,\pi], \phi\in [0,2\pi)$. The line-element is given by
\beq
ds^2=-d\mathcal{T}^2+a^2\cosh^2\left(\frac{\mathcal{T}}{a}\right)d\psi^2+a^2d\Omega_2^2.
\eeq
From this line-element one can see that the spacetime has the following features:
(1) it has geometry $dS_2\times \mathbb{S}^2$ and topology $\mathbb{R}\times\mathbb{S}^1\times\mathbb{S}^2$ (the radius of the 1-sphere diminishes with time down
to a value $a$ at $\mathcal{T}=0$ and then increases monotonically with time $\mathcal{T}$,
whereas the 2-spheres have {\it constant} radius $a$),
(2) it is symmetric (ie, $R_{\mu\nu\rho\sigma;\tau}=0$), with $R_{\mu \nu} = \Lambda g_{\mu \nu}$, and constant Ricci scalar, $R = 4\Lambda$, 
where $\Lambda=1/a^2$ is the value of the cosmological constant,
(3) it is spherically symmetric (though not isotropic), homogeneous and locally (not globally) static,
(4) its conformal structure can be obtained by noting the Kruskal-like coordinates defined via 
$U= -(1-\Lambda UV)(Z_0+Z_1)/2,\ V= -(1-\Lambda UV)(Z_0-Z_1)/2$, for which the line-element is then
\beq
ds^2=-\frac{4dUdV}{\left(1-\Lambda UV\right)^2}+d\Omega_2^2.
\eeq
Its two-dimensional conformal {\it Penrose diagram} is shown in Fig.~\ref{fig:penrose} (see, e.g.,~\cite{Ortaggio:2002}),
where we have defined the conformal time $\zeta\equiv 2\exp\left(\mathcal{T}/a\right)\in(0,\pi)$.
Its Penrose diagram differs from that of de Sitter spacetime in that here each point represents a 2-sphere of {\it constant} radius;
note also that the corresponding angular coordinate $\psi$ in de Sitter spacetime has a different range, $\psi\in  [0,\pi)$,
as corresponds to its $\mathbb{R}\times\mathbb{S}^3$ topology.
Past and future timelike infinity $i^{\pm}$ coincide with past and future null infinity $\mathcal{I}^{\pm}$, respectively,
and they are all spacelike hypersurfaces. A consequence of the latter is the existence of 
{\it `past/future (cosmological) event horizons'}~\cite{Hawking:Ellis,Gibbons:Hawking:1977,Ortaggio:2002}:
not all events in the spacetime will be influentiable/observable by a geodesic observer;
the boundary of the future/past of the worldline of the observer is its past/future (cosmological) event horizon.

In this paper, we consider the {\it static} region of the Nariai spacetime which is covered by the coordinates $\{t,\R,\theta,\phi\}$, where 
$\R \equiv a\tanh (\Rs/a)\in(-a,+a),\ \Rs\equiv (v-u)/2\in (-\infty,+\infty),\ t\equiv (v+u)/2 \in (-\infty,+\infty)$ and the null coordinates $\{u,v\}$ are given via $U=ae^{-u/a},\ V=-ae^{v/a}$. 
This coordinate system, $\{t,\R,\theta,\phi\}$, covers the diamond-shaped region in the Penrose diagram (Fig.~\ref{fig:penrose}) around the hypersurface, say, $\psi=\pi$ 
(because of homogeneity we could choose any other $\psi=constant$ hypersurface).
We denote by $\mathcal{H^-_{\pm}}$ the past cosmological event horizon at $\R=\pm a$ of an observer moving along $\psi=\pi$;
similarly, $\mathcal{H^+_{\pm}}$ will denote its future cosmological event horizon at $\R=\pm a$.
Interestingly, Ginsparg and Perry~\cite{Ginsparg:Perry:1983} showed that this static region is obtained from the 
Schwarzschild-deSitter black hole spacetime
as a particular limiting procedure in which the event and cosmological horizons coincide 
(see also~\cite{Dias:Lemos:2003,Cardoso:Dias:Lemos:2004,Bousso:Hawking:1995,Bousso:Hawking:1996}).

Note that there are three hypersurfaces $\rho=0$, only two of which (those corresponding to $\psi=0$ and $2\pi$) are identified
(the one corresponding to $\psi=\pi$ is not).
Without loss of generality, we will take $\Lambda=1=a$.
The line-element corresponding to this static coordinate system is given in (\ref{Nariai-le}).

\subsection{Geodesics on Nariai spacetime}\label{subsec:geodesics}
Let us now consider geodesics on the Nariai spacetime. Our chief motivation is to find the orbiting geodesics, the analogous rays to those shown in Fig.~\ref{fig:circ_orbits} for the Schwarzschild spacetime. We wish to find the coordinate times $t - t'$ for which two angularly-separated points \emph{at the same `radius'}, $\R$, may be connected by a null geodesic. We expect the Green function to be singular at these times $t-t'$.

We will assume that particle motion takes place within the central diamond of the Penrose diagram in Fig.~\ref{fig:penrose}; that is, the region $-1 < \R < 1$ (notwithstanding the fact that timelike geodesics may pass through the future horizons $\mathcal{H}^+_+$ and $\mathcal{H}^+_-$ in finite proper time). Without loss of generality, let us consider motion in the equatorial plane $(\theta = \pi / 2)$ described by the world line $z^\mu(\lam) = [t(\lam), \R(\lam), \pi/2, \phi(\lam)]$ with tangent vector $u^\mu = [\dot{t}, \dot{\R}, 0, \dot{\phi}]$, where the overdot denotes differentiation with respect to an affine parameter $\lam$. Symmetry implies two constants of motion, $k = f(\R) \dot{t}$ and $h = \dot{\phi}$. The radial equation is $\dot{\R} = \pm H (\R^2 - \R_0^2)^{1/2}$ where $\R_0 = \sqrt{1 - k^2/H^2}$ is the closest approach point and $H^2 = h^2 + \kappa w^2$. Here, $w$ is the scaling of the affine parameter and $\kappa = +1$ for timelike geodesics, $\kappa = 0$ for null geodesics, and $\kappa = -1$ for spacelike geodesics. We still have the freedom to rescale the affine parameter, $\lam$, by choosing a value for $w$. It is conventional to rescale so that $\lam$ corresponds to proper time or distance, that is, set $w = 1$. Instead, we will rescale so that $\lam = \phi$, that is, we set $h=1$.

Let us consider a geodesic that starts at $\R = \R_1$, $\phi = 0$ which returns to `radius' $\R = \R_1$ after passing through an angle of $\phif$ (N.B. $\phif$ is unbounded, as opposed to $\gamma\in [0,\pi]$). The geodesic distance in this case is $s = -\kappa (H^2 - 1)^{1/2} \phif$. It is straightforward to show that
\beq
\R(\phi) = \R_1 \frac{\cosh \left(H\phi - H\phif/2\right)}{\cosh \left(H \phif / 2 \right)} ,  \label{yphi}
\eeq
hence 
\beq
\R_0 = \R_1 \sech(H \phif / 2)   \label{y0} .
\eeq
The coordinate time it takes to go from $\R=\R_1$, $\phi = 0$ to $\R=\R_1$, $\phi = \phif$ is
\beq
\Delta t_1 = 2 \rho_{*_1} + \ln \left( \frac{\R_1 - \R_0^2 + \sqrt{(1-\R_0^2)(\R_1^2-\R_0^2)}}{\R_1 + \R_0^2 - \sqrt{(1-\R_0^2)(\R_1^2 - \R_0^2)}} \right)
\label{tbar-def1}
\eeq
where $\rho_{*_1}  = \tanh^{-1}(\R_1)$.  Substituting (\ref{y0}) into (\ref{tbar-def1}) yields $\Delta t_1$ as a function of the angle $\phif$,
\beq
\Delta t_1 = 2 \rho_{*_1} + \ln \left( \frac{1-\R_1 \sech^2(H \phif/2) + \tanh(H \phif/2) \sqrt{1 - \R_1^2 \sech^2(H \phif/2)}}{ 1+ \R_1 \sech^2(H \phif/2) - \tanh(H \phif/2) \sqrt{1 - \R_1^2 \sech^2(H \phif/2) }}  \right)  \label{geodesic-time}
\eeq
This takes a particularly simple form as $\R_1 \rightarrow 1$,
\beq
\Delta t_1 \sim 2 \rho_{*_1} + \ln\left( \sinh^2( H \phif / 2 ) \right),\quad\text{for}\quad\ \R_1 \rightarrow 1.  \label{sing-time-inf}
\eeq
As $\phif \rightarrow \infty$, the geodesic coordinate time increases linearly with the orbital angle $\phif$
\beq
\Delta t_1 \sim 2 \rho_{*_1} +  H \phif ,\quad \text{for}\quad\phif \rightarrow \infty, \R_1 \rightarrow 1.
\label{sing-time-periodic}
\eeq
In other words, for fixed spatial points near $\R= 1$, the geodesic coordinate times $\Delta t_1$ are very nearly periodic, with period $2 \pi H$. Results (\ref{yphi}), (\ref{geodesic-time}), (\ref{sing-time-inf}) and (\ref{sing-time-periodic}) will prove useful when we come to consider the singularities of the Green function in Secs. \ref{subsec:Poisson} and \ref{subsec:Hadamard}.

\section{The Scalar Green Function}\label{sec:scalarGF}



\subsection{Retarded Green function as a Mode Sum} \label{subsec:FG}

The retarded Green function for a scalar field on the Nariai spacetime is defined by Eq.~(\ref{eq:gf-waveeq}), together with appropriate causality conditions. As described in Sec.~\ref{sec:matched-expansions} the Green function may be defined through an integral transform and a mode sum, 
\beq
\Gret(t, \Rs; t^\prime, \Rs^\prime; \gam) = 
\frac{1}{2\pi}\int_{-\infty+ic}^{+\infty+ic} d\omega\sum_{l=0}^{+\infty} \tilde{g}_{l\omega}(\Rs, \Rs')
(2l+1)P_l(\cos\gamma)e^{-i\omega (t-t')}   \label{eq:Gret:mode-sum}
\eeq
where $\Rs$ and $t$ are the `tortoise' and `time' coordinates in the line element (\ref{Nariai-le}), $c$ is a positive real constant, $t - t^\prime$ is the coordinate time difference, and $\gam$ is the spatial angle separating the points. 
The remaining ingredient in this formulation is the one-dimensional (radial) Green function $\Grad_{l\omega}(\R, \R^\prime)$ which satisfies
\beq
\left[ \frac{d^2}{d \Rs^2} + \omega^2 - \frac{U_0}{\cosh^2 \Rs} \right] \Grad_{l\omega}(\Rs, \Rs^\prime) = - \delta( \Rs - \Rs^\prime )  \label{rad-gf}
\eeq
The radial Green function may be constructed from two linearly-independent solutions of the radial equation (\ref{rad-eq-nar}).
To ensure a \emph{retarded} Green function we apply causal boundary conditions: no flux may emerge from the past horizons $\mathcal{H}^-_-$ and $\mathcal{H}^-_+$ (see Fig.~\ref{fig:penrose}). To this end, we will employ a pair of solutions denoted $u_{l \omega}^{\text{in}}$ and $u_{l \omega}^{\text{up}}$, in analogy with the Schwarzschild case. These solutions are defined in the next subsection.

\subsubsection{Radial Solutions}\label{subsec:radial-solns}
The homogeneous radial equation (\ref{rad-eq-nar}) may be rewritten as the Legendre differential equation
\beq
\frac{d}{d\R} \left( (1-\R^2) \frac{d u_{l\omega} }{d \R}  \right) +  \left( \beta (\beta + 1) - \frac{\mu^2}{1 - \R^2} \right) u_{l\omega}= 0  \label{legendre-de}
\eeq
where
\begin{eqnarray}
\mu = \pm i \omega, \quad \quad \beta = -1/2 + i \lam ,  \label{mu-nu-def} \\
 \lambda = \pm \sqrt{(l+1/2)^2 + d},   \quad \quad d = 4\xi - 1/2  .  \label{lam-def}
\end{eqnarray}
We choose $\mu = i \omega$, $\lam = \sqrt{(l+1/2)^2 + d}$ and note that the choice of signs will not have a bearing on the result.
The value of the constant $\xi$ in the conformally-coupled case in a $D$-dimensional spacetime is: $(D-2)/(4(D-1))$. 
Note that for conformal coupling in 4-D ($\xi = 1/6$) the constant is $d = 1/6$, and for minimal coupling ($\xi = 0$) we have $d = -1/2$. For the special value $\xi = 1/8$ we have $d = 0$. The possible significance of the value $\xi = 1/8$, the conformal coupling factor in three dimensions, was recently noted in a study of the self-force on wormhole spacetimes \cite{Bezerra:2009}. 

The solutions of Eq.~(\ref{legendre-de}) are Legendre functions of complex order, which are defined in terms of hypergeometric functions as follows (Ref.~\cite{GradRyz} Eq.~(8.771)),
\beq
P^{\mu}_\beta(\R) = \frac{1}{\Gamma(1-\mu)} \left( \frac{1+\R}{1-\R} \right)^{\mu/2} \, {}_2 F_1 \left(-\beta, \beta+1; 1 - \mu ; \frac{1-\R}{2} \right)   .  \label{LegP-def}
\eeq
In the particular case $\mu = 0$, the solutions belong to the class of conical functions (Ref.~\cite{GradRyz}  Eq.~(8.84)). We define the pair of linearly-independent solutions to be
\begin{eqnarray}
\uin(\R) &=& \Gamma(1 - \mu) P_{\beta}^\mu (-\R),  \label{uin-def} \\
\uup(\R) &=& \Gamma(1 - \mu) P_{\beta}^\mu (\R).   \label{uup-def}
\end{eqnarray}
These solutions are labelled ``in'' and ``up'' because they obey analogous boundary conditions to the ``ingoing at horizon'' and ``outgoing at infinity'' solutions that are causally appropriate in the Schwarzschild case \cite{Andersson:1997}. 
It is straightforward to verify that the ``in'' and ``up'' solutions obey
\begin{eqnarray}
\uin  &\sim& e^{-i \omega \Rs}  \quad \quad \text{as} \quad \Rs \rightarrow -\infty \\ 
\uup &\sim& e^{+i \omega \Rs}  \quad \quad \text{as} \quad \Rs \rightarrow +\infty
\end{eqnarray}
To find the asymptotes of $\uin$ near $\R = 1$, we may employ the series expansion
\begin{eqnarray}
{}_2 F_1 (a, b; c; z) &=& \frac{\Gamma(c) \Gamma(a+b-c)}{\Gamma(a) \Gamma(b)} (1-z)^{c-a-b} \left[ \sum_{k=0}^\infty \frac{(c-a)_k(c-b)_k}{(c+1-a-b)_k} \frac{(1-z)^k}{k!} \right]  \nn \\
 && + \, \frac{\Gamma(c) \Gamma(c-a-b)}{\Gamma(c-a) \Gamma(c - b)} \left[ \sum_{k=0}^\infty \frac{a_k b_k}{(1+a+b-c)_k} \frac{(1-z)^k}{k!}  \right]  \label{2F1-power-series}
\end{eqnarray}
where $(z)_k \equiv \Gamma(z + k) / \Gamma(z)$ is the Pochhammer symbol. In our case $a = -\beta$, $b = \beta + 1$, $c = 1 - \mu$ and $1 - z = (1 - \R) / 2$. It is straightforward to show that
\beq
\uin (\Rs) \sim \left \{  \begin{array}{ll} 
e^{-i \omega \Rs}, & \Rs \rightarrow -\infty ,  \\
\Aout_{l \omega} e^{i \omega \Rs} + \Ain_{l \omega} e^{-i \omega \Rs},
 \quad & \Rs \rightarrow + \infty ,
 \end{array} \right.    \label{bc-in}
\eeq
where
\begin{eqnarray}
 \Ain_{l\omega}  &=&  \frac{ \Gamma(1 - i \omega ) \Gamma(- i \omega) }{ \Gamma(1 + \beta - i\omega) \Gamma( - \beta - i \omega) } , \label{Ain-defn} \\
 \Aout_{l \omega}  &=& \frac{ \Gamma(1 - i\omega) \Gamma(i\omega)}{ \Gamma(1 + \beta) \Gamma(- \beta) } ,  \label{Aout-defn}
\end{eqnarray}
with $\beta$ as defined in (\ref{mu-nu-def}).
The ``up'' solution is found from the ``in'' solution via spatial inversion $\R \rightarrow - \R$; hence
\beq
\uup (\Rs) \sim \left \{  \begin{array}{ll} 
\Aout_{l \omega} e^{- i \omega \Rs} + \Ain_{l \omega} e^{i \omega \Rs}, \quad \quad & \Rs \rightarrow -\infty,  \\
 \quad  e^{i \omega \Rs}, & \Rs \rightarrow + \infty .
 \end{array} \right.    \label{bc-up}
\eeq


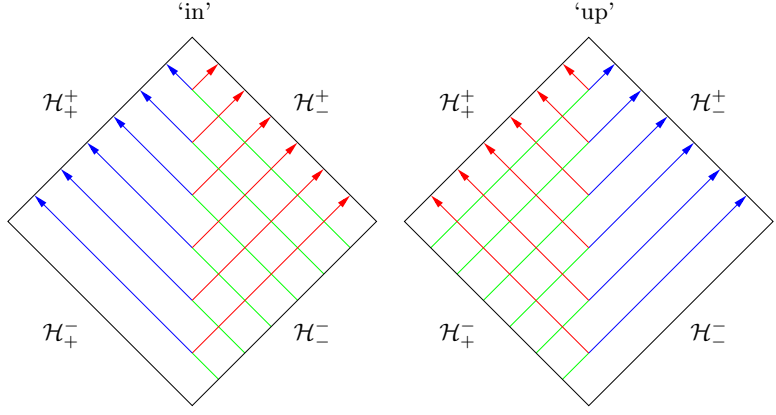
\begin{figure}

\hspace{-13cm}
\begin{picture}(0,150)
\put(98,163){\makebox(0,0)[tl]{`in'}}
\put(47,43){\makebox(0,0)[tl]{$\mathcal{H}_+^-$}}
\put(47,130){\makebox(0,0)[tl]{$\mathcal{H}_+^+$}}
\put(142,43){\makebox(0,0)[tl]{$\mathcal{H}_-^-$}}
\put(142,130){\makebox(0,0)[tl]{$\mathcal{H}_-^+$}}
\put(30,10){
\includegraphics[width=35mm,angle=45]{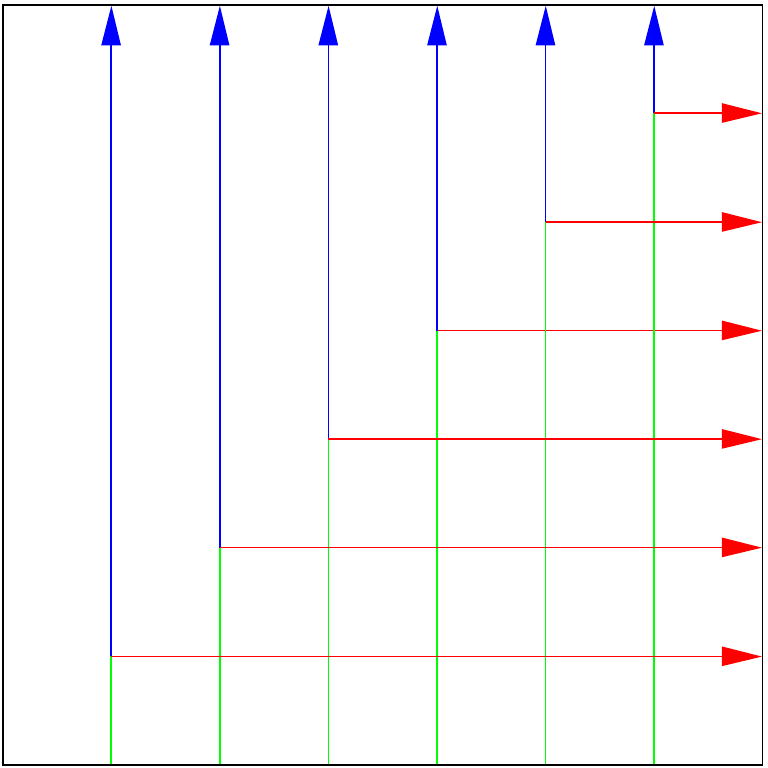}
}
\put(248,163){\makebox(0,0)[tl]{`up'}}
\put(197,43){\makebox(0,0)[tl]{$\mathcal{H}_+^-$}}
\put(197,130){\makebox(0,0)[tl]{$\mathcal{H}_+^+$}}
\put(292,43){\makebox(0,0)[tl]{$\mathcal{H}_-^-$}}
\put(292,130){\makebox(0,0)[tl]{$\mathcal{H}_-^+$}}
\put(180,10){
\includegraphics[width=35mm,angle=45]{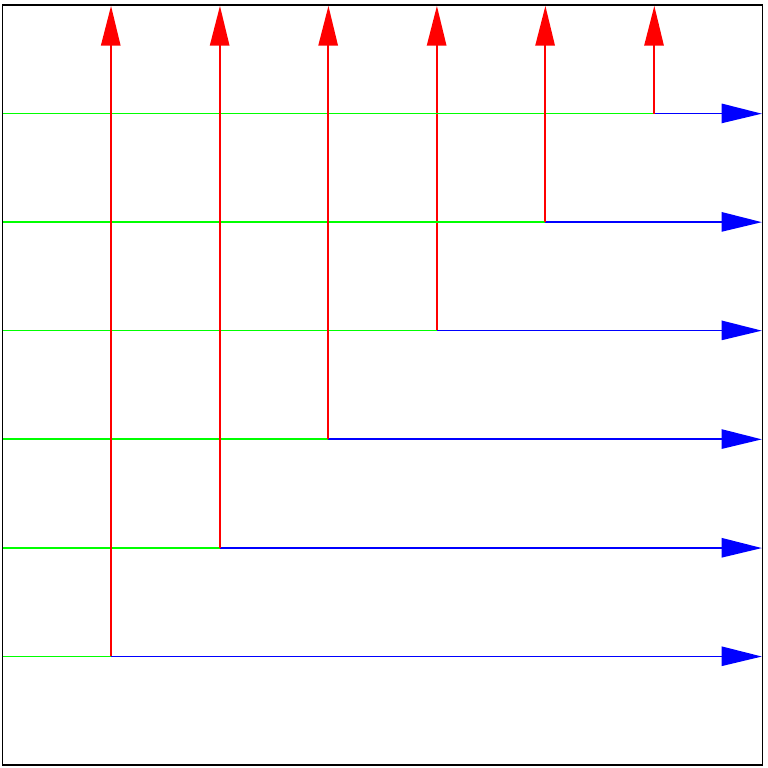}}
\end{picture}

\caption{\emph{Penrose diagrams for IN and UP radial solutions}.}
 \label{fig:inup}
\end{figure}

The Wronskian $W$ of the two linearly-independent solutions $u_{l \omega}^{\text{in}}(\Rs)$ and $u_{l \omega}^{\text{up}}(\Rs)$ can be easily obtained:
\beq
W = u_{l \omega}^{\text{in}}(\Rs) \frac{d u_{l \omega}^{\text{up}}}{d \Rs} - u_{l \omega}^{\text{up}}(\Rs) \frac{d u_{l \omega}^{\text{in}}}{d \Rs} = 2 i \omega A_{l \omega}^{\text{(in)}} 
 \label{Wronskian}
\eeq
The one-dimensional Green function $\Grad_{l \omega}(\Rs, \Rs^\prime)$ is then given by
 \begin{eqnarray}
\Grad_{l \omega}(\Rs, \Rs^\prime) &=& - \frac{1}{W} \left\{   \begin{array}{ll} u_{l \omega}^{\text{in}}(\Rs)   u_{l \omega}^{\text{up}}(\Rs^\prime) , \quad \quad & \Rs < \Rs^\prime , \\  u_{l \omega}^{\text{up}}(\Rs) u_{l \omega}^{\text{in}}(\Rs^\prime),  & \Rs > \Rs^\prime , \end{array} \right. \nonumber \\
 &=&
 \tfrac{1}{2} \Gamma(1+\beta-\mu) \Gamma(-\beta - \mu) P_{\beta}^{\mu}(-\R_<)P_{\beta}^{\mu}(\R_>)
, \label{eq:g-radial} \quad 
\end{eqnarray}
where $\R_<\equiv \text{min}(\R,\R')$ and $\R_>\equiv \text{max}(\R, \R')$.
The four-dimensional retarded Green function can thus be written as 
 \begin{align}  \label{full ret Green,mode sum}
\Gret (x,x')=
\frac{1}{4 \pi }
\sum_{l=0}^{\infty} & (2l+1)P_l(\cos\gamma)  \nonumber \\
&\times \int_{-\infty + ic}^{+\infty + ic}d\omega e^{-i\omega (t-t')}
\Gamma\left(\frac{1}{2}+ i \lambda -i\omega \right) \Gamma\left(\frac{1}{2} - i \lambda - i \omega \right)
P_{ -1/2 + i \lam}^{i \omega}(-\R_<)P_{ -1/2 + i \lam}^{i \omega}(\R_>)
\end{align}

 \subsection{Distant Past Green Function: The Quasinormal Mode Sum} \label{subsec:DP}

\begin{figure}
 \begin{center}
  \includegraphics[width=8cm]{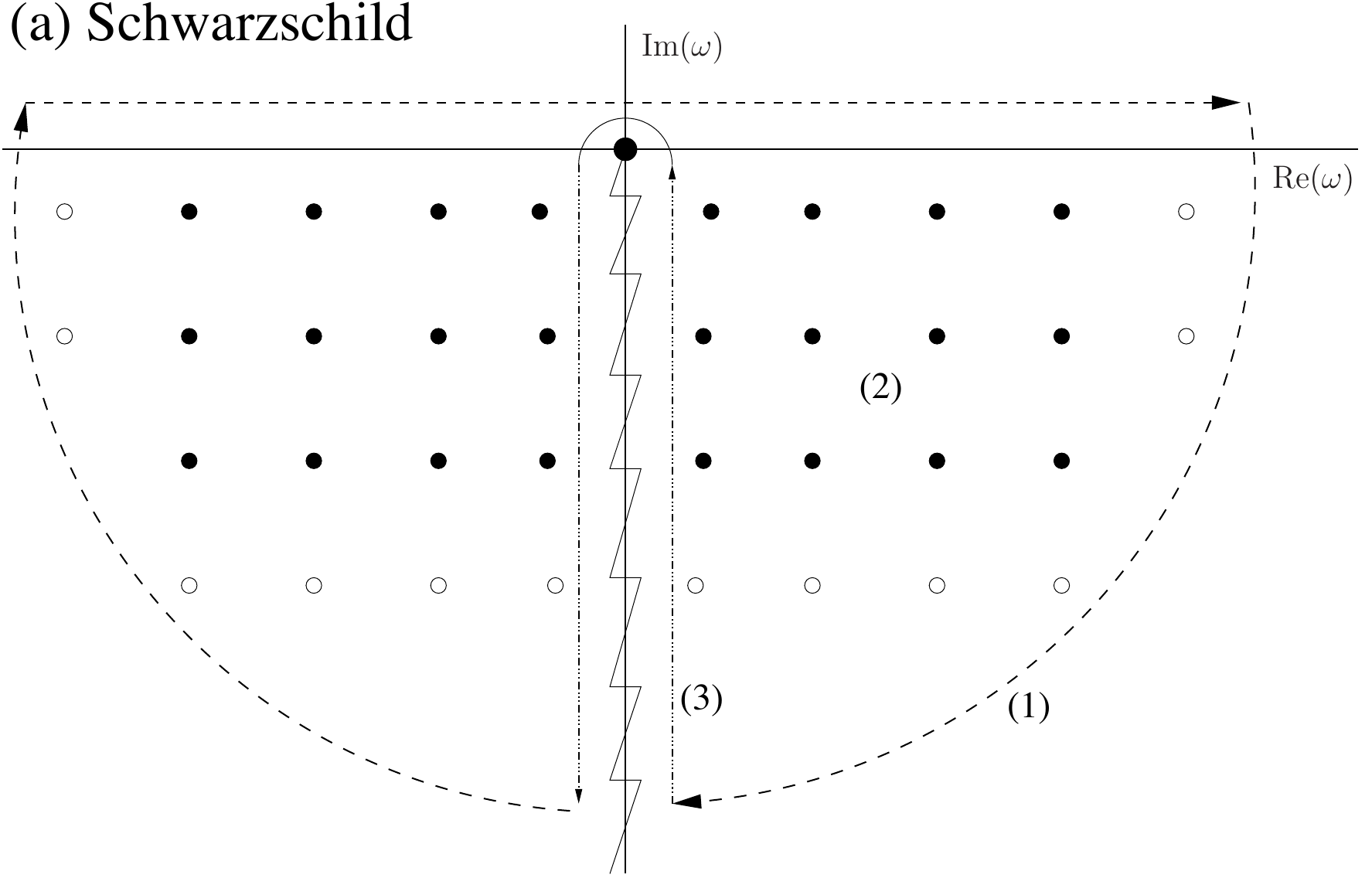}
  \includegraphics[width=8cm]{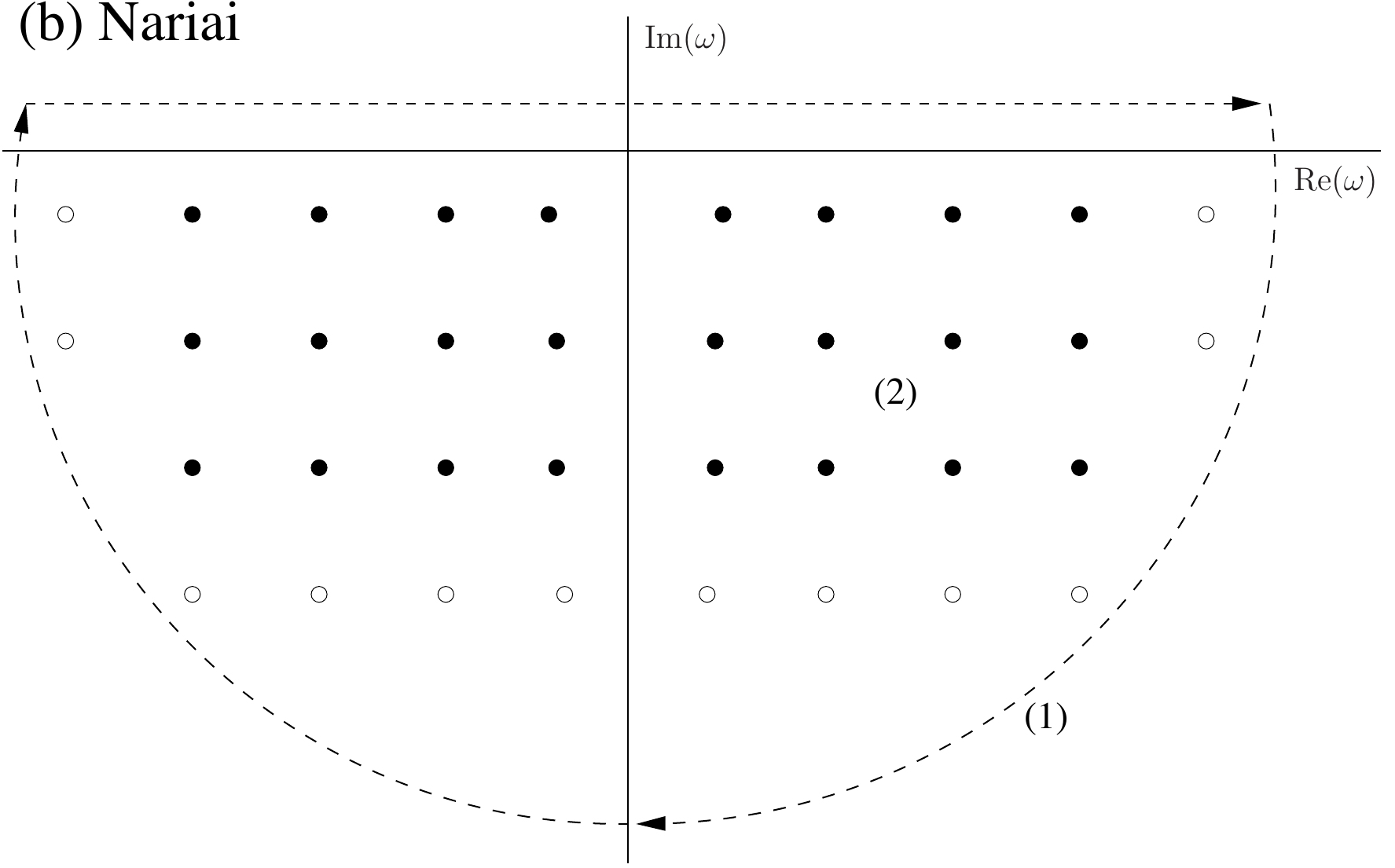}
 \end{center}
 \caption{\emph{Contour Integrals}. These plots show the deformation of the integral over frequency to include the poles of the Green function (quasinormal modes), for two spacetimes: (a) Schwarzschild [left] and (b) Nariai [right].  
}
 \label{fig:contours}
\end{figure}

As discussed in Sec.~\ref{sec:matched-expansions}, the integral over frequency in Eq.~(\ref{eq:Gret:mode-sum}) may be evaluated by deforming the contour in the complex plane \cite{Leaver:1986, Andersson:1997}. The deformation is shown in Fig.~\ref{fig:contours}. The left plot (a) shows the Schwarzschild case, and the right plot (b) shows the Nariai case. 

On the Schwarzschild spacetime, it is well-known that a `power-law tail' arises from the frequency integral along a branch cut along the (negative) imaginary axis (Fig. \ref{fig:contours}, part (3)). In the Schwarzschild case, the branch cut is necessary due to a branch point in $\Grad_{l \omega}(r,r^\prime)$ at $\omega = 0$ \cite{Hartle:Wilkins:1974, Leaver:1986}. In contrast, for the Nariai case with $\xi > 0$, the Wronskian (\ref{Wronskian}) is well-defined and non-zero in the limit $\omega \rightarrow 0$. For minimal coupling ($\xi = 0$), we find that $\omega = 0$ is a simple pole of the Green function. In either case, $\omega =0$ is not a branch point and hence power law decay does not arise on the Nariai spacetime.

The simple poles of the Green function (shown as dots in Fig.~\ref{fig:contours}) occur in the lower half-plane of the complex frequency plane. The poles correspond to the zeros of the Wronskian (\ref{Wronskian}). The Wronskian is zero when the ``in'' and ``up'' solutions are linearly-dependent. This occurs at a discrete set of (complex) \emph{Quasinormal Mode} (QNM) frequencies $\omega_q$. Beyer \cite{Beyer:1999} has shown that, for the P\"oschl-Teller potential, the corresponding QNM radial solutions form a \emph{complete basis} at sufficiently late times ($t > t_c$, to be defined below). Completeness means that \emph{any} wavefunction obeying the correct boundary conditions at $\Rs \to \pm \infty$ can be represented as a sum over quasinormal modes, to arbitrary precision. Intuitively, we may expect this to mean that, at sufficiently late times, the Green function itself can be written as a sum over the residues of the poles.

\subsubsection{Quasinormal Modes}
Quasinormal Modes (QNMs) are solutions to the radial wave equation (\ref{rad-eq-nar})
which are left-going ($e^{-i \omega \Rs}$) at $\Rs \rightarrow -\infty$ and right-going ($e^{+ i \omega \Rs}$) as $\Rs \rightarrow +\infty$ . QNMs occur at discrete complex frequencies $\omega = \omega_q$ for which $\Ain_{l \omega_q} = 0$. At QNM frequencies, the ``in'' and ``up'' solutions ($u_{l \omega_q}^{\text{in}}$ and $u_{l \omega_q}^{\text{up}}$) are linearly-dependent and the Wronskian 
(\ref{Wronskian}) is zero. 

The QNMs of the Schwarzschild black hole have been studied in much detail \cite{Leaver:1985, Nollert:1999, Kokkotas:Schmidt:1999, Cardoso:Witek:2008}. QNM frequencies are complex, with the real part corresponding to oscillation frequency, and the (negative) imaginary part corresponding to damping rate. QNM frequencies $ \omega_{ln}$ are labelled by two integers: $l$, the angular momentum, and $n = 0,1, \ldots \infty$, the \emph{overtone number}. For every multipole $l$, there are an infinite number of overtones. In the asymptotic limit $l \gg n$, the Schwarzschild QNM frequencies approach \cite{Ferrari:Mashhoon:1984, Iyer:1987, Konoplya:2003}
\beq
M \omega_{ln}^{(S)} \approx \frac{1}{\sqrt{27}} \left[ \pm(l+1/2) - i (n+1/2) \right]
\eeq
In general, the damping increases with $n$. The $n=0$ (`fundamental') modes are the least-damped.

The quasinormal modes of the Nariai spacetime are found from the condition $\Ain_{l \omega_{ln}} = 0$. Using (\ref{Ain-defn}), we find
\begin{equation}
1 + \beta - i \omega_{ln} = -n \quad \quad \text{or} \quad \quad -\beta - i\omega_{ln} = -n
\end{equation}
where $n$ is a non-negative integer. These conditions lead to the QNM frequencies
\begin{equation}
\omega_{ln} =  - \lambda - i (n+1/2) ,
\label{QNM freq}
\end{equation}
where $\lam$ is defined in (\ref{lam-def}). (Note we have chosen the sign of the real part of frequency here for consistency with previous studies \cite{Ferrari:Mashhoon:1984, Berti:Cardoso:2006} which use $\sig = -\omega$ as the frequency variable.)

\subsubsection{The Quasinormal Mode Sum}
The quasinormal mode is constructed from (\ref{modesum1}) by taking the sum over the residues of the poles of $\Grad_{l\omega}(\R,\R^\prime)$ in the complex-$\omega$ plane. Applying Leaver's analysis \cite{Leaver:1986} to (\ref{modesum1}), with the radial Green function (\ref{eq:g-radial}), we obtain 
\beq
\GQNM(t, \R; t^\prime, \R^\prime; \gam)  = 2 \, \text{Re} \sum_{n=0}^{\infty} \sum_{l=0}^\infty (2l+1) P_l(\cos \gam) \Bef_{ln} \tilde{u}_{ln}(\R) \tilde{u}_{ln}(\R^\prime) e^{ - i \omega_{ln} (t - t^\prime - \Rs - \Rs^\prime)}
\label{G-QNM}
\eeq
where the sum is taken over \emph{either} the third or fourth quadrant of the frequency plane only. Here, $\R$ and $t$ are the coordinates in line element (\ref{Nariai-le}), $\Rs$ is defined in (\ref{eq:rhostar}), $\omega_{ln}$ are the QNM frequencies and $\Bef_{ln}$ are the \emph{excitation factors}, defined as
\beq
\Bef_{ln} \equiv \frac{\Aout_{l \omega_{ln}}}{2 \omega_{ln} \left.\frac{d \Ain_{l \omega}}{d \omega}\right|_{\omega_{ln}} } , \label{excitation-factors}
\eeq
and $\unorm_{ln}(\R)$ are the QNM radial functions, defined by
\beq
\unorm_{ln}(\Rs) = \frac{u^{\text{in}}_{l \omega_{ln}} (\Rs)}{\Aout_{l \omega_{ln}} e^{i \omega_{ln} \Rs}} .   \label{unorm-def}
\eeq
The QNM radial functions are normalised so that $\unorm_{ln}(\R) \rightarrow 1$ as $\R \rightarrow 1$ ($\Rs \rightarrow \infty$).

The QNM frequencies have a negative imaginary part; hence the exponentials in (\ref{G-QNM}) diverge with $n$ at `early' times $t - t' < t_c$ and $t_c = \Rs + \Rs^\prime$. The exponentials converge with $n$ at `late' times $t - t' > t_c$. Beyer \cite{Beyer:1999} has shown that, for late times $t - t' > t_c$, QNMs form a \emph{complete basis}. Physically, for the QNM sum to be appropriate, sufficient coordinate time must elapse for a light ray to propagate inwards from $\R^\prime$, reflect off the potential barrier near $\R = 0$, and propagate outwards again to $\R$. 

The excitation factors, defined in (\ref{excitation-factors}), may be shown to be
\begin{eqnarray}
\mathcal{B}_{ln} &=&
\frac{1}{2 \, n !}  \frac{ \Gamma(n + 1 - 2i\omega_{ln} ) }{ [ \Gamma(1 - i \omega_{ln} ) ]^2} , \nonumber \\
&=& \frac{1}{2 \, n !} \frac{\Gamma(-n + 2 i \lam ) }{[\Gamma(-n+1/2 + i \lam)]^2} . \label{EF2}
\end{eqnarray}
The steps in the derivation are given in Appendix A.3 of \cite{Berti:Cardoso:2006}.

The Green function (\ref{G-QNM}) now takes the form of a double infinite sum, taken over both angular momentum $l$ and overtone number $n$. The convergence of the sum over $l$ is by no means guaranteed. For example, the magnitude of $\Bef_{ln}$ is proportional to $(l+1/2)^{n-1/2}$ in the large-$l$ limit, for fixed $n$. Hence, the magnitude of each term in the series grows with $l$. Nevertheless, we will show that well-defined and meaningful values can be extracted from the sums.

\subsubsection{Green Function at Arbitrary `Radii'}
Unfortunately, it is not straightforward to perform the sum over $n$ explicitly for general values of $\R$ and $\R'$. Instead we must include the QNM radial functions $\unorm_{ln}(\Rs)$, defined by (\ref{uin-def}) and (\ref{unorm-def}). We note that the Legendre function appearing in (\ref{uin-def}) can be expressed in terms of a hypergeometric function (see Eq.~(\ref{LegP-def})), and 
the hypergeometric function can be written as power series about $\R = 1$ (Eq.~\ref{2F1-power-series}). 
At quasinormal mode frequencies, the second term in Eq. (\ref{2F1-power-series}) is zero, and hence the wavefunction is purely outgoing at infinity, as expected. Combining results (\ref{LegP-def}), (\ref{uin-def}), (\ref{2F1-power-series}) and (\ref{unorm-def}) we find the normalised wavefunctions to be 
\beq
\unorm_{ln}(\R) = \left( \frac{2}{1+\R} \right)^{i \omega_{ln}} \mathcal{S}_{ln}(\R) ,   \label{unorm-ser}
\eeq
where $\R$ is the radial coordinate in line element (\ref{Nariai-le}) and $\mathcal{S}_{ln}$ is a finite series with $n$ terms,
\beq
\mathcal{S}_{ln}(\R) =  \sum_{k=0}^n \frac{1}{k!} \frac{ (- n + 2i\lambda)_k (-n)_k }{ (-n+1/2 + i\lambda)_k } \left( \frac{1-\R}{2} \right)^{k}  =  {}_2F_{1}\left(-n+2 i \lambda, -n; -n + \frac{1}{2} + i \lambda; \frac{1-\R}{2} \right)  ,
\eeq
where we have adopted the sign convention $\omega_{ln} =  -\lam - i(n+1/2)$ of (\ref{QNM freq}).
Hence the Green function at arbitrary `radii' may be written as the double sum,
\begin{eqnarray}
\GQNM(x,x^{\prime}) &=& 2 \, \text{Re} \sum_{l=0}^\infty (2l+1)  P_l(\cos \gam)  e^{ i \lam \left[ T - \ln(2/(1+\R)) - \ln(2/(1+\R^\prime)) \right] } \nn \\ && \quad \quad \quad \times \sum_{n=0}^{\infty} \Bef_{ln} \left[ \frac{4 e^{-T} }{ (1+\R)(1+\R^\prime) } \right]^{n+1/2} \mathcal{S}_{ln}(\R) \mathcal{S}_{ln}(\R^\prime)   \label{G-arbitrary}
\end{eqnarray}

\subsubsection{Green Function near Spatial Infinity, $\Rs, \Rs^\prime \rightarrow +\infty$}
In the limit that both radial coordinates $\Rs$ and $\Rs^\prime$ tend to infinity ($\R,\R' \rightarrow 1$), the QNM sum (\ref{G-QNM}) may be rewritten
\beq
\lim_{\Rs,\Rs^\prime \rightarrow +\infty} \GQNM(T, \gam) = 2 \, \text{Re} \sum_{l=0}^\infty (2l+1) P_l(\cos \gam)  e^{ i \lam T} \sum_{n=0}^{\infty} \Bef_{ln} e^{- (n+1/2) T}
\eeq
where 
\beq
T \equiv t - t^\prime - \Rs - \Rs^\prime   \label{T-def}
\eeq
and $\lam$ was defined in (\ref{lam-def}). Using the expression for the excitation factors (\ref{EF2}), the sum over $n$ can be evaluated explicitly, as follows,
\begin{eqnarray}
\sum_{n=0}^{\infty} \Bef_{ln} e^{- (n+1/2) T} &=& \frac{z^{1/2}}{2}  \sum_{n=0}^\infty \frac{\Gamma(-n + 2 i \lam)}{[\Gamma(-n + 1/2 + i \lam)]^2} \frac{z^n}{n!} \nonumber \\
&=&  \frac{z^{1/2}}{2} \frac{\Gamma(2i \lam)}{[\Gamma(1/2 + i\lam)]^2} \sum_{n=0}^\infty \frac{(2 i \lam)_{-n}}{[(1/2 + i \lam)_{-n}]^2} \frac{z^n}{n!}
\end{eqnarray}
where $z = e^{- T}$ and $(\cdot)_k$ is the Pochhammer symbol. Using the identity $(x)_{-n} = (-1)^n / (1-x)_n$ and the duplication formula $\Gamma(z) \Gamma(z+1/2) = 2^{1-2z} \sqrt{\pi} \, \Gamma(2z)$ we find
\begin{eqnarray}
\sum_{n=0}^{\infty} \Bef_{ln} e^{- (n+1/2) T} &=&  \frac{e^{- T / 2}}{4 \sqrt{\pi}}  \frac{2^{2 i \lam} \Gamma(i \lam)}{\Gamma(1/2 + i\lam)} \, {}_2F_1(-\beta, -\beta; -2\beta; -e^{-T})
\end{eqnarray}
where ${}_2F_1$ is the hypergeometric function, and $\beta$ was defined in (\ref{mu-nu-def}).
Hence the Green function near spatial infinity ($\R, \R^\prime \rightarrow 1$) is 
\beq
\lim_{\Rs,\Rs^\prime \rightarrow +\infty} \GQNM(T, \gam) = 
\frac{ e^{-T / 2} }{\sqrt{\pi}} \, \text{Re} \sum_{l=0}^\infty \frac{(l+1/2) \Gamma(i \lam)}{\Gamma(1/2 + i\lam)} P_l(\cos \gam) e^{i \lam (T + 2 \ln 2)} {}_2F_1(-\beta, -\beta; -2\beta; -e^{-T}).  \label{GF-inf}
\eeq

\subsubsection{Green Function Approximation from Fundamental Modes}\label{subsec:fundamental-modes}
Expression (\ref{G-arbitrary}) is complicated and difficult to analyse as it involves a double sum. It would be useful to have a simple approximate expression, with only a single sum, which captures the essence of the physics. At late times, we might expect that the Green function is dominated by the least-damped modes, that is, the $n=0$ fundamental quasinormal modes. If we discard the higher modes $n > 0$, we are left with an approximation to the Green function which does indeed seem to capture the essential features and singularity structure. However, as we show in section \ref{subsec:Poisson}, it does not correctly predict the singularity times. 

The $n=0$ approximation to the Green function is
\beq
\Gret^{(n=0)}(x, x^{\prime}) =  \left( \frac{e^{-T} }{\pi(1+\R)(1+\R^\prime)} \right)^{1/2} \sum_{l=0}^\infty \frac{ (2l+1) P_l(\cos \gam) \Gamma(i \lam)}{\Gamma(1/2 + i\lam)} e^{i \lam \left[ T + \ln(1+\R)(1+\R^\prime) \right]} \mathcal{S}_{l0}(\R) \mathcal{S}_{l0}(\R^\prime).  \label{G-n0}
\eeq
Note that the sum over $l$ is approximately periodic  in $T$ (exactly periodic in the case $\xi = 1/8$), with period $4 \pi$.

 \subsection{Quasilocal Green Function: Hadamard-WKB Expansion} \label{subsec:QL}

\label{sec:ql}
We now consider the Green function in the quasilocal region, which is needed for the calculation of the quasilocal contribution to the scalar self-force $\Phi_\mu^{\text{(QL)}}$ given in Eq. (\ref{eq:fld-QL-DP}).
When spacetime points $x$ and $x'$ are sufficiently close together (within a \emph{convex normal neighbourhood}), the retarded Green function may be expressed in the Hadamard parametrix \cite{Hadamard,Friedlander},
\begin{equation}
\label{eq:Hadamard}
G_{ret}\left( x,x' \right) = \theta_{-} \left( x,x' \right) \left\lbrace U \left( x,x' \right) \delta \left( \sigma \left( x,x' \right) \right) - V \left( x,x' \right) \theta \left( - \sigma \left( x,x' \right) \right) \right\rbrace ,
\end{equation}
where $\theta_{-} \left( x,x' \right)$ is analogous to the Heaviside step-function (unity when $x'$ is in the causal past of $x$, zero otherwise), $\delta \left( \sigma\left( x,x' \right) \right)$ is the standard Dirac delta function, $U \left( x,x' \right)$ and $V \left( x,x' \right)$ are symmetric bi-scalars having the benefit that they are regular for $x' \rightarrow x$, and $\sigma \left( x,x' \right)$ is the Synge \cite{Synge,Poisson:2003,DeWitt:1965} world function. Clearly, the term involving $U(x,x')$, the `direct' part, will not contribute to the quasilocal integral in Eq.~(\ref{eq:fld-QL-DP}) since it has support only on the light-cone, while the integral is internal to the light-cone. We will therefore only concern ourselves with the calculation of the function $V(x,x')$, the `tail' part, which has support inside the light-cone.

The fact that $x$ and $x'$ are close together suggests that an expansion of $V(x,x')$ in powers of the separation of the points may give a good representation of the function within the quasilocal region. In Ref.~\cite{QL} we use a WKB method (based on that of Refs.~\cite{Anderson:2003,Howard:1985,Winstanley:2007}) to derive such a coordinate expansion and we also give estimates of its range of validity. Referring to the results therein, we have $V(x,x')$ as a power series in $(t-t')$ and $(\cos \gamma - 1)$,
\begin{equation}
\label{eq:CoordGreen}
V\left( x,x' \right) = \sum_{i,j=0}^{+\infty}  v_{ij}(\R) ~ \left( t - t' \right)^{2i} \left( \cos \gamma - 1 \right)^j,
\end{equation}
where $\gamma$ is the angular separation of the points. In general this expression also includes a third index, $k$, corresponding to the $k$-th power of the radial separation of the points, $(\R-\R')^k$. However, for the non-radial motion considered in the present work, we will only need the terms of order  $O\left[(\R-\R')^0\right]$ and $O\left[(\R-\R')^1\right]$. The $k=0$ terms, $v_{ij0}=v_{ij}$, are given by Eq.~(\ref{eq:CoordGreen}) and Ref.~\cite{QL} and the $k=1$ terms, $v_{ij1}$, are easily calculated from the $k=0$ terms using the identity \cite{Ottewill:Wardell:2008}
\begin{equation}
v_{ij1}(\R) = - \frac{1}{2} v_{ij0,\R}(\R).
\end{equation}
Eq.~(\ref{eq:CoordGreen}) therefore gives the quasilocal contribution to the retarded Green function as required in the present context.

\section{Singular Structure of the Green Function}\label{sec:singularities}
In this section we investigate the singular structure of the Green function.
We note that one expects the Green function to be singular when its two argument points are connected by a null geodesic, on account of the `Propagation of Singularities' theorems of Duistermaat and H\"ormander \cite{Duistermaat:Hormander:1972,Hormander:1985} and their application to the Hadamard elementary function 
(which is, except for a constant factor, the imaginary part of the Feynman propagator defined below in Eq.(\ref{eq:Hadamard G_F}))
for the Klein-Gordon equation by, e.g., Kay, Radzikowski and Wald~\cite{Kay:Radzikowski:Wald:1997}:
``if such a distributional bisolution is singular for sufficiently nearby pairs of
points on a given null geodesic, then it will necessarily remain singular for
all pairs of points on that null geodesic."

We begin in Sec.~\ref{subsec:large-l} by exploring the large-$l$ asymptotics of the quasinormal mode sum expressions (\ref{GF-inf}), (\ref{G-arbitrary}) and (\ref{G-n0}). The large-$l$ asymptotics of the mode sums are responsible for the singularities in the Green function. We argue that the Green function is singular whenever a `coherent phase' condition is satisfied. The coherent phase condition is applied to find the times at which the Green function is singular. We show that the `singularity times' are exactly those predicted by the geodesic analysis of Sec.~\ref{subsec:geodesics}. In Sec.~\ref{subsec:watson-and-poisson} we introduce two methods for turning the sum over $l$ into an integral. We show in Sec.~\ref{subsec:watson} that the Watson transform can be applied to extract meaningful values from the QNM sums, away from singularities. We show in Sec.~\ref{subsec:Poisson} that the Poisson sum formula may be applied to study the behaviour of the Green function near the singularities. We show that there is a four-fold repeating pattern in the singular structure of the Green function, and use uniform asymptotics to improve our estimates. In Sec.~\ref{subsec:Hadamard} we rederive the same effects by computing the Van Vleck determinant along orbiting geodesics, to find the `direct' part of the Green function arising from the Hadamard form. The two approaches are shown to be consistent.



\subsection{Singularities of the Green function: Large-l Asymptotics}\label{subsec:large-l}
We expect the Green function $\Gret(x,x^\prime)$ 
to be `singular' if the spacetime points $x$ and $x^\prime$ are connected by a null geodesic. By `singular' we mean that $\Gret(x,x^\prime)$ does not take a finite value, although it may be well-defined in a distributional sense. Here we show that the Green function is `singular' in this sense if the large-$l$ asymptotics of the terms in the sum over $l$ satisfy a \emph{coherent phase condition}. 

\subsubsection{Near spatial infinity $\R,\R'\to+\infty$}
Insight into the occurrence of singularities in the Green function may be obtained by examining the large-$l$ asymptotics of the terms in the series (\ref{GF-inf}). Let us write
\beq
\lim_{\Rs,\Rs^\prime \rightarrow +\infty} \GQNM(x,x^{\prime}) = \text{Re} \sum_{l=0}^{\infty} \mathcal{G}_l(T, \gam)
\eeq
The asymptotic behaviour of the gamma function ratio is straightforward: $\Gamma(i \lam) / \Gamma(1/2 + i\lam) \sim \lam^{-1/2} e^{- i \pi / 4},\quad \lam\to +\infty$. 
The large-$l$ asymptotics of the hypergeometric function are explored in Appendix \ref{appendix-hypergeom}. We find
\beq
{}_2 F_1(-\beta, -\beta; -2\beta; -e^{- T}) \sim \left( 1 + e^{-T} \right)^{-1/4} \exp \left( i \lam \left[ \ln \left\{ \frac{\sqrt{1+e^{-T}} + 1}{\sqrt{1+e^{-T}} - 1} \right\} - 2 \ln 2 - T \right] \right),\quad  \lam\to +\infty .  \label{eq:hypergeom-asymp}
\eeq
For simplicity, let us consider the special case of spatial coincidence $\gam = 0$ (near spatial infinity $\R, \R^\prime \to 1$),
\beq
\mathcal{G}_l(T, \gam=0) \sim \frac{e^{-T / 2}}{\sqrt{\pi} (1+e^{-T})^{1/4}} \frac{(l+1/2)}{\lam^{1/2}} \exp \left( i \lam \ln \left[ \frac{\sqrt{1+e^{- T}} + 1}{\sqrt{1+e^{-T}} - 1} \right] - i \pi / 4 \right),\quad \lam \rightarrow \infty  . 
\eeq
Asymptotically, the magnitude of the terms in this series grows with $(l+1/2)^{1/2}$. Hence the series is not absolutely convergent. Nevertheless, due to the oscillatory nature of the series, well-defined values can be extracted (see Sec.~\ref{subsec:watson}), provided that the \emph{coherent phase condition},
\beq
\lim_{l \rightarrow +\infty} \arg \left( \mathcal{G}_{l+1} / \mathcal{G}_{l} \right) = 2 \pi N, \quad \quad N \in \mathbb{Z} , \label{coherent-phase}
\eeq
is {\it not} satisfied.
In other words, the Green function is `singular' in our sense if Eq.~(\ref{coherent-phase}) is satisfied. 
In this case, 
\beq
 \ln \left( \frac{\sqrt{1+e^{-T}} + 1}{\sqrt{1+e^{-T}} - 1} \right) =  2 \pi N, \quad \quad N \in \mathbb{Z}  . \label{coherent-phase-1}
\eeq
Rearranging, we see that the Green function (\ref{GF-inf}) with $\gam = 0$ is `singular' if
\beq
T =  t -t' - \Rs - \Rs^\prime  = \ln \left[ \sinh^2 ( \pi N ) \right] .
\eeq
Note that the coherent phase condition (\ref{coherent-phase}) implies that the Green function is singular at precisely the null geodesic times (\ref{sing-time-inf}) (with $H=1$ and $\Delta\phi=2\pi N$), derived in Sec.~\ref{subsec:Poisson}. 

For the more general case where the spacetime points $x$, $x^\prime$ are separated by an angle $\gam$ on the sphere, it is straightforward to use the asymptotics of the Legendre polynomials to show that the Green function (\ref{GF-inf}) is singular when
\beq
T = t-t' - \Rs - \Rs^\prime  = \ln \left[ \sinh^2 ( \phif / 2 ) \right] ,  \quad \quad \text{where} \quad \phif = 2\pi N \pm \gam
\eeq
again in concordance with (\ref{sing-time-inf}).

\subsubsection{`Fundamental mode' $n=0$}
In Section \ref{subsec:fundamental-modes} we suggested that a reasonable approximation to the Green function may be found by neglecting the higher overtones $n > 0$. The `fundamental mode' series (\ref{G-n0}) also has singularities arising from the coherent phase condition (\ref{coherent-phase}), but they occur at slightly different times; we find that these times are
\beq
T_{\text{(n=0)}} =  \Delta \phi  - \ln[(1+\R)(1+\R^\prime)]    \quad \quad \text{where} \quad \Delta \phi = 2 \pi N \pm \gam .  \label{T-periodic}
\eeq
Note that the singularity times $T_{\text{(n=0)}}$ are periodic. 
Towards spatial infinity, (\ref{T-periodic}) simplifies to $T_{\text{(n=0)}} = \Delta \phi - 2 \ln 2$, which should be compared with the `null geodesic time' given in (\ref{sing-time-inf}). 
Clearly, the periodic times $T_{(n=0)}$ are not quite equal to the geodesic times. Nevertheless, the latter approaches the former as $N \rightarrow \infty$. In Sec. \ref{subsec:Poisson} we compare the singularities of the approximation (\ref{G-n0}) with the singularities of the exact solution (\ref{GF-inf}) at spatial infinity.

\vspace{0.5cm}

To investigate the form of the Green function near, and away from, the null cone, we now introduce two methods for converting a sum over $l$ into an integral in the complex $l$-plane.

 \subsection{Watson Transform and Poisson Sum}\label{subsec:watson-and-poisson}
In Sec.~\ref{subsec:DP}, the distant--past Green function was expressed via a sum over $l$ of the form
\beq
I\equiv \sum\limits_{l=0}^{+\infty} \mathcal{F} (l+\textstyle{\frac12}) P_l(\cos \gamma).    \label{eq:lsum}
\eeq
Here, $\mathcal{F} (l +\textstyle{\frac12})$ may be immediately read off from (\ref{GF-inf}), (\ref{G-arbitrary}) and (\ref{G-n0}).   
The so-called \emph{Watson Transform} \cite{Watson:1918} and \emph{Poisson Sum Formula} \cite{Aki:Richards} provide two closely-related ways of transforming a sum over $l$ into an integral in the complex $l$-plane. 
The two methods provide complementary advantages in understanding the sum over $l$. 

\begin{figure}
 \begin{center}
  \includegraphics[width=8cm]{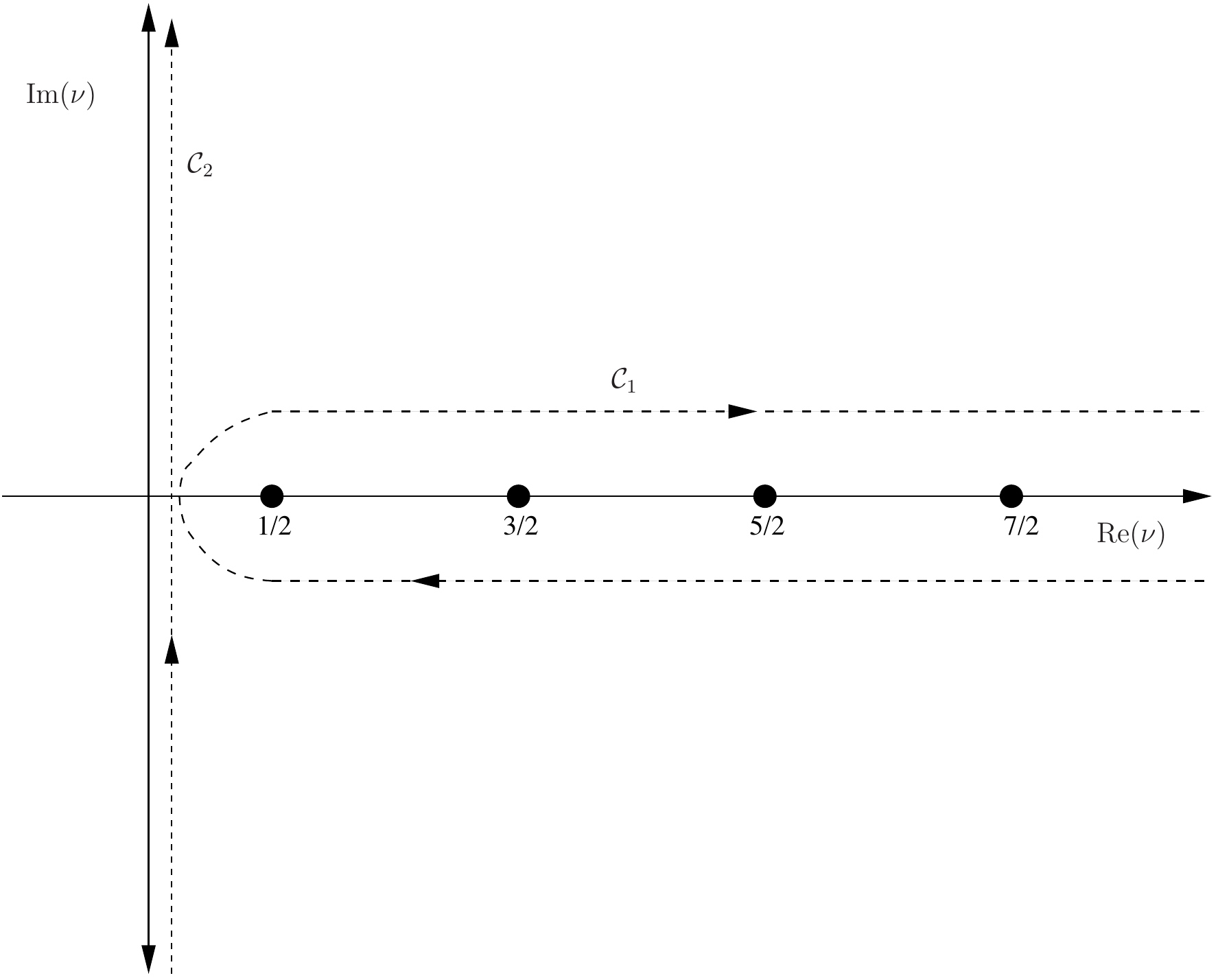}
 \end{center}
 \caption{\emph{The Watson Transform}. The plot shows the contour $\mathcal{C}_1$ that defines the Watson transform in Eq.~(\ref{eq:watson-integral}). Provided the integrand is convergent in both quadrants $I$ and $IV$, the contour may be deformed onto $\mathcal{C}_2$ .
}
 \label{fig:contours2}
\end{figure}

A key element of the Watson transform is that, when extending the Legendre polynomial to non-integer $l$, the function with the appropriate behaviour is $P_l(-\cos \gamma)$. This is obscured by the fact that
 $P_l(-\cos \gamma)= (-1)^l P_l(\cos \gamma)$ when $l$ is an integer. Our first step is then to rewrite the sum (\ref{eq:lsum}) as
 \beq
I=\sum\limits_{l=0}^{+\infty} e^{i(2N+1)\pi l} \mathcal{F} (l+\textstyle{\frac12}) P_l(- \cos \gamma), 
\eeq
where we have also introduced an integer $N$ for later convenience.
Using the Watson transform, we may now express the sum (\ref{eq:lsum}) as a contour integral
 \beq
I=\frac{(-1)^N}{2 i } \int_{\mathcal{C}_1}  \, e^{i 2N \pi \nu} \mathcal{F} (\nu) P_{\nu-1/2}(- \cos \gamma)\frac{d\nu}{\cos(\pi \nu)},   \label{eq:watson-integral}
\eeq
where $\nu = l+1/2$. The contour $\mathcal{C}_1$ starts just below the real axis at $\infty$ encloses the points $\nu=\frac12, 1+\frac12, 2+\frac12, \dots $ which are
poles of the integrand and returns to just above the real axis at $\infty$. The contour $\mathcal{C}_1$ is shown in Fig.~\ref{fig:contours2}. If the integrand is exponentially convergent in both quadrants $I$ and $IV$, the contour may be deformed in the complex-$l$ plane onto a contour $\mathcal{C}_2$ parallel to the imaginary axis (see Fig.~\ref{fig:contours2}). Note that `Regge' poles are not present inside quadrants $I$ and $IV$ in this case.
  
To study the asymptotic behaviour of the Green function near singularities it is convenient to use the alternative respresentation of the sum 
obtained by writing
\begin{align}
\frac{1}{\cos(\pi \nu)}&=
\begin{cases}
\displaystyle 2i\sum_{l=0}^{\infty}e^{i\pi (2l+1)(\nu-1/2)}, & \text{Im}(\nu)>0,\\
\displaystyle -2i\sum_{l=0}^{\infty}e^{-i\pi (2l+1)(\nu-1/2)} = -2i\sum_{l=-\infty}^{-1}e^{i\pi (2l+1)(\nu-1/2)}, & \text{Im}(\nu)<0 .
\end{cases} \label{eq:cospi-rep}
\end{align} 
Inserting representation (\ref{eq:cospi-rep}) into (\ref{eq:watson-integral}) leads to the Poisson sum formula,
\beq
I=\sum_{s = -\infty}^{+\infty} (-1)^s \text{Re} \int_0^\infty d \nu e^{2\pi i s \nu}  \mathcal{F} (\nu) P_{\nu-1/2}(\cos \gam) .  \label{eq:poisson-sum}
\eeq
The Poisson sum formula is applied in Sec.~\ref{subsec:Poisson} to study the form of the singularities. 

\subsection{Watson Transform: Computing the Series}\label{subsec:watson}

Let us now show how the Watson transform may be applied to extract well-defined values from sums over $l$, even though the terms in the series are not absolutely convergent as $l \rightarrow \infty$.  
We will illustrate the approach by considering the $n$th-overtone QNM contribution to the Green function Eq.~(\ref{G-arbitrary}) for which case
\begin{eqnarray}
\mathcal{F}(x,x^{\prime};\nu) &=& \text{Re}\>  \frac{1}{ n !} \left[ \frac{4 e^{-T} }{ (1+\R)(1+\R^\prime) } \right]^{n+1/2} 2 \nu e^{ i \lam \left[ T - \ln(2/(1+\R)) - \ln(2/(1+\R^\prime)) \right] }   \frac{\Gamma(-n + 2 i \lam ) }{[\Gamma(-n+1/2 + i \lam )]^2} \\&& \quad \times  {}_2F_1\left(-n-2 i \lam ,-n,-n+{\textstyle\frac12}+ i \lam; (1-\R)/2\right) \times{}_2F_1\left(-n-2 i \lam,-n,-n+{\textstyle\frac12}+ i \lam; (1-\R^\prime)/2\right), \nn
\end{eqnarray}
where $\lambda$ was defined in (\ref{lam-def}). 
We will choose the integer $N$ so that this contour  may be deformed into the complex plane (as shown in Fig.~\ref{fig:contours2}) to a contour on which the integral converges more rapidly. First, we note that the Legendre function may be be written as the sum of waves propagating clockwise and counterclockwise, $P_{\nu-1/2}(\cos \gam) = \mathcal{Q}^{(+)}_{\nu-1/2}(\cos \gam) + \mathcal{Q}^{(-)}_{\nu-1/2}(\cos \gam)$, where 
\beq
\mathcal{Q}^{(\pm)}_{\mu} (z) = \frac{1}{2} \left[ P_{\mu}(z) \pm \frac{2i}{\pi} Q_{\mu}(z) \right]  \label{Qpm-def}
\eeq
and here $Q_{\mu}(z)$ is a Legendre function of the second kind. The functions $\mathcal{Q}^{(\pm)}_{\nu-1/2}$ have exponential asymptotics in the limit  $\nu \gam \gg 1$,
\begin{eqnarray}
\mathcal{Q}^{(\pm)}_{\nu-1/2} (\cos \gam) 
&\sim& \left( \frac{1}{2 \pi \nu \sin \gam} \right)^{1/2} e^{\pm i \pi / 4} e^{\mp i \nu\gam}.   \label{exp-approx}
\end{eqnarray}
With these asymptotics, and with the large-$l$ asymptotics (\ref{eq:hypergeom-asymp}) of the hypergeometric functions, one finds that the contour may be rotated to run, for example, along a line $\text{Re}(\nu) = c$ with $c$ a constant between  $0$ and $\frac{1}{2}$ \emph{provided that we choose}
\beq
N=  \begin{cases}
     \left[(T+\log \left((\R+1) (\R^\prime+1)\right)+\gamma )/(2 \pi )\right] & \text{for} \ \mathcal{Q}^{(+)}\\
     \left[(T+\log \left((\R+1) (\R^\prime+1)\right)+2 \pi - \gamma )/(2 \pi )\right]& \text{for}\  \mathcal{Q}^{(-)}\\
     \end{cases}
     \label{eq:critical}
\eeq
where here $[x]$ denotes the greatest integer less than or equal to $x$.

We performed the integrals along $\text{Re}(\nu) = \frac14$ and found rapid convergence of the integrals except near the critical times defining the jumps in $N$ given by Eq.~(\ref{eq:critical}), when the integrands fall to 0 increasingly slowly. An alternative method for extracting meaningful values from series which are not absolutely convergent is described in Sec.~\ref{subsec:nummeth}.

 \subsection{The Poisson sum formula: Singularities and Asymptotics}\label{subsec:Poisson}
In this section we study the singularity structure of the Green function by applying the Poisson sum formula (\ref{eq:poisson-sum}). The first step is to group the terms together so that
\begin{equation}
I= \sum_{N=0}^\infty \II_N \quad \quad \text{where} \quad \quad \II_N \equiv \text{Re} \int_0^{+\infty} d \nu \mathcal{F}(\nu) R_N(\nu, \gam)    \label{InRN}
\end{equation}
and
\begin{equation} \label{eq:RN}
R_N = 
\begin{cases}
 \displaystyle  (-1)^{N/2} \left[ \mathcal{Q}_{\nu-1/2}^{(-)}(\cos \gam) e^{iN\pi\nu} + \mathcal{Q}_{\nu-1/2}^{(+)}(\cos \gam) e^{-iN\pi\nu}  \right] , \quad \quad & N \; \text{even} , \\
 \displaystyle (-1)^{(N+1)/2} \left[ \mathcal{Q}_{\nu-1/2}^{(+)}(\cos \gam) e^{i(N+1)\pi\nu} + \mathcal{Q}_{\nu-1/2}^{(-)}(\cos \gam) e^{-i(N+1)\pi\nu}  \right] , \quad \quad & N \; \text{odd} .
  \end{cases}
\end{equation}
and $\mathcal{Q}_{\nu - 1/2}^{(\pm)}$ were defined in Eq.~(\ref{Qpm-def}). We can now use the exponential approximations for $\mathcal{Q}_{\nu - 1/2}^{(\pm)}$ given in (\ref{exp-approx}) to establish
\beq
R_N \sim \frac{1}{(2\pi \nu \sin \gam)^{1/2}}  \begin{cases}
\displaystyle (-1)^{N/2} \left[ e^{-i \pi / 4} e^{ i \nu ( N \pi + \gam ) } + \text{c.c.} \right] ,  & \quad N \; \text{even}   , \\
\displaystyle (-1)^{(N+1)/2}  \left[ e^{i \pi / 4} e^{ i \nu ( (N+1) \pi - \gam ) } + \text{c.c.} \right] ,  & \quad N \; \text{odd}  . 
  \end{cases} \label{RN-asymp}
\eeq
It should be borne in mind that the exponential approximations (\ref{exp-approx}) are valid in the limit $\gam \nu \gg 1$. Hence the approximations are not suitable in the limit $\gam \rightarrow 0$ case. Below, we use alternative asymptotics (\ref{Olver-approx}) to investigate this case. 

\subsubsection{`Fundamental mode' $n=0$}
Let us apply the method to the `fundamental mode' QNM series (\ref{G-n0}). In this case we have
\beq
\mathcal{F} (\nu) = \left( \frac{4 e^{- T} }{\pi(1+\R)(1+\R^\prime)} \right)^{1/2} \frac{ \nu \, \Gamma(i \lam)}{\Gamma(1/2 + i\lam)} e^{i \lam \chi} 
\label{F n=0}
\eeq 
where $T$ was defined in (\ref{T-def}), $\lam$ was defined in (\ref{lam-def}), and
\beq
\chi = T + \ln[(1+\R)(1+\R^\prime)]  .  \label{eq:chi}
\eeq
Taking the asymptotic limit $\nu \rightarrow \infty$ we find
\beq
\mathcal{F} (\nu) \sim \left( \frac{4 \nu \, e^{-T} }{i \pi(1+\R)(1+\R^\prime)} \right)^{1/2} \, e^{i \nu \chi}, \quad \nu \rightarrow \infty.
\eeq
Now let us combine this with the `exponential approximations' (\ref{RN-asymp}) for $R_N$,
\beq
\mathcal{F}(\nu) R_N(\nu, \gam) \sim \left( \frac{2 \, e^{-T} }{\pi^2 \sin\gam (1+\R)(1+\R^\prime)} \right)^{1/2} 
\begin{cases}
\displaystyle (-1)^{N/2} \left[-i e^{i \nu (\chi + N\pi + \gam)} + e^{i  \nu (\chi - N\pi - \gam)} \right] , \quad & N \; \text{even} , \\
\displaystyle (-1)^{(N+1)/2} \left[ e^{i \nu (\chi + (N+1)\pi - \gam)} - i e^{i \nu (\chi - (N+1)\pi + \gam)} \right], \quad & N \; \text{odd} , 
\end{cases} \label{Poisson1}
\eeq
for $\nu \rightarrow \infty$.
It is clear that the integral in (\ref{InRN}) will be singular if the phase factor in either term in (\ref{Poisson1}) is zero. In other words, each wave $R_N$ gives rise to two singularities, occurring at particular `singularity times'. We are only interested in the singularities for $T > 0$; hence we may neglect the former terms in (\ref{Poisson1}). Now let us note that
\beq
\lim_{\eps \rightarrow 0^+} \int_{0}^{\infty} e^{i \nu (\zeta + i \eps)} d \nu = \lim_{\eps \rightarrow 0^+} \left( \frac{i}{\zeta + i\eps} \right) = i / \zeta + \pi \delta( \zeta )
\eeq
Upon substituting (\ref{Poisson1}) into (\ref{InRN}) and performing the integral, we find
\beq
\II_N^{(0)} \sim  \left( \frac{2 \, e^{-T } }{\sin\gam (1+\R)(1+\R^\prime)} \right)^{1/2} 
\begin{cases}
\displaystyle (-1)^{N/2} \delta( t-t' - t^{(0)}_{N} )  , \quad & N \; \text{even} , \\
\displaystyle \frac{(-1)^{(N+1)/2}}{\pi (t-t' - t^{(0)}_{N} )}  , \quad & N \; \text{odd} ,
\end{cases}
\label{sing-Poisson}
\eeq
where $\II_N^{(0)}$ is $\II_N$ in (\ref{InRN}) with $\mathcal{F} (\nu)$ given by (\ref{F n=0}), and 
\beq
t^{(0)}_N = \Rs + \Rs^\prime - \ln \left( (1+\R)(1+\R^\prime) \right) +
\begin{cases}
\displaystyle N  \pi + \gam   , \quad & N \; \text{even} , \\
\displaystyle (N+1) \pi - \gam   , \quad & N \; \text{odd} .
\end{cases}
\eeq
These times $t_{N}^{(0)}$ are equivalent to the `periodic' times identified in Sec.~\ref{subsec:large-l}  (Eq.~\ref{T-periodic}) and Sec.~\ref{subsec:geodesics} (Eq.~\ref{sing-time-periodic}). 

Let us consider the implications of Eq.~(\ref{sing-Poisson}) carefully. Let us fix the spatial coordinates $\R$, $\R^\prime$ and $\gam$ and consider variations in $t - t'$ only. Each term $\II_N$ corresponds to a particular singularity in the mode sum expression (\ref{G-n0}) for the
($n=0$)-Green function. The $N$th singularity  occurs at $t-t' = t_{N}^{(0)}$. For times close to $t_{N}^{(0)}$, we expect the term $\II_N$ to give the dominant contribution to the ($n=0$)-Green function. Eq.~(\ref{sing-Poisson}) suggests that the ($n=0$)-Green function has a repeating four-fold singularity structure. The `shape' of the singularity alternates between a a delta-distribution ($\pm \delta( t-t' - t_N^{(0)} )$, $N$ even) and a singularity with antisymmetric `wings' ($\pm 1/(t- t' - t_N^{(0)})$,  $N$ odd).
 
The $N$th wave may be associated with the $N$th orbiting null geodesic shown in Fig.~\ref{fig:circ_orbits}. Note that `even N' and `odd N' geodesics pass in opposite senses around $\R=0$ (see, for example, Fig.~\ref{fig:circ_orbits}). Now, $N$ has a clear geometrical interpretation: it is the number of \emph{caustics} through which the corresponding geodesic has passed. Caustics are points where neighbouring geodesics are focused, and in a spherically-symmetric spacetime caustics occur whenever a geodesic passes through angles $\Delta \phi = \pi$, $2\pi$, $3\pi, $ etc. Equation (\ref{sing-Poisson}) implies that the singularity structure of the Green function changes each time the wavefront passes through a caustic \cite{Ori1}. 

More accurate approximations to the singularity structure may be found by using the uniform asymptotics established by Olver \cite{Olver:1974}
(as an improvement on the `exponential asymptotics' (\ref{exp-approx})),
\begin{eqnarray}
\mathcal{Q}^{(\pm)}_{\nu-1/2} (\cos \gam) &\sim& \frac{1}{2} \left( \frac{\gam}{\sin \gam} \right)^{1/2} H_0^{(\mp)} (\nu \gam),  \label{Olver-approx}  
\end{eqnarray}
where $H_0^{(\pm)}(\cdot) = J_0(\cdot) \pm i Y_0(\cdot)$ are Hankel functions of the first $(+)$ and second $(-)$ kinds. This approximation (\ref{Olver-approx}) is valid in the large-$\nu$ limit for angles in the range $0 \le \gam < \pi$. With these asymptotics, we replace (\ref{RN-asymp}) with
\beq
R_N \sim \frac{1}{2} \left( \frac{\gam}{\sin \gam} \right)^{1/2} 
\begin{cases}
\displaystyle (-1)^{N/2} \left[ H_0^{(+)}(\nu \gam) e^{iN\pi\nu} + H_0^{(-)}(\nu \gam) e^{- iN\pi\nu}  \right] ,  & \quad N \; \text{even}   , \\
\displaystyle (-1)^{(N+1)/2} \left[  H_0^{(-)}(\nu \gam) e^{i(N+1)\pi\nu} + H_0^{(+)}(\nu \gam) e^{- i(N+1)\pi\nu}  \right] ,  & \quad N \; \text{odd}  . 
  \end{cases}  \label{RN-asymp-Hankel}
\eeq
In Appendix \ref{appendix:poisson-sum} we derive the following asymptotics for the `fundamental mode'  ($n=0$) Green function,
\begin{eqnarray}
 \II^{(0)}_1 &\sim& 
 \begin{cases}
 \displaystyle \frac{2\Agam \, }{\sqrt{\pi} [(2\pi - \gam) - \chi] \, [(2\pi + \gam) - \chi]^{1/2} }  E\left(2\gam/[(2\pi + \gam) - \chi]\right) , \quad \quad & \chi < 2 \pi - \gam , \\
 \displaystyle \frac{-\Agam \sqrt{\pi}}{2 [\chi - (2\pi - \gam)]^{3/2}} \, {}_2 F_1\left( 3/2, 1/2; 2; \frac{\chi - 2\pi - \gam}{\chi - 2\pi + \gam} \right) , \quad \quad & \chi > 2 \pi - \gam  ,
 \end{cases} \label{eq:I1-asymp}
 \\
\II^{(0)}_2 &\sim& \begin{cases}
\displaystyle  - \sqrt{\frac{2 \pi}{\gam}} \Agam \delta \left( \chi - (2\pi + \gam) \right) ,  \quad \quad & \chi \le 2 \pi + \gam  , \\
\displaystyle   \frac{\Agam \sqrt{\pi}}{2 [\chi - (2\pi - \gam)]^{3/2}} \, {}_2 F_1\left( 3/2, 1/2; 2; \frac{\chi - 2\pi - \gam}{\chi - 2\pi + \gam} \right)  , \quad \quad & \chi > 2 \pi + \gam  ,
   \end{cases}   \label{eq:I2-asymp}
\end{eqnarray}
where $E$ is the complete elliptic integral of the second kind, $\chi$ was defined in (\ref{eq:chi}) and
\beq
\Agam = \left( \frac{\gam}{\sin \gam} \right)^{1/2} \left( \frac{e^{-T}}{\pi (1+\R)(1+\R^\prime)} \right)^{1/2}.  \label{eq:A-gam}
\eeq

\begin{figure}
 \begin{center}
  \includegraphics[width=8cm]{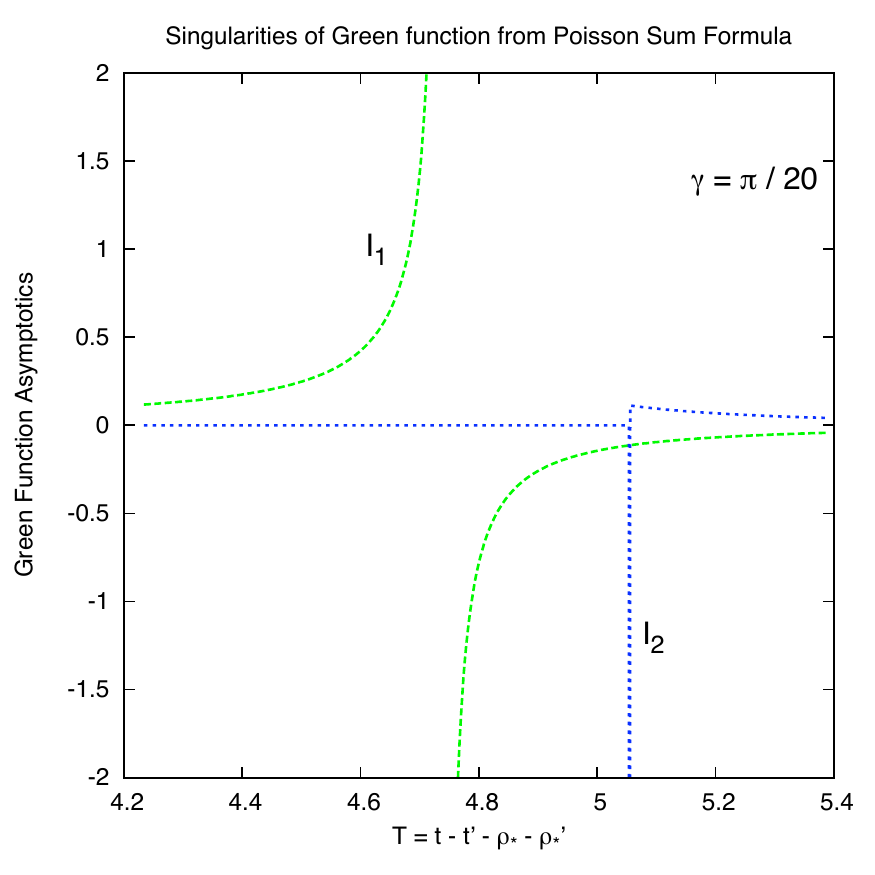}
  \includegraphics[width=8cm]{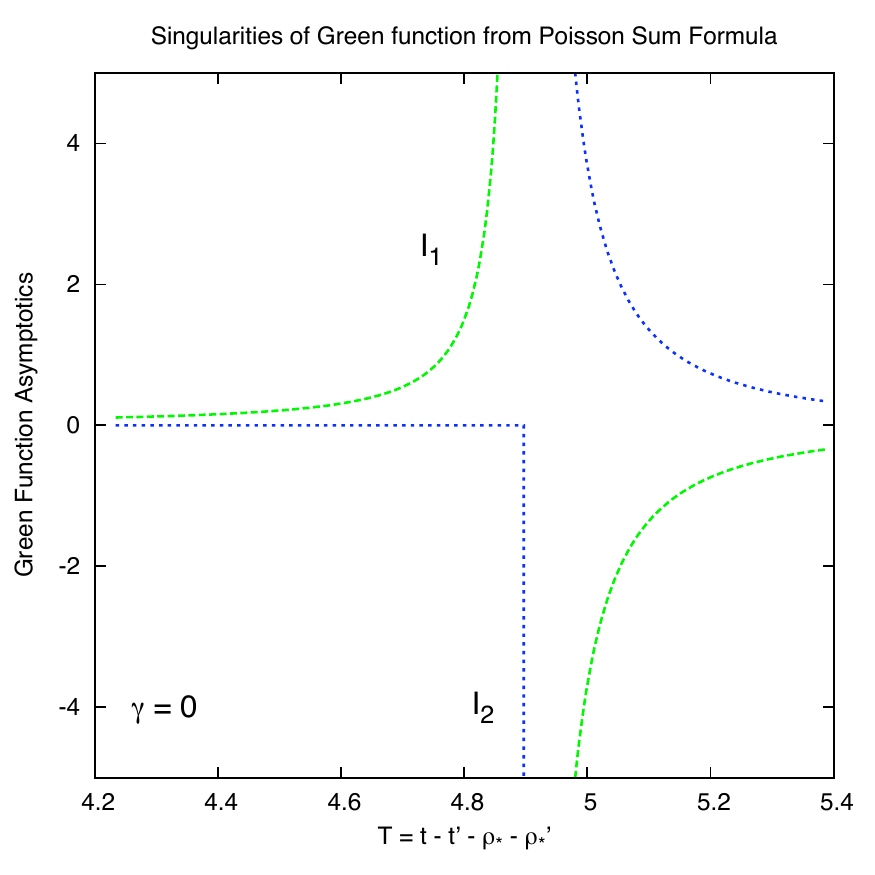}
 \end{center}
 \caption{\emph{Singularities of the `fundamental mode' Green function (\ref{G-n0}) near the caustic at $2\pi$}. These plots show the $\II_1$  (dashed) and $\II_2$ (dotted) contributions to the Poisson sum, given explicitly by (\ref{eq:I1-asymp}) and (\ref{eq:I2-asymp}). The left plot shows an angular separation $\gam = \pi/20$ and the right plot shows angular coincidence $\gam =0$. Note that, for $T > 2 \pi + \gam - 2\ln 2$, the $I_1$ and $I_2$ integrals are equal and opposite and will exactly cancel out (see text).}
 \label{fig:I1I2}
\end{figure}

The asymptotics (\ref{eq:I1-asymp}) and (\ref{eq:I2-asymp}) provide insight into the singularity structure near the caustic at $\Delta\phi = 2 \pi$. Figure \ref{fig:I1I2} shows the asymptotics (\ref{eq:I1-asymp}) and (\ref{eq:I2-asymp}) for two cases: (i) $\gam = \pi / 20$ (left) and (ii) $\gam = 0$ (right). In the left plot, the $\II_1^{(0)}$ integral has a (nearly) antisymmetric form. The $\II_2^{(0)}$ integral is a delta function with a `tail'. However, the `tail' is exactly cancelled by the $\II_1^{(0)}$ integral in the regime $\chi > 2\pi + \gam$. The cancellation creates a step discontinuity in the Green function at $\chi = 2\pi + \gam$.
The form of the divergence shown in the right plot ($\gam = 0$) may be understood by substituting $\gam = 0$ into (\ref{eq:I1-asymp}) to obtain
\beq
\II_1^{(0)}(\gam = 0) \sim \left( \frac{e^{-T}}{(1+\R) (1+\R^\prime)} \right)^{1/2} ( 2 \pi - \chi )^{-3/2} .
\eeq

\subsubsection{Near spatial infinity, $\R,\R^\prime\to 1$}

It is straightforward to repeat the steps in the above analysis for the closed-form Green function (\ref{GF-inf}), valid for $\R,\R^\prime\to 1$. We reach a result of the same form as (\ref{sing-Poisson}), but with modified singularity times, 
\beq
t^{(\infty)}_{N} = \Rs + \Rs^\prime + 
\begin{cases}
\displaystyle \ln \left( \sinh^2 \left([N \pi + \gam] / 2 \right) \right) , \quad & N \text{ even} , \\
\displaystyle \ln \left( \sinh^2 \left([(N+1) \pi - \gam] / 2 \right) \right) , \quad & N \text{ odd} ,
\end{cases}
\eeq
corresponding to the geodesic times (\ref{sing-time-inf}).  
For instance, with the exponential asymptotics (\ref{RN-asymp}) applied to (\ref{GF-inf}) we obtain
\beq
\II^{(\infty)}_N \sim  \left( \frac{ e^{-T} }{ 2 \sin \gam \sqrt{1 + e^{-T}} } \right)^{1/2} 
\begin{cases}
\displaystyle (-1)^{N/2} \, \delta\left( t-t'- t^{(\infty)}_{N} \right)  , \quad & N \; \text{even} , \\
\displaystyle \frac{(-1)^{(N+1)/2}}{\pi \left(t-t' - t^{(\infty)}_{N} \right)}  , \quad & N \; \text{odd} .
\end{cases}
\label{sing-Poisson-2}
\eeq
In Sec.~\ref{sec:results} the asymptotic expressions derived here are compared against numerical results from the mode sums.

\vspace{0.5cm}

We believe that the 4-fold cycle in the singularity structure of the Green function which we have just unearthed using tricks we picked up from
seismology~\cite{Aki:Richards} is
characteristic of the $ \mathbb{S}^2$ topology (different types of cycle arising in different cases).
This cycle may thus also appear in
the more astrophysically interesting case of the Schwarzschild spacetime. Since this cycle does not seem to be
widely known in the field of General Relativity (with the notable exception of \cite{Ori1}),
in Appendix \ref{appendix-Green on S2} we apply the large-$l$
asymptotic analysis of this section to the simplest case with $ \mathbb{S}^2$
topology: the spacetime of $T\times  \mathbb{S}^2$, where the same cycle
blossoms in a clear manner.

\subsection{Hadamard Approximation and the Van Vleck Determinant}\label{subsec:Hadamard}
In this section, we rederive the singularity structure found in (\ref{sing-Poisson}) and (\ref{sing-Poisson-2}) using a `geometrical' argument based on the Hadamard form of the Green function. 
In Sec.~\ref{subsec:QL} we used the Hadamard parametrix of the Green function to find the quasilocal contribution to the self-force. Strictly speaking, the Hadamard parametrix of Eq.~(\ref{eq:Hadamard}), is only valid when $x$ and $x^\prime$ are within a convex normal neighborhood \cite{Friedlander}. Nevertheless, it is plausible (particularly in light of the previous sections) that the Green function near the singularities may be adequately described by a Hadamard-like form, but with contributions from \emph{all appropriate orbiting geodesics} (rather than just the unique timelike geodesic joining $x$ and $x^\prime$). 

We first introduce 
the Feynman propagator $G_F(x,x')$
(see, e.g.,~\cite{DeWitt:1960,Birrell:Davies})
which satisfies
the inhomogeneous scalar wave equation (\ref{eq:gf-waveeq}). 
The Hadamard form, which in principle is only valid for points
$x'$ within the normal neighbourhood of $x$, for the Feynman propagator in 4-D is~\cite{DeWitt:1960,Decanini:Folacci:2005b}
\beq\label{eq:Hadamard G_F}
G_F(x,x')=\frac{i}{2\pi}\left[\frac{U(x,x')}{\sigma+i\epsilon}+V(x,x')\ln(\sigma+i\epsilon)+W(x,x')\right],
\eeq
where $U(x,x')$, $V(x,x')$ (already introduced in Sec.\ref{sec:ql}) and $W(x,x')$ are bitensors which are regular at coincidence ($x\to x'$) and
$\sig(x,x^\prime)$ is Synge's world function: half the square of the geodesic distance \emph{along a specific geodesic} joining $x$ and $x^\prime$.
Note the {\it Feynman prescription} `$\sigma\to\sigma+i\epsilon$' in (\ref{eq:Hadamard G_F}),
where $\epsilon$ is an infinitesimally small positive value.

The retarded Green function is readily obtained from 
the Feynman propagator by
\begin{equation}
G_{\text{ret}}(x,x')=
2\theta_-(x,x')\ \text{Re}\left(G_F(x,x')\right),
\end{equation}
which yields (\ref{eq:Hadamard}) inside the convex normal neighbourhood, since $U(x,x')$, $V(x,x')$ and $W(x,x')$ 
are real-valued there.
We posit that the `direct' part of the Green function 
remains in Hadamard form,
\beq
\Gret^{\text{dir.}}(x, x^\prime) = \lim_{\eps \rightarrow 0^+} \frac{1}{\pi} \text{Re} \left[ i \frac{U(x,x^\prime)}{\sig + i \eps} \right] = \text{Re} \left[ U(x,x^\prime) \left( \delta(\sig)+\frac{i}{\pi \sig}\right) \right],  \label{Gdir1}
\eeq
even {\it outside} the convex normal neighbourhood [note that outside the convex normal neighbourhood we still use the term `direct' part to refer to the contribution from the $U(x,x')$ term, even if its support may not be restricted on the null cone anymore]. It is plausible that the Green function near the $N$th singularity (see previous section) is dominated by the `direct' Green function (\ref{Gdir1}) calculated along geodesics near the $N$th orbiting null geodesic. To test this assertion, we will calculate the structure and magnitude of the singularities and compare with (\ref{sing-Poisson-2}).

In four dimensional spacetimes, the symmetric bitensor $U(x,x^\prime)$ is given by
\beq
U(x,x^\prime) = \Delta^{1/2}(x, x^\prime),
\eeq
where $\Delta(x,x^\prime)$ is the Van Vleck determinant \cite{VanVleck:1928, Morette:1951, Visser:1993}. The Van Vleck determinant can be found by integrating a system of transport equations along the appropriate geodesic joining $x$ and $x^\prime$. The first of these \cite{Poisson:2003},
\beq
\label{eq:vv-transport}
\lambda \frac{d \ln \Delta }{ d \lam } = 4 - {\sig^\alp}_\alp
\eeq
is a transport equation for the Van Vleck determinant itself, with the initial condition $\Delta(\lambda = 0) = 1$. Here, $\lambda$ is an affine parameter along the geodesic joining $x$ and $x^\prime$, and ${\sig^\alp}_\bet = \nabla_\bet  \nabla^\alp \sig$ is the second covariant derivative (taken with respect to spacetime point $x$) of Synge's world function, which in turn is found from the coupled system of transport equations \cite{Avramidi:2000, Hawking:Ellis}
\beq
\label{eq:z-transport}
\lam \frac{d {\sig^{\alp}}_\bet}{d \lam} = {\sig^\alp}_\bet - {\sig^\alp}_\mu {\sig^\mu}_\bet + \lam u^\mu \left( {\Gamma^\nu}_{\bet \mu} {\sig^\alp}_\nu - {\Gamma^\alp}_{\nu \mu} {\sig^\nu}_\bet \right)  - \lam^2 {R^\alp}_{\mu \bet \nu} u^\mu u^\nu
\eeq
and the boundary condition ${\sig^\alp}_\bet (\lambda=0) = \delta^\alp_\bet$. 

In principle, transport equations (\ref{eq:vv-transport}) and (\ref{eq:z-transport}) may be integrated numerically to determine the Van Vleck determinant along any given geodesic on any given spacetime. This is the approach that we might take on, for example, in Schwarzschild. A numerical approach is not necessary for the Nariai spacetime, however. This spacetime is the Cartesian product of a two-sphere with a 2-D de Sitter spacetime. On a product spacetime $\mathcal{M} = \mathcal{M}_1 \times \mathcal{M}_2$ we may make the following decomposition:
\beq
\sig = \sig_1 + \sig_2 ,  \quad \quad \quad \Delta = \Delta_1 \Delta_2,
\eeq
where $\sig_i$ and $\Delta_i$ ($i=1,2$) are, respectively, Synge's world function and the Van Vleck determinant on the manifold $\mathcal{M}_i$.
It will be shown in a forthcoming work \cite{Nolan:2009} that the Van Vleck determinant on the Nariai spacetime when the two points $x$ and $x'$
are within the normal neighbourhood is simply
\beq
\Delta (x,x')= \left( \frac{\gam}{\sin \gam} \right) \left( \frac{ \eta }{ \sinh \eta}  \right)  \label{eq:vanVleckDet}
\eeq
where $\gam \in [0, \pi)$ is the geodesic distance traversed on the two-sphere and $\eta$ is the geodesic distance traversed in the two-dimensional de Sitter subspace. Hence the Van Vleck determinant is singular at the angle $\gam = \pi$. 


This may be seen another way. Using the spherical symmetry, let us assume without loss of generality that the motion is in the plane $\phi = \text{const}$, from which it follows that the equation for ${\sig^\phi}_\phi$ in (\ref{eq:z-transport}) decouples from the remainder; it is
\beq
\theta \frac{d {\sig^\phi}_\phi}{d \theta} + \theta^2 - {\sig^\phi}_\phi (1 - {\sig^\phi}_\phi ) = 0
\label{eq:V-V transport eq}
\eeq
Here we have rescaled the affine parameter $\lambda$ to be equal to the angle $\theta$ subtended by the geodesic. Note that here we let $\theta$ take values greater than $\pi$. 
It is straightforward to show that the solution of Eq.~(\ref{eq:V-V transport eq}) is ${\sig^\phi}_\phi = \theta \cot \theta$. Hence ${\sig^\phi}_\phi$ is singular at the angles $\theta = \pi$, $2\pi$, $3\pi$, etc.  In other words, the Van Vleck determinant is singular at the antipodal points, where neighbouring geodesics are focused: the \emph{caustics}. The 
Van Vleck determinant may be separated in the following manner: $\Delta = \Delta_{\phi} \Delta_{ty}$, where
\begin{eqnarray}
\theta \frac{d \ln \Delta_{\phi}}{d \theta} &=& 1 - {\sig^\phi}_{\phi},   \label{Delta_pp}  \\
\theta \frac{d \ln \Delta_{ty} }{d \theta} &=& 2 - \Ztt - \Zyy  \label{Delta_ty} .
\end{eqnarray}
Eq.~(\ref{Delta_pp}) yields
\beq \label{eq:V-V sing}
\ln  \Delta_{\phi}=\ln \left(\frac{\theta}{\theta_0}\right)
-\int_{\theta_0}^{\theta}d\theta'\cot\theta'
\eeq
which can be integrated analytically by following a Landau contour in the complex $\theta'$-plane around 
the (simple) poles of the integrand (located at $\theta'=k \pi$, $k\in \mathbb{Z}$), which are the caustic points.
Following the Feynman prescription `$\sigma\to\sigma+i\epsilon$', we choose 
the Landau contour so that the poles lie {\it below} the contour.
We then obtain (setting $\theta_0=0$ without loss of generality)
\beq \label{eq:Delta_phi}
\Delta_{\phi} = \left| \frac{\theta}{\sin{\theta}} \right| e^{-i N \pi} .
\eeq
Here, $N$ is the number of caustic points the geodesic has passed through. The phase factor, obtained by continuing the contour of integration past the singularities at $\theta = \pi, 2\pi, $ etc., is crucial. Inserting the phase factor $e^{-i N \pi / 2}$ in (\ref{Gdir1}) leads to exactly the four-fold singularity structure predicted by the large-$l$ asymptotics of the mode sum (\ref{sing-Poisson-2}). That is:
\beq
 G_{N}^{dir} \sim  \left( \frac{\eta}{\sinh \eta } \right)^{1/2} 
 \left( \frac{\theta}{\sin \theta } \right)^{1/2}
 \begin{cases}
 \displaystyle (-1)^{N/2} \delta(\sig), \quad \quad &  N \text{ even}, \\ 
 \displaystyle \frac{(-1)^{(N-1)/2}}{\pi \sig}, \quad & N \text{ odd}.
 \end{cases} 
 \label{GNdir-Hadamard}
\eeq

The accumulation of a phase of `$-i$' on passing through a caustic, and the alternating singularity structure which results, is well-known to researchers in other fields involving wave propagation -- for example, in acoustics~\cite{Kravtsov:1968},
seismology~\cite{Aki:Richards},
symplectic geometry \cite{Arnold} and quantum mechanics~\cite{B&M}  the integer $N$ is known as the Maslov index~\cite{Maslov'65,MaslovWKB}.

We would expect to find an analogous effect in, for example, the Schwarzschild spacetime. The four-fold structure has been noted before by at least one researcher \cite{Ori1}. 
Nevertheless, the effect of caustics on wave propagation in four-dimensional spacetimes does not seem to have received much attention in the gravitational literature (see \cite{Friedrich:Stewart:1983, Ehlers:Newman:2000} for exceptions). 

To compare the singularities in the mode-sum expression (\ref{sing-Poisson-2}) with the singularities in the Hadamard form (\ref{GNdir-Hadamard}), let us consider the `odd-$n$' singularities of $1/\sig$ form. We will rearrange (\ref{sing-Poisson-2}) into analogous form by expanding $\sigma$ to first-order in $t - t_N^{(\infty)}$, where $t_N^{(\infty)}$ is the $N$th singularity time for orbiting geodesics starting and finishing at $\R\to 1$.
For the orbiting geodesics described in Sec. \ref{subsec:geodesics} we have $\sig = - \tfrac{1}{2} (H^2 - 1) \theta^2$. 
At $\R\to 1$, expanding to first order and using (\ref{sing-time-inf}) yields
\beq
 \sig \sim  - (H \theta) \tanh( H \theta / 2) \left(t - t_N^{(\infty)}\right) .
\eeq
The mode-sum expression (\ref{sing-Poisson-2}) may then be rewritten in analogous form to the `$N$ odd' expression in (\ref{GNdir-Hadamard}),
\beq
\GQNM \sim (-1)^{(n-1)/2} \frac{\left| \Delta_{\text{(QNM)}} \right|^{1/2}}{\pi \sig} 
\quad \quad \text{where} \quad \quad 
 \left| \Delta_{\text{(QNM)}} \right|^{1/2} =  \left( \frac{H \eta \sinh(\eta/2)}{2 \cosh^3(\eta/2)} \right)^{1/2}
 \left| \frac{\theta}{\sin \theta} \right|^{1/2} .
\label{Delta-QNM}
\eeq
Here $\eta = H \theta$, where $H$ is the constant of motion introduced in Sec.~\ref{subsec:geodesics}. We find very good agreement between (\ref{eq:vanVleckDet}) and (\ref{Delta-QNM}) in the $\theta \gtrsim \pi$ regime. The disagreement at small angles is not unexpected as the QNM sum is invalid at early times (or equivalently, for orbiting geodesics which have passed through small angles $\theta$).



\section{Self-Force on the Static Particle}
\label{sec:sf}


In this section we turn our attention to a simple case: the self-force acting on a static scalar particle in the Nariai spacetime. By `static' we mean a particle with constant spatial coordinates. It is not necessarily at rest, since its worldline may not be a geodesic, and it may require an external force to keep it static.  In Sec.~\ref{subsec:static} we review previous calculations for the static self-force on a range of spacetimes and in Sec.~\ref{subsec:static full Green} we explore one such analytic method for computing the static self-force in Nariai spacetime.  This method, based on the massive field approach of Rosenthal \cite{Rosenthal:2004} provides an independent check on the matched-expansion approach. In Sec.~\ref{subsec:matched-static} we describe how the method of matched expansions may be applied to the static case. To compute the self-force, we require robust numerical methods for evaluating the quasinormal mode sums such as (\ref{G-arbitrary}); two such methods are outlined in Sec. \ref{subsec:nummeth}. The results of all methods are validated and compared in Sec. \ref{sec:results}. 

\subsection{The Static Particle} \label{subsec:static}
A static particle -- a particle with constant spatial coordinates -- has been the focus of several scalar self-force calculations, in particular for the Schwarzschild spacetime \cite{SW:1980,Wiseman:2000,Rosenthal:2004,Anderson:Wiseman:2005,Anderson:Eftekharzadeh:Hu:2006,CTW:2007,Ottewill:Wardell:2008}. Although it may not be a particularly physical case, it is frequently chosen because it involves relatively straightforward calculations and has an exact solution for the Schwarzschild spacetime. It therefore provides a good testing ground for new approaches to the calculation of the self-force.

Smith and Will~\cite{SW:1980} calculated the self-force on a static {\it electric} charge in the Schwarzschild background and found
it to be non-zero. In~\cite{Wiseman:2000}, Wiseman considered the analogous case of a static {\it scalar} charge in the case of 
{\it minimal-coupling} (i.e., $\xi=0$) in Schwarzschild. Using isotropic coordinates, he managed to sum the Hadamard series for the Green function in the static case (i.e., the ``Helmholtz"-like equation in Schwarzschild with zero-frequency, $\omega=0$) and thus obtain in closed form the field created by the static charge in the scalar and also electrostatic (already found in~\cite{Copson:1928,Linet:1976} using a different method) cases. He then found the self-force to be zero in the scalar, minimally-coupled case.

In~\cite{CTW:2007}, the calculation of the self-force on a static scalar charge in Schwarzschild is extended to the case of {\it non-minimal coupling} ($\xi\neq 0$) and is found to be zero as well. The fact that the value of the scalar self-force in Schwarzschild is the same (zero) independently of the value of the coupling constant is in agreement with the Quinn-Wald axioms~\cite{Quinn:Wald:1997,Quinn:2000}:
their method relies only on the field equations, and these are independent of the coupling constant in a Ricci-flat spacetime such as Schwarzschild.
The calculation (without using the Quinn-Wald axioms) is by no means trivial, however, since the effect of the coupling constant might be felt through the stress-energy tensor (in fact,~\cite{CTW:2007} corrected a previous result in~\cite{Zelnikov:Frolov:1982a,Zelnikov:Frolov:1982b}, where the self-force had been incorrectly found to be non-zero). Rosenthal \cite{Rosenthal:2004} has also considered this case of a static particle in Schwarzschild and used it as an example application of the massive field approach \cite{Rosenthal:2003} to self-force calculations.

On the other hand, Hobbs~\cite{Hobbs:1968b} showed that, in a {\it conformally-flat} spacetime, the ``tail" contribution
to the self-force on an {\it electric} charge (on any motion, static or not) is zero.
The only possible contribution to the self-force might then come from the local Ricci-terms, which are zero in cases of physical interest
such as in de Sitter universe.

Noting the conformal-invariance of Maxwell's equations, one would then expect the ``tail" contribution to the 
{\it scalar} self-force to also be zero in the two following cases: 
(1)  for a charge undergoing any motion in a conformally-flat 4-D spacetime with conformal-coupling (i.e., $\xi=1/6$), and 
(2)  for a static charge (where the time-independence effectively reduces the problem to a 3-D spatial one), in a spacetime such that its 3-D spatial section is conformally-flat and with conformal-coupling in 3-D (i.e., $\xi=1/8$).
Indeed, in a recent article~\cite{Bezerra:Khus} it was shown that the scalar self-force on a massless static particle in a {\it wormhole spacetime} (with non-zero Ricci scalar and where the 3-D spatial section is conformally-flat) changes sign at $\xi=1/8$ and it is equal to zero at this 3-D conformal value.

The Nariai spacetime, not being Ricci-flat and being conformal to a wormhole spacetime (and so with conformally-flat 3-D spatial section), suggests a very interesting playground for calculating the self-force: What role does the coupling constant $\xi$ play? Do particular values such as $\xi=1/6$ (4-D conformal-coupling) and
$\xi=1/8$ (3-D conformal-coupling, so a particular value in the case of a static charge) yield particular values for the self-force? Do they support the Quinn-Wald axioms?

\subsection{Static Green Function Approach} \label{subsec:static full Green}

The conventional approach to calculating the self-force on a static particle due to Wiseman \cite{Wiseman:2000} uses
the `scalarstatic' Green function. 
Following Copson\cite{Copson:1928}, Wiseman was able to obtain this Green function by summing the Hadamard series. 
Only by performing the full sum was he able to verify that his Green function satisfied the appropriate boundary conditions.
Linet \cite{Linet:2005} has classified all spacetimes in which the scalarstatic equation is solvable by the Copson ansatz
and the Nariai metric does not fall into any of the classes given. Therefore, instead we work with the mode form 
for the static Green function. This corresponds to the integrand at $\omega=0$ of Eq.~(\ref{full ret Green,mode sum}) (with integral measure $\frac{d\omega}{2\pi}$), 
 \beq \label{Green static}
G_{static}(\R,\Omega; \R',\Omega') =
\sum_{l=0}^{\infty}
(2l+1)P_l(\cos\gamma)
\frac{\pi P_{ -1/2 + i \lam}(-\R_<)P_{ -1/2 + i \lam}(\R_>)} {2 \cosh (\pi \lambda )}
\eeq
where, as before, $\lam = \sqrt{(l+\frac12)^2 +d}$. This equation, having only one infinite series, is amenable to numerical computation.

To regularise the self-force we follow the method of Rosenthal \cite{Rosenthal:2004}, who used a massive field approach to calculate 
the static self-force in Schwarzschild.  Following his prescription, we calculate the derivative of the scalar field and of a massive scalar field. 
In the limit of the field mass going to infinity, we obtain the derivative of the radiative field which is regular.
This method can be carried through to Nariai spacetime, where it  yields the expression
\beq
m a^\R = q^2(1-\R^2)^\frac32  \lim_{\R'\to \R^-} \left [ \partial_\R  G_{static}(\R,\Omega;\R',\Omega) +   \frac{1}{(\R-\R')^2} - \frac{2 (\xi - \frac16)}{1-\R^2} \right]
\eeq

The singular subtraction term may be expressed in a convenient form using the identity~\cite{Erdelyi:1953,Candelas:Jensen:1986},
\beq \label{reg terms}
\int_0^{+\infty}d\lambda 
\lambda \tanh (\pi\lambda)
\frac{\pi P_{ -1/2 + i \lam}(-\R_<)P_{ -1/2 + i \lam}(\R_>)}{\cosh (\pi\lambda)}=\frac{1}{\R_>-\R_<}
\eeq
Subtracting  (\ref{reg terms}) from (\ref{Green static}), we can express the regularised Green function as 
a sum of two well-defined and easily calculated sums/integrals
\beq
G_{static}(\R,\Omega; \R',\Omega) - \frac{1}{\R_>-\R_<} = \mathcal{I} (\R, \R') + \mathcal{J} (\R, \R') 
\eeq
 where
 \beq
\mathcal{I} (\R, \R') = 
   \int_0^{+\infty}d\lambda ~
\lambda \left(1-\tanh (\pi\lambda)\right)
\frac{\pi P_{ -1/2 + i \lam}(-\R_<)P_{ -1/2 + i \lam}(\R_>)}{\cosh (\pi\lambda)}
\eeq
and
\beq
\mathcal{J} (\R,\R')  = \sum_{l=0}^{+\infty}
(l+{\textstyle{\frac12}})
\frac{\pi P_{ -1/2 + i \lam}(-\R_<)P_{ -1/2 + i \lam}(\R_>)} {\cosh (\pi \lambda )} - 
 \int_0^{+\infty}d\lambda \> \lambda
\frac{\pi P_{ -1/2 + i \lam}(-\R_<)P_{ -1/2 + i \lam}(\R_>)}{\cosh (\pi\lambda)}\>.
\eeq
$\mathcal{J} (\R,\R') $  may either be calculated directly as a sum or by 
using the Watson-Sommerfeld transform to write
\beq
\sum\limits_{l=0}^\infty g\left(l+\frac12\right) =  \Re e \left[ \frac{1}{i} \int_\gamma dz \> \tan (\pi z) g(z) \right]  ,
\eeq
where $\gamma$ runs from $0$ to $\infty$ just above the real axis, and for us
\beq
g(z) = z
\frac{\pi P_{ -1/2 + i \sqrt{z^2 +d}}(-\R_<)P_{ -1/2 + i  \sqrt{z^2 +d}}(\R_>)} {\cosh (\pi  \sqrt{z^2 +d} )} .
\eeq
Writing 
\beq
 \tan (\pi z) = i - \frac{2 i} {1 + e^{- 2 \pi i z}},
\eeq
the first term yields 
\beq
\int_{0}^\infty  dx\>g(x) = \int_{\sqrt{d}}^{\infty}d\lambda \> \lambda
\frac{\pi P_{ -1/2 + i \lam}(-\R_<)P_{ -1/2 + i \lam}(\R_>)}{\cosh (\pi\lambda)}.
\eeq
The contribution from the second term can be best evaluated by rotating the original contour  to a contour
 $\gamma'$, running from $0$ to $i \infty$ just to the right of the imaginary axis.
 This is permitted since the Legendre functions are analytic functions of their parameter
and the contribution from the arc at infinity vanishes for our choice of $g(z)$.
From the form of $g(z)$ it is clear that it possesses poles along the contour $\gamma'$ but these give a 
purely imaginary contribution to the integral.  We conclude that
\begin{align}
   \mathcal{J} (\R, \R') &= - \int_0^{\sqrt{d}}   t dt\>  \tanh(\pi \sqrt{d-t^2})\frac{\pi P_{ -1/2 +i t }(-\R_<)P_{ -1/2 +i t }(\R_>)}{\cosh(\pi t)} 
   + \nonumber \\ & \qquad + \mathcal{P} \int_0^{+\infty} \frac{2\pi  t dt} { (1+e^{2\pi \sqrt{d+t^2}})\cos(\pi t)} P_{ -1/2 -t }(-\R_<)P_{ -1/2 -t }(\R_>)
\end{align}
where $\mathcal{P}$ denotes the Principal Value. These integrals and that defining $\mathcal{I} (\R,\R')$ and their derivatives 
with respect to $\R$ are very rapidly convergent and easily calculated.


\subsection{Matched Expansions for Static Particle} \label{subsec:matched-static}
In Sec.~\ref{sec:matched-expansions}, we outlined the method of matched expansions. In this subsection, we show how to apply the method to a specific case: the computation of the self-force on a static particle in the Nariai spacetime. 

The four velocity of the static particle is simply 
\beq
u^\R = u^\theta = u^\phi = 0, \quad u^t = (1-\R^2)^{-1/2}
\eeq
and hence $d \tau^\prime = (1-\R^2)^{1/2} dt^\prime$. 
We find from Eqs.~(\ref{eq:ma}), (\ref{eq:dmdtau}) and (\ref{eq:rad-field-deriv}) that $ma^t = ma^\theta = ma^\phi = 0$ and 
\begin{align}
 ma^\R &= q^2\left( \frac{1}{3} \dot{a}^\R + \lim_{\eps \rightarrow 0^+} \int_{-\infty}^{\tau-\eps}  g^{\R\R} \partial_\R \Gret (z(\tau), z(\tau^\prime)) d \tau^\prime\right) \label{rad-acc-eq} \\
 \frac{dm}{d\tau} &= -q^2 \left( \frac{1}{12}(1 - 6\xi) R + \left( 1-\R^2 \right)^{-1/2} \lim_{\eps \rightarrow 0^+} \int_{-\infty}^{\tau-\eps} \partial_t \Gret (z(\tau), z(\tau^\prime)) d \tau^\prime\right)  \label{mass-loss-eq}
\end{align}
where $\partial_{\mu}$ denotes partial differentiation with respect to the coordinate $x^\mu$. 
We note that in the tail integral of the mass loss equation, (\ref{mass-loss-eq}),  the time derivative $\partial_t$ may be replaced with $-\partial_{t^\prime}$ since the retarded Green function is a function of $(t-t')$. Hence we obtain a total integral,
\begin{align}
\left( 1-\R^2 \right)^{-1/2} \lim_{\eps \rightarrow 0^+} \int_{-\infty}^{\tau-\eps} \partial_t \Gret (z(\tau), z(\tau^\prime)) d \tau^\prime 
&= - \lim_{\eps \rightarrow 0^+} \int_{-\infty}^{t - \eps} \partial_{t^\prime} \Gret (z(\tau), z(\tau^\prime)) d t^\prime \nonumber \\
&= - \lim_{\eps \rightarrow 0^+}\left[ \Gret (x,x') \right]_{t'=-\infty}^{t'=t-\eps}
\end{align}
The total integral depends only on the values of the Green function at the present time and in the infinite past ($t'\rightarrow \infty$). The QNM sum expressions for the Green function (e.g. Eq.~(\ref{GF-inf})) are zero in the infinite past, as the quasinormal modes decay exponentially. The value of the Green function at coincidence ($t^\prime \rightarrow t$) is found from the coincidence limit of the function $-V(x,x')$ in the Hadamard form (\ref{eq:Hadamard}). It is
$
\frac{1}{12}(1 - 6\xi) R
$, 
which exactly cancels the local contribution in the mass loss equation (\ref{mass-loss-eq}). It is no surprise to find that this cancellation occurs -- the local terms were originally derived from the coincidence limit of the Green function. In fact, because $\frac{d \Phi_R}{d \tau} = 0$ due to time-translation invariance, we can see directly from the original equation (\ref{eq:dmdtau}) that the mass loss is zero in the static case. 

Now let us consider the radial acceleration (\ref{rad-acc-eq}). The acceleration keeping the particle in a static position is constant ($\dot{a}^\R = 0$). The remaining tail integral may be split into two parts,
\begin{align}
ma^\R &= 
 q^2  (1-\R^2)^{3/2}  \left( - \lim_{\eps \rightarrow 0^+} \int_{t-\Delta t}^{t - \eps} \partial_{\R} V  (z(t), z(t^\prime)) d t^\prime + \int_{-\infty}^{t- \Delta t} \partial_{\R} \Gret  (z(t), z(t^\prime)) d t^\prime \right)
 \label{eq:sf-matched}
\end{align}

For the first part of (\ref{eq:sf-matched}), we use the quasilocal calculation of $V(x,x')$ from Sec.~\ref{subsec:QL}. As $V(x,x'$) is given as a power series in $(\R-\R')$ and $(t-t')$, the derivatives and integrals can be done termwise and are straightforward. The quasilocal integral contribution is therefore simply
\beq
\lim_{\eps \rightarrow 0} \int_{t-\Delta t}^{t - \eps} \partial_{\R} V (z(t), z(t^\prime)) d t^\prime = \frac{1}{2} \sum_{k=0}^{\infty} \frac{1}{(2k+1)!} \delta_\R v_{k0} ( \Delta t )^{2k+1}.
\label{eq:sf-ql}
\eeq

The second part of (\ref{eq:sf-matched}) can be computed using the QNM sum (\ref{G-arbitrary}). To illustrate the approach, let us rewrite (\ref{G-arbitrary}) as 
\beq
\GQNM (\R,t;\R',t;)= \text{Re} \sum_{l n} \mathcal{G}_{ln}(\R^\prime) e^{- i \omega_{ln} (t - t^\prime - \Rs - \Rs^\prime)} \unorm_{ln}(\R).
\eeq
Applying the derivative with respect to $\R$ and taking the integral with respect to $t^\prime$ leads to
\begin{eqnarray}
\int_{-\infty}^{t - \Delta t} \partial_{\R} \Gret  (z(t), z(t^\prime)) d t^\prime &=&
\left( \frac{d\R}{d\Rs} \right)^{-1}  \int_{-\infty}^{t - \Delta t} \partial_{t^\prime} \GQNM dt^\prime
+ \text{Re} \sum_{ln} \int_{-\infty}^{t-\Delta t} \mathcal{G}_{ln} e^{- i \omega_{ln} (t - t^\prime - \Rs - \Rs^\prime)}  \frac{d \unorm_{ln}}{d \R} dt^\prime \nn
 \\
&=& \left( 1-\R^2 \right)^{-1} \left[ \Gret \right]^{t'=t - \Delta t} + \sum_{ln} \frac{ i \mathcal{G}_{ln} }{ \omega_{ln} } e^{- i \omega_{ln} (\Delta t - \Rs - \Rs^\prime)} \frac{d \unorm_{ln}}{d \R}
\label{eq:sf-dp}
\end{eqnarray}
It is straightforward to find the derivative of the radial wavefunction from the definition (\ref{unorm-ser}). 
In Sec.~\ref{subsec:nummeth} we outline two methods for numerically computing mode sums such as (\ref{eq:sf-dp}).

The self-force computed via (\ref{eq:sf-matched}), (\ref{eq:sf-ql}) and (\ref{eq:sf-dp}) should be independent of the choice of the matching time (we verify this in Sec.~\ref{subsec:results:SF}). This invariance provides a useful test of the validity of our matched expansions. Additionally, through varying $\Delta \tau$ we may estimate the numerical error in the self-force result. 

\subsection{Numerical Methods for Computing Mode Sums}\label{subsec:nummeth}
The static-self-force calculation requires the numerical calculation of mode sums like (\ref{eq:sf-dp}). 
We used two methods for robust numerical calculations: (1) `smoothed sum', and (2) Watson transform (described previously in Sec.~\ref{subsec:watson}). We see in Sec. \ref{sec:results} that the results of the two methods are consistent.

The `smoothed sum' method is straightforward to describe and implement.  Let us suppose that we wish to extract a numerical value from an infinite sum
\beq
 \sum_{l=0}^\infty a_l    \label{sum1}
\eeq
which may not be absolutely convergent (i.e. $|a_{l+1} / a_l| \ge 1$). We may instead compute the finite sum
\beq
S(\lmax) = \sum_{l=0}^{l_\infty} a_l e^{-l^2 / 2 \lmax^2}  \label{num-meth1}
\eeq
where $l_\infty$ is large enough to suppress any high-$l$ oscillations in the result (typically $l_\infty > 4\lmax$). We find that (\ref{num-meth1}) is a good approximation to (\ref{sum1}) provided we are not within $\delta t \sim 1/\lmax$ of a singularity of the Green function. Increasing the cutoff $\lmax$ therefore improves the resolution of the singularities.



\section{Results}\label{sec:results}
We now present a selection of results from our numerical calculations. 
In Sec. \ref{subsec:results:GF} the distant past Green function is examined. We plot the Green function as a function of coordinate time $t - t'$ for fixed spatial points. A four-fold singularity structure is observed. In Sec.~\ref{subsec:results-asymptotics} we test the asymptotic approximations of the singular structure, derived in Secs. \ref{subsec:Hadamard} and \ref{subsec:Poisson} (Eqs. \ref{sing-Poisson-2} and \ref{GNdir-Hadamard}). We show that the `fundamental mode' ($n=0$) series (\ref{G-n0}) is a good approximation of the exact result (\ref{GF-inf}), if a `time-offset' correction is applied. In Sec. \ref{subsec:results:matched} the quasilocal and distant past expansions for the Green function are compared and matched. We show that the two methods for finding the Green function are in excellent agreement for a range of matching times $\Delta \tau$. In Sec. \ref{subsec:results:SF} we consider the special case of the static particle. We present the Green function, the radiative field and the self-force in turn. The radial self-force acting on the static particle is computed via the matched expansion method (described in Secs. \ref{sec:matched-expansions} and \ref{subsec:matched-static}), and plotted as a function of coordinate $\R$, and compared with the result derived in Sec.~\ref{subsec:static full Green}.

 \subsection{The Green Function Near Infinity from Quasinormal Mode Sums}\label{subsec:results:GF}
 


Let us begin by looking at the Green function for fixed points near spatial infinity, $\R = \R^\prime \rightarrow 1$ (i.e. $\Rs = \Rs^\prime \rightarrow +\infty$). The Green function may be computed numerically by applying either the Watson transform (Sec.~\ref{subsec:watson}) or the `smoothed sum' method (Sec.~\ref{subsec:nummeth}) to the QNM sum (\ref{GF-inf}).

\begin{figure}
 \begin{center}
  \includegraphics[width=10cm]{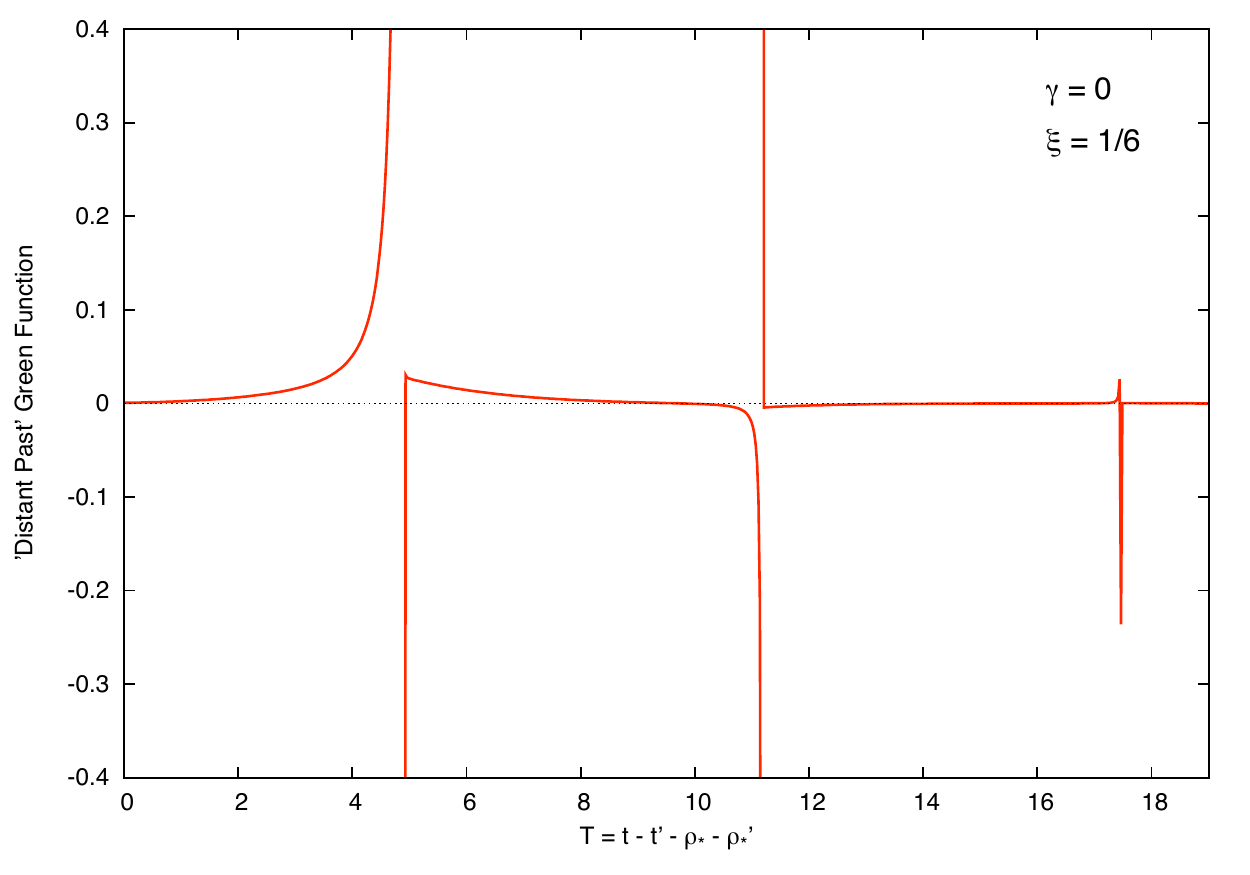}
 \end{center}
 \caption{\emph{Distant Past Green Function for spatially-coincident points near infinity ($\R=\R^\prime \rightarrow 1$, $\gam = 0$)}. The Green function was calculated from mode sum (\ref{GF-inf}) numerically using the smoothed sum method (\ref{num-meth1}) with $\lmax = 200$ and curvature coupling factor $\xi = 1/6$.}
 \label{gf-inf-gam0}
\end{figure}

Figure \ref{gf-inf-gam0} shows the Green function for fixed spatially-coincident points near infinity ($\R = \R^\prime \rightarrow 1$, $\gam = 0$). The Green function has been calculated from series (\ref{GF-inf}) using the `smoothed sum' method. It is plotted as a function of QNM time, $T =t- t^\prime - (\Rs + \Rs^\prime)$. We see that singularities occur at the times (\ref{sing-time-inf}) predicted by the geodesic analysis of Sec. \ref{subsec:geodesics}. In this case, $T_{C} =  \ln[\sinh^2 (N \pi)] \approx 4.893$, $11.180$, $17.463$, etc. At times prior to the first singularity at $T \approx 4.893$, the Green function shows a smooth power-law rise. At the singularity itself, there is a feature resembling a delta-distribution, with a negative sign. Immediately after the singularity the Green function falls close to zero (although there does appear a small `tail'). This behaviour is even more marked in the case $\xi = 1/8$ (not shown). A similar pattern is found close to the second singularity at $T \approx 11.180$, but here the Green function takes the opposite sign, and its amplitude is smaller.

\begin{figure}
 \begin{center}
  \includegraphics[width=10cm]{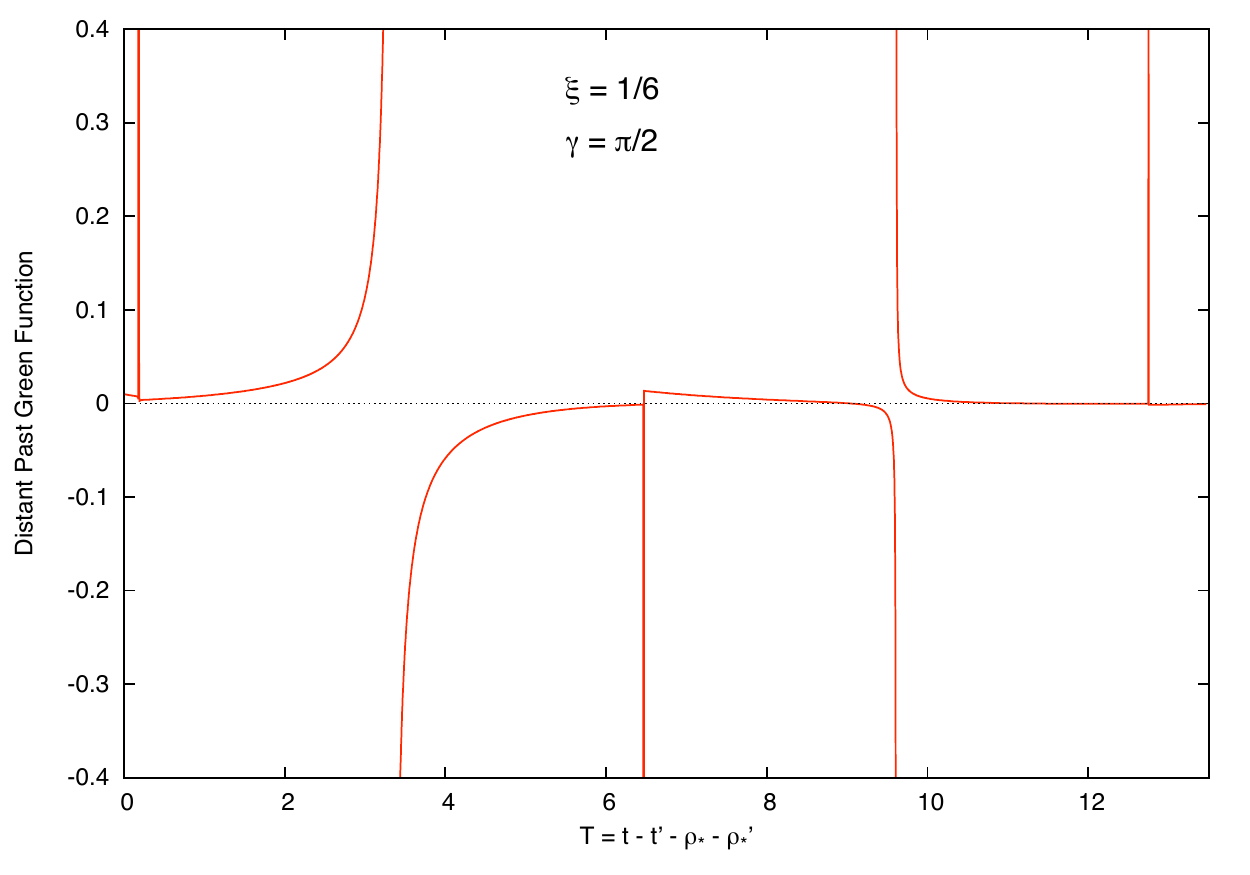}
 \end{center}
 \caption{\emph{Distant Past Green Function near spatial infinity ($\R=\R^\prime \rightarrow 1$) for points separated by angle $\gam = \pi /2 $}. The Green function was calculated from the `fundamental mode' approximation (\ref{G-n0}) numerically using the smoothed sum method with $\lmax = 1000$. Note the four-fold singularity structure (see text). }
 \label{gf-inf-piby2}
\end{figure} 

Figure \ref{gf-inf-piby2} shows the Green function for points near infinity ($\R=\R^\prime \rightarrow1$) separated by an angle of $\gam = \pi / 2$. The Green function shown here is computed from the `fundamental mode' approximation (\ref{G-n0}), again using the `smoothed sum' method. In this case, the singularities occur at `periodic' times (\ref{sing-time-periodic}), given by $T = \Delta \phi - 2 \ln 2 \approx 0.1845, 3.326, 6.468,$ etc., where $\Delta \phi = \pi / 2, 3 \pi /2, 5 \pi /2, \ldots$. As discussed, there is a one-to-one correspondence between singularities and orbiting null geodesics, and the four-fold singularity pattern predicted in Sec. \ref{subsec:Poisson} (\ref{GNdir-Hadamard}) and Sec.~\ref{subsec:Hadamard} (\ref{sing-Poisson-2}) is clearly visible. Every `even' singularity takes the form of a delta distribution. Numerically, the delta distribution is manifest as a Gaussian-like spike whose width (height) decreases (increases) as $\lmax$ is increased. By contrast (for $\gam \neq 0, \pi$), every `odd' singularity diverges as $1/(T-T_c)$; it has  antisymmetric wings on either side. The singularity amplitude diminishes as $T$ increases.

\subsection{Asymptotics and Singular Structure}\label{subsec:results-asymptotics}
\begin{figure}
 \begin{center}
  \includegraphics[width=14cm]{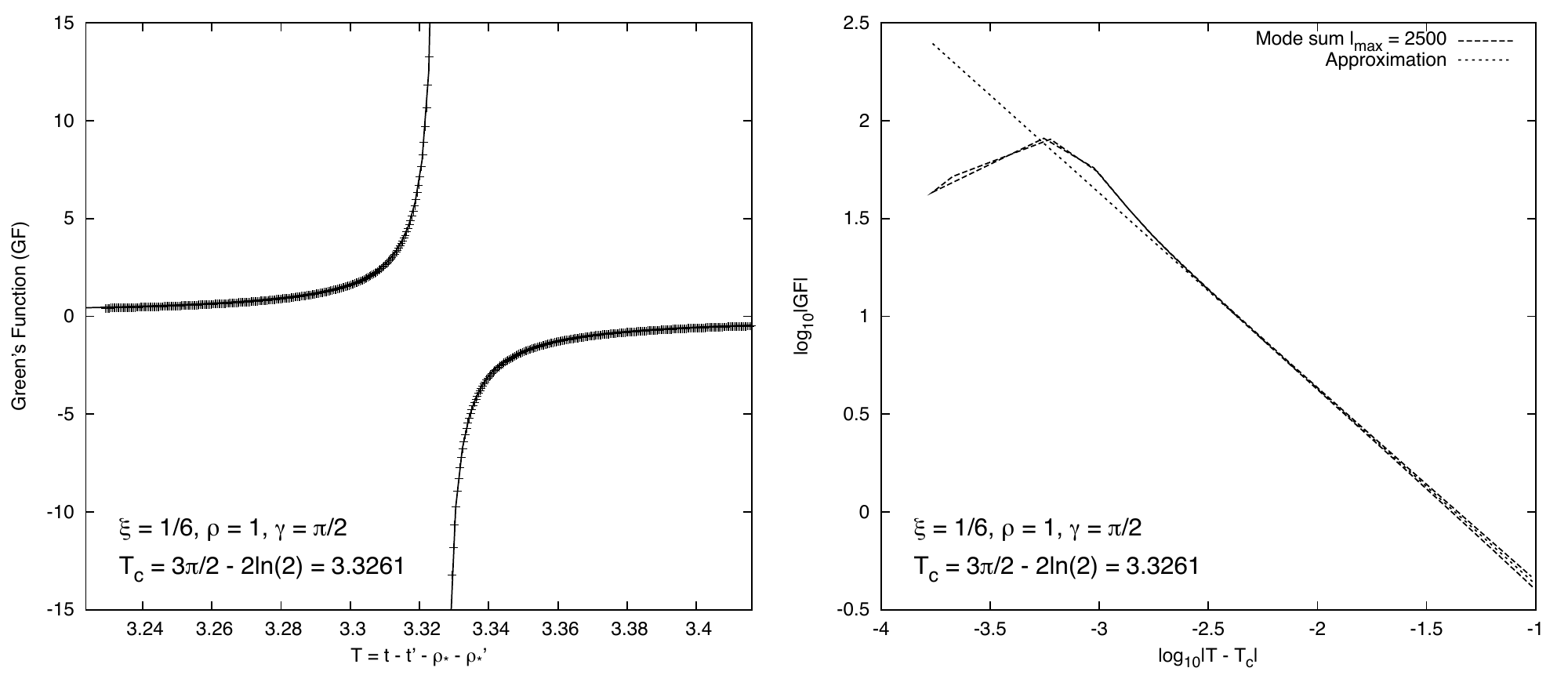}
 \end{center}
 \caption{\emph{Green function near the singularity arising from a null geodesic passing through an angle $\Delta \phi = 3\pi/2$ and with $\R = \R^\prime \rightarrow \infty$}. The `fundamental mode' ($n=0$) Green function (\ref{G-n0}) (with $\lmax = 2500$) is compared with approximations (\ref{sing-Poisson-2}) and (\ref{GNdir-Hadamard}) from considering high-$l$ asymptotics. The approximations give $\Gret \sim -0.04268 / (T - T_c)$ and $\Gret \sim -0.04344 / (T-T_c)$, respectively.  The left panel shows the Green function in the vicinity of the (`periodic') singularity at $T_c  = 3\pi/2 - 2 \ln 2 \approx 3.3261$. The right panel shows the same data on a log-log scale, and compares the mode sum (dashed) with the approximation (dotted). The discrepancy close to the singularity may be improved by increasing $\lmax$.}
 \label{gf-sing-3piby2}
\end{figure}

The analyses of Sec.~\ref{subsec:Poisson} and Sec.~\ref{subsec:Hadamard} yielded approximations for the singularity structure of the Green function. In particular, Eq.~(\ref{sing-Poisson-2}) gives an estimate for the amplitude of the `odd' singularities as $\R =\R^\prime \rightarrow \infty$. We tested our numerical computations against these predictions. Figure \ref{gf-sing-3piby2} shows the Green function near the singularity associated with the null geodesic passing through an angle $\Delta \phi = 3\pi / 2$. The left plot compares the numerically-determined Green function (\ref{G-n0}) with the asymptotic prediction (\ref{sing-Poisson-2}). The right plot shows the same data on a log-log plot. The asymptotic prediction (\ref{sing-Poisson-2}) is a straight line with gradient $-1$, and it is clear that the numerical data is in excellent agreement.

\begin{figure}
 \begin{center}
  \includegraphics[width=8cm]{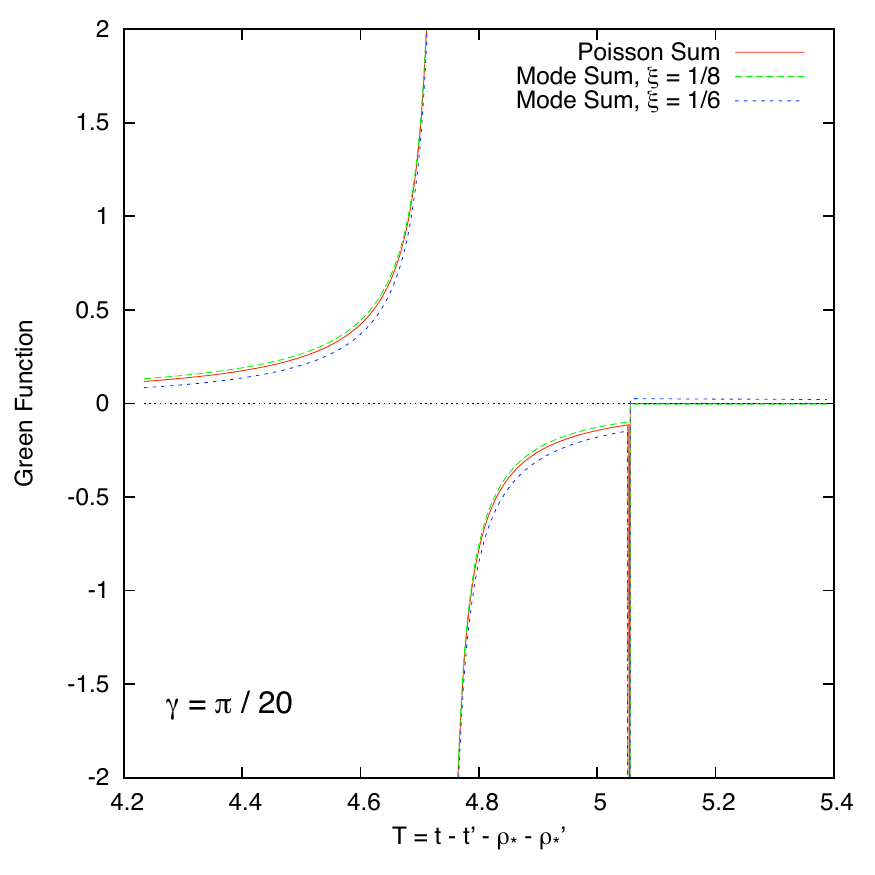}
  \includegraphics[width=8cm]{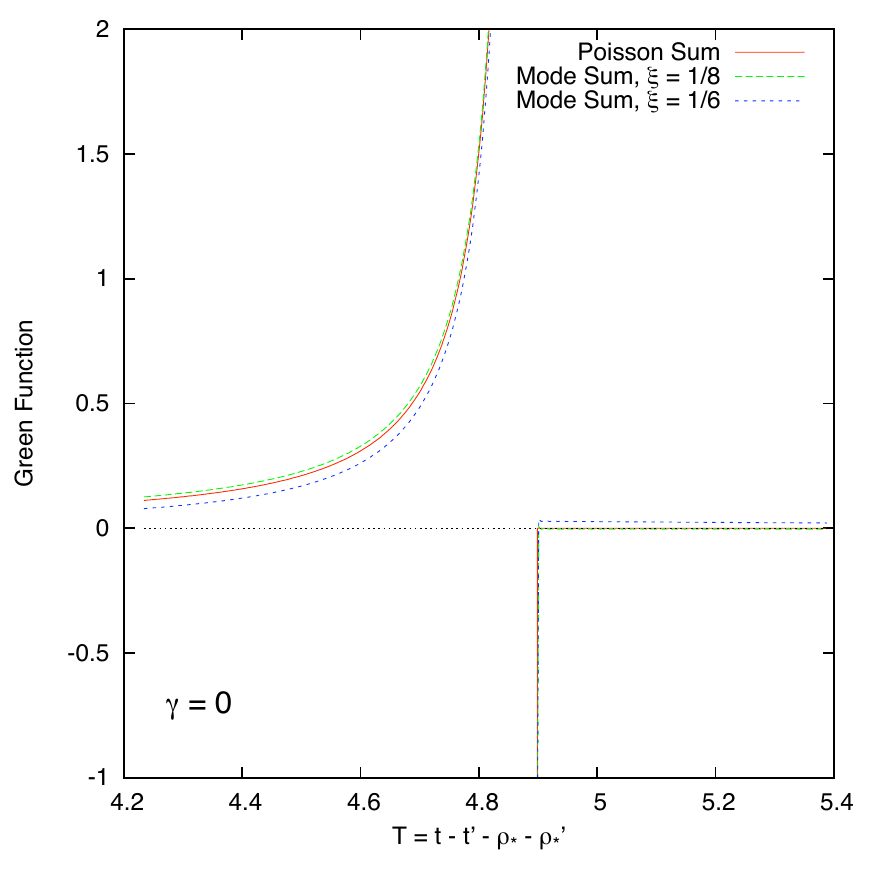}
 \end{center}
 \caption{\emph{Singularities of the `Fundamental Mode' Green function (\ref{G-n0}) compared with asymptotics from the Poisson sum (\ref{eq:I1-asymp}) and (\ref{eq:I2-asymp})}. The left plot shows a small angular separation $\gam = \pi / 20$, and the right plot shows coincidence $\gam = 0$, for $\rho,  \rho' \rightarrow 1$.}
 \label{fig:I1I2-numerical}
\end{figure}

Improved asymptotic expressions for the singular structure of the fundamental mode Green function were given in (\ref{eq:I1-asymp}) and (\ref{eq:I2-asymp}). These asymptotics are valid all the way up to $\gam = 0$. Figure \ref{fig:I1I2-numerical} compares the asymptotic expressions (\ref{eq:I1-asymp}) and (\ref{eq:I2-asymp})(solid line) with numerical computations (broken lines) from the mode sum (\ref{G-n0}). It is clear that the asymptotics  (\ref{eq:I1-asymp}) and (\ref{eq:I2-asymp}) are in excellent agreement with the numerically-determined Green function. Closest agreement is found near the singular times, but the asymptotics provide a remarkably good fit over a range of $t$. 

\begin{figure}
 \begin{center}
  \includegraphics[width=10.0cm]{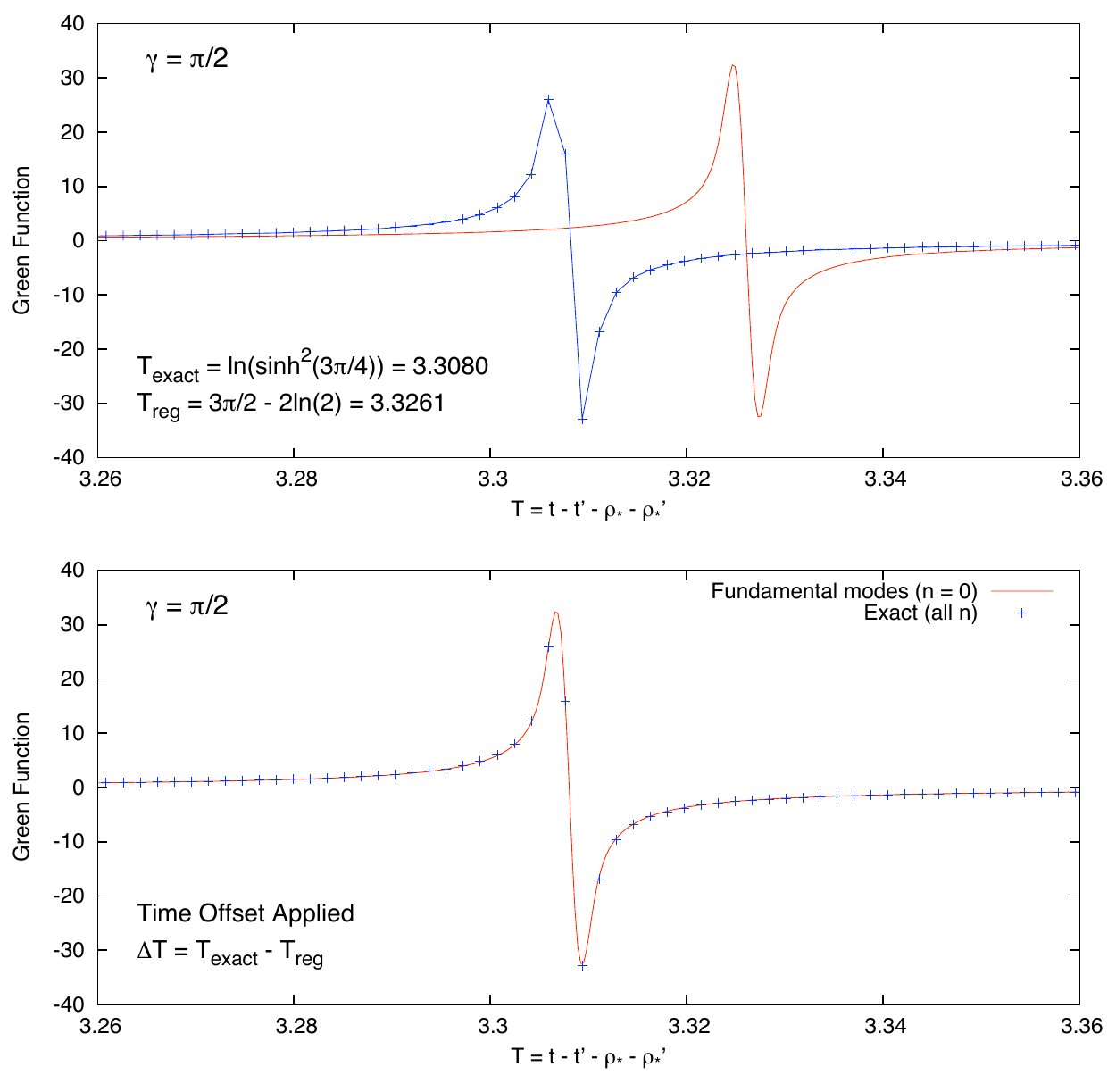}
 \end{center}
 \caption{\emph{Singularities of `Fundamental Mode' approximation}.  The top plot shows the Green function near spatial infinity ($\R = \R^\prime\rightarrow 1$, $\Rs = \Rs^\prime \rightarrow \infty$ ), for an angular separation $\gam = \pi / 2$. It compares the exact solution (\ref{GF-inf}) [red] and the approximation from the fundamental modes [blue], as a function of time $T = t - t^\prime - \Rs - \Rs^\prime$. The singularities occur at two distinct times, marked $T_{exact}$ and $T_{reg}$ respectively. If a time offset is applied to the `Fundamental Mode' approximation (see text), then we find the singularities look remarkably similar (lower plot). }
 \label{fig:singularities}
\end{figure}

In Fig.~\ref{fig:singularities}, the `fundamental mode' ($n=0$) approximation (\ref{G-n0}) is compared with the exact QNM Green function (\ref{GF-inf}). Away from singularities, the former is found to be a good approximation to the latter. However, close to singularities this is not the case. The singularities of the `fundamental mode' approximation (\ref{G-n0}) occur at slightly different times to the singularities of the exact solution (\ref{GF-inf}), as discussed in Sec. \ref{subsec:large-l}. For the fundamental mode series (\ref{G-n0}), the singularity times $T_{reg}^N$ given in Eq.~(\ref{T-periodic}) are periodic. For the exact solution (\ref{GF-inf}), the singularity times $T_{exact}^{N}$ are precisely the `null geodesic times' given in Eq.~(\ref{sing-time-inf}). Remarkably, if we apply a \emph{singularity time offset} to the `fundamental mode' approximation ($T \rightarrow T + \Delta T$  where $\Delta T = T_{exact}^{N} - T_{reg}^{N}$) we find that the `fundamental mode' Green function is an almost perfect match to the exact Green function. This is clearly shown in the lower plot of Fig.~\ref{fig:singularities}. Comparing the series (\ref{G-n0}) and (\ref{GF-inf}) we see that, in both cases, the magnitude of the terms in the series increases as $(l+1/2)^{1/2}$ in the large-$l$ limit. This observation raises the possibility that the $n=0$ modes may give the essential features of the full solution; if true, this would certainly aid the analysis of the Schwarzschild case, where it is probably not feasible to perform a sum over $n$ analytically.

 \subsection{Matched Expansions: Quasilocal and Distant Past}\label{subsec:results:matched}

Let us now turn our attention to the match between quasilocal and distant past Green functions. The 
quasilocal expansion (\ref{eq:CoordGreen}) is valid within the convergence radius of the series, $t-t^\prime<  t_{QL}$, while the QNM sum is convergent at `late' times, $t-t' > \Rs + \Rs'$. 
Hence, a matched expansion method will only be practical if the quasilocal and distant past Green functions overlap in an intermediate regime $\Rs+\Rs^\prime < t-t' < t_{QL}$. It is expected that the convergence radius of the quasilocal series, $ t_{QL}$, will lie within the normal neighbourhood, $t_{NN}$, of spacetime point $x$. The size of the normal neighborhood is limited by the earliest time at which spacetime points $x$ and $x'$ may be connected by more than one non-spacelike geodesic. Typically this will happen when a null geodesic has orbited once, taking a time $t_{NN} > \Rs + \Rs^\prime$, so we can be optimistic that an intermediate regime will exist. To test this idea, we computed the quasilocal Green function using (\ref{eq:CoordGreen}), and the distant past Green function (\ref{G-arbitrary}) for a range of situations.

\begin{figure}
 \begin{center}
  \includegraphics[width=10cm]{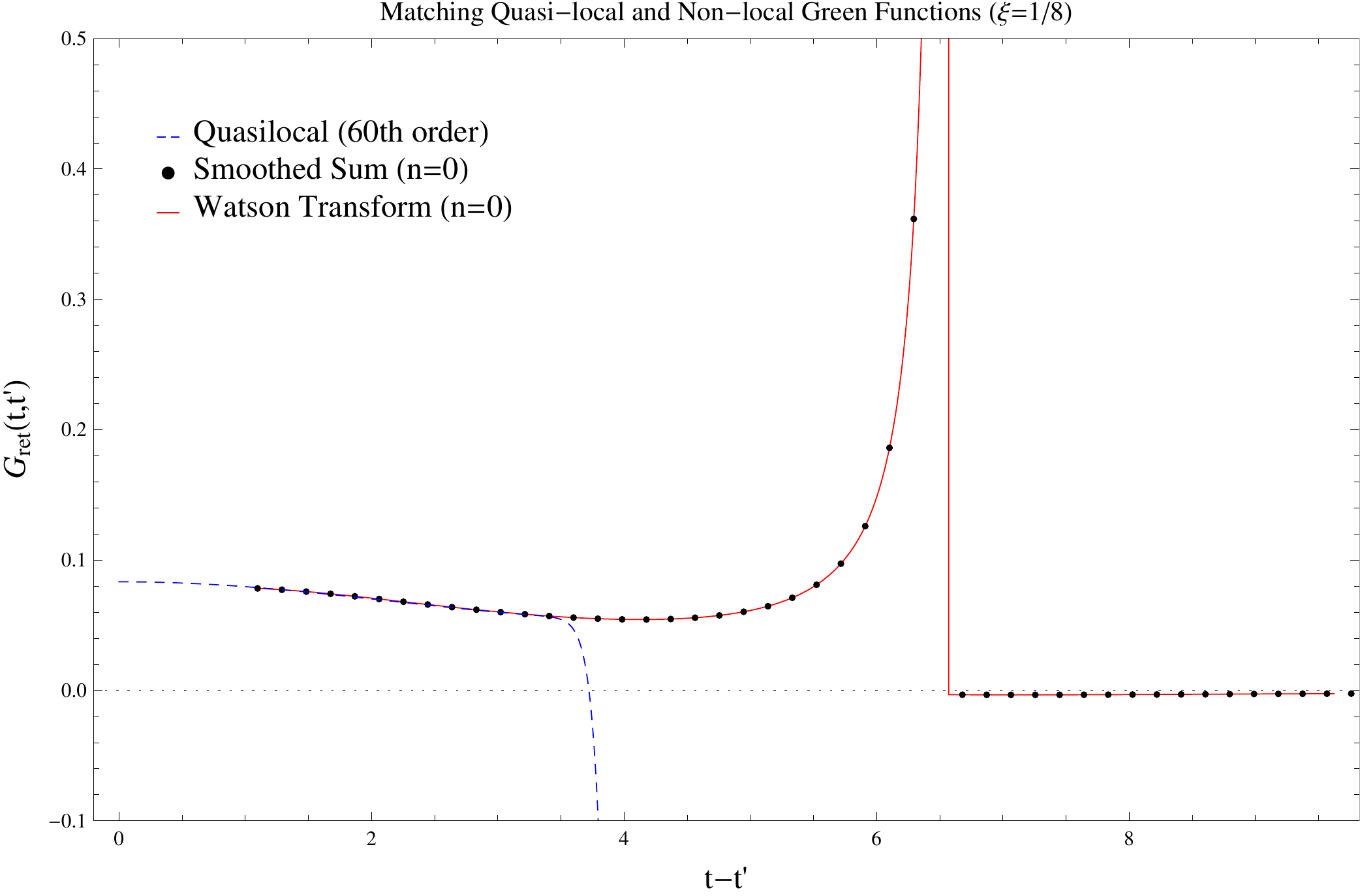}
 \end{center}
 \caption{\emph{Matching of the Quasilocal and Distant Past Green Functions} for $\xi=1/8$ and for a static particle at $\R = \R^\prime = 0.5$. Here, the Green functions are plotted as functions of coordinate time, $t - t^\prime$ (\emph{not} `QNM time' as in most other plots). The black dots and red line show the results of the `smoothed sum' and `Watson transform' methods applied to compute the QNM sum (\ref{G-n0}) (see text). The blue dashed line shows the quasilocal series expansion taken to order $(t- t^\prime)^{60}$. The distant past Green function cannot be computed for early times $t-t^\prime< \Rs + \Rs^\prime = 2 \tanh^{-1}(1/2) = 1.099$, whereas the quasilocal series diverges at large $t-t^\prime \gtrsim 3.46$. In the intermediate regime, we find excellent agreement (see also Figs. \ref{matching} and \ref{matcherr}). Note that the quasilocal Green function tends to $\frac{1}{12}(1 - 6 \xi)R = \frac{1}{12}$ in the limit $t^\prime \rightarrow t$.}
 \label{matching2}
\end{figure}

Figure \ref{matching2} shows the retarded Green function as a function of coordinate time $t - t'$ for a static particle at $\R = \R^\prime = 0.5$. At early times, the quasilocal Green function is well-defined, but the distant past Green function is not. Conversely, at late times the quasilocal series is not convergent. At intermediate times $1.099 < \Delta t \lesssim 3.45$, we find an excellent match. Figure \ref{matching2} also shows that the results of the two numerical methods for evaluating QNM sums are equivalent. That is, the Green function found from the Watson transform (Sec.~\ref{subsec:watson}, red line) coincides with the Green function calculated by the method of smoothed sums (Sec.~\ref{subsec:nummeth}, black dots).

\begin{figure}
 \begin{center}
  \includegraphics[width=8cm]{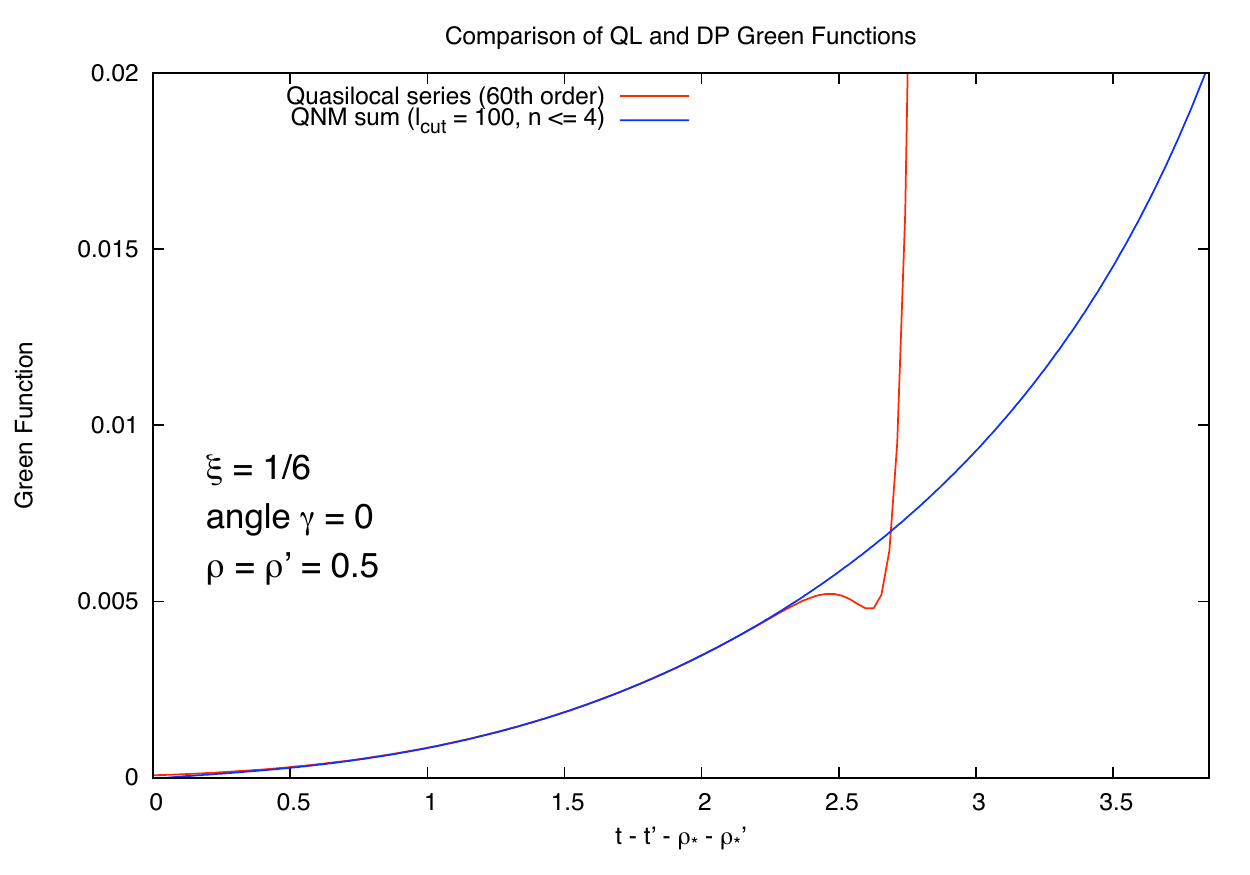}
  \includegraphics[width=8cm]{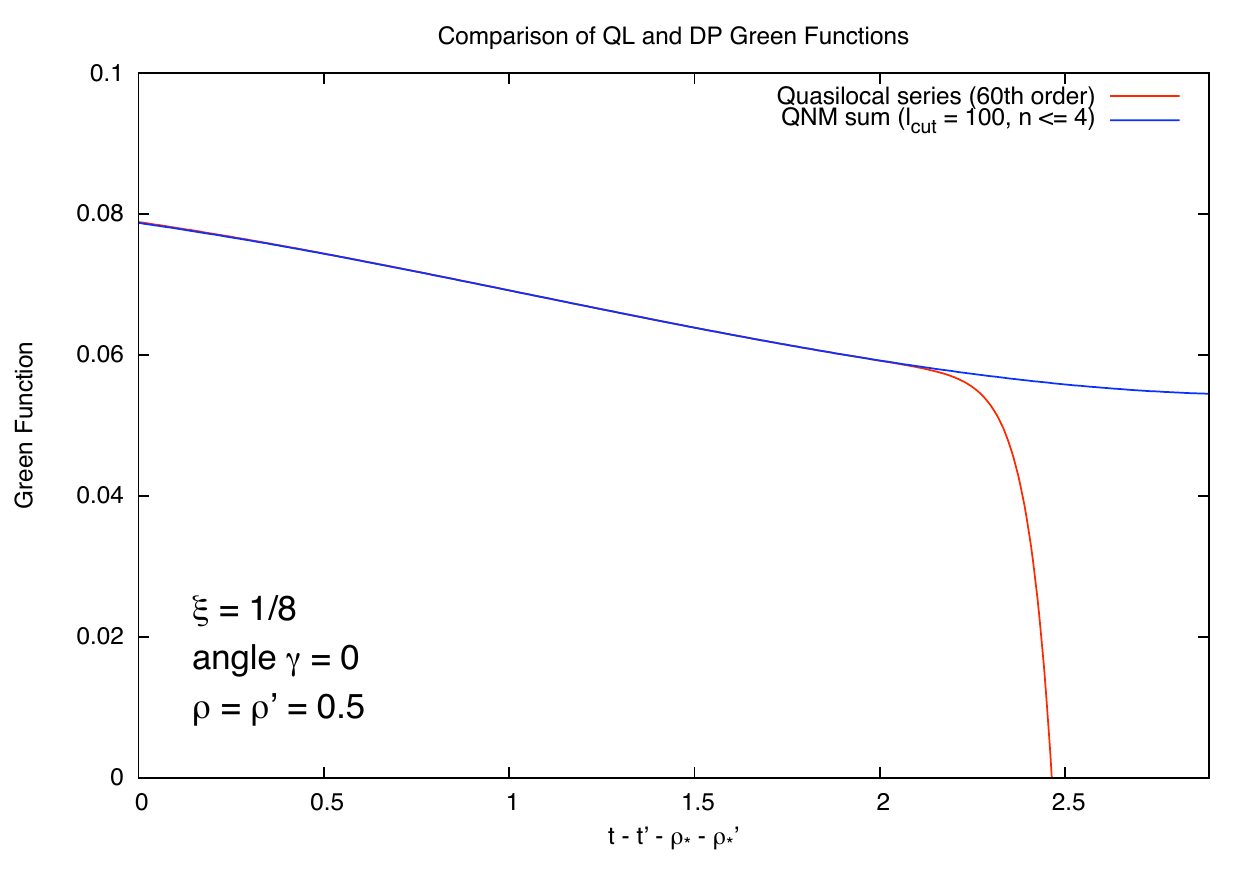}
 \end{center}
 \caption{\emph{Matching of the Quasilocal and Distant Past Green Functions.} The left plot shows curvature coupling $\xi = 1/6$ and the right plot shows $\xi = 1/8$, for a static particle at $\R = \R^\prime = 0.5$. Note the timescale on the horizontal axis, $T = t - t^\prime - \Rs - \Rs^\prime$. The distant past Green function cannot be computed for $T < 0$, whereas the quasilocal series clearly diverges at large $T$. In the intermediate regime, we find excellent agreement.}
 \label{matching}
\end{figure}

Let us now examine the matching procedure in more detail. Figure \ref{matching} shows the match between the distant past and quasilocal Green functions, computed from (\ref{G-arbitrary}) and (\ref{eq:CoordGreen}), in the case $\R = \R^\prime = 0.5$. 
In Fig. \ref{matching}, the left plot shows the case for conformal coupling $\xi = 1/6$ and the right plot shows the case for $\xi = 1/8$. Note that Green function tends to the constant value $\frac{1}{12}(1 - 6 \xi)R$ in the limit $\Delta t \rightarrow 0^+$. In both cases, we find that the fit between `quasilocal' and `distant past' Green functions is good up to nearly the radius of convergence of the quasilocal series. 

\begin{figure}
 \begin{center}
  \includegraphics[width=10cm]{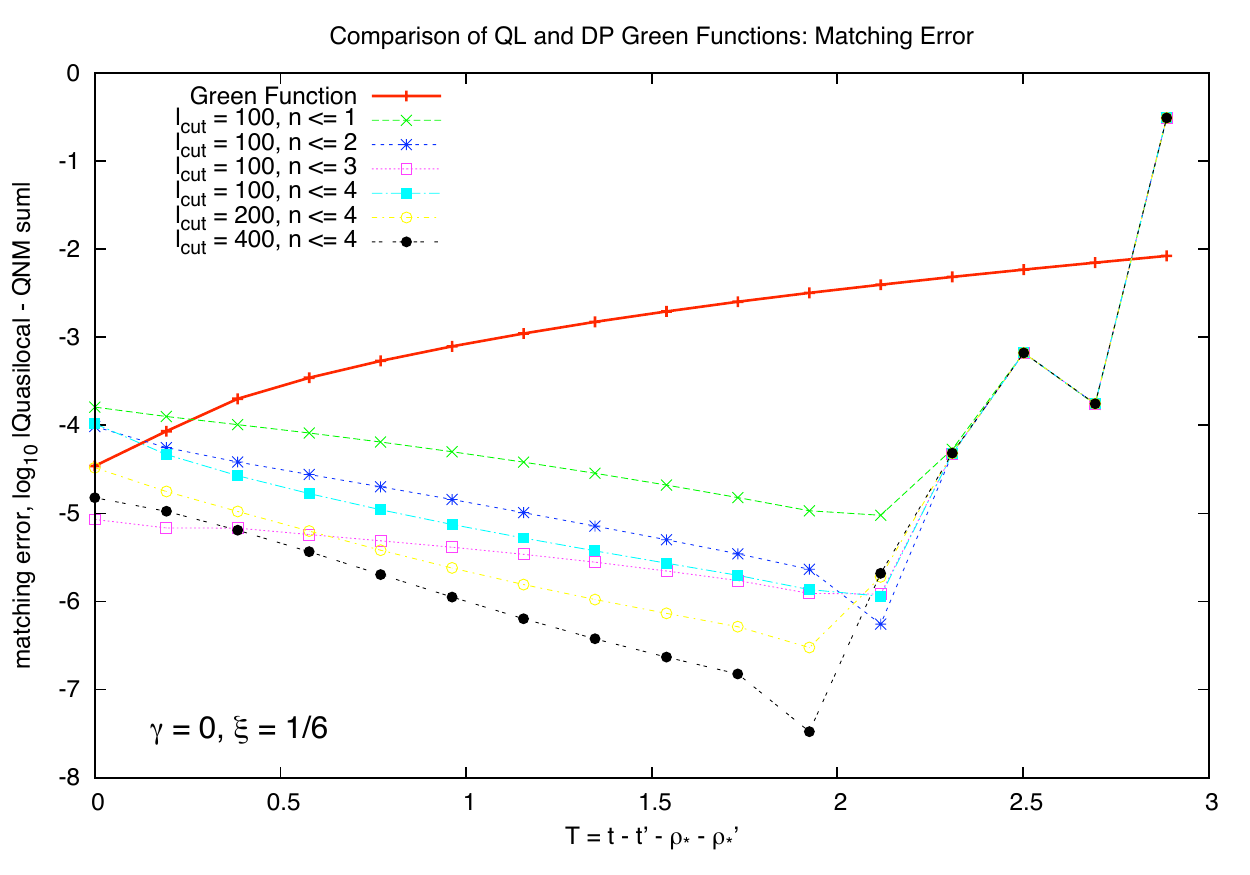}
 \end{center}
 \caption{\emph{Error in Matching the Quasilocal and Distant Past Green Functions.} This plot shows the difference between the quasilocal and distant past Green functions in the matching regime. The magnitude of the Green function as a function of time $T = t - t^\prime - \Rs - \Rs^\prime$ is shown as a solid red line. The broken lines show the `matching error': the difference between the quasilocal and QNM sum Green functions, for various $\lmax$ and $\nmax$ in (\ref{num-meth1}) (see Sec.~\ref{subsec:nummeth}). Note the logarithmic scale on the vertical axis. It is clear that the matching error is reduced by increasing $\nmax$ and $\lmax$, and that the best agreement is found close to the radius of convergence of the quasilocal series (at $T \sim 1.9$). The plot shows that matching accuracy of above one part in $10^4$ is achievable.}
 \label{matcherr}
\end{figure}

Figure \ref{matcherr} quantifies the accuracy of the match between quasilocal and distant past Green functions. Here, we have used the `smoothed sum' method (Sec.~\ref{subsec:nummeth}) to compute the distant past Green function from (\ref{G-arbitrary}). To apply this method, we must choose appropriate upper limits for $l$ (angular momentum) and $n$ (overtone number).  We have experimented with various cutoffs $\lmax$ and $\nmax$. As expected, better accuracy is obtained by increasing $\lmax$ and $\nmax$, although the run time for the code increases commensurately. With care, a relative accuracy of one part in $10^4$ to $10^5$ is possible. This accuracy is sufficient for confidence in the self-force values computed via matched expansions, presented in Sec. \ref{subsec:results:SF}.   

 \subsection{The Self-Force on a Static Particle}\label{subsec:results:SF}
In this section, we present a selection of results for a specific case: a `static' particle at fixed spatial coordinates. Our goal is to compute the self-force as a function of $\R$, to demonstrate the first practical application of the Poisson-Wiseman-Anderson method of matched expansions \cite{Poisson:Wiseman:1998, Anderson:Wiseman:2005}.

\begin{figure}
 \begin{center}
  \includegraphics[width=10cm]{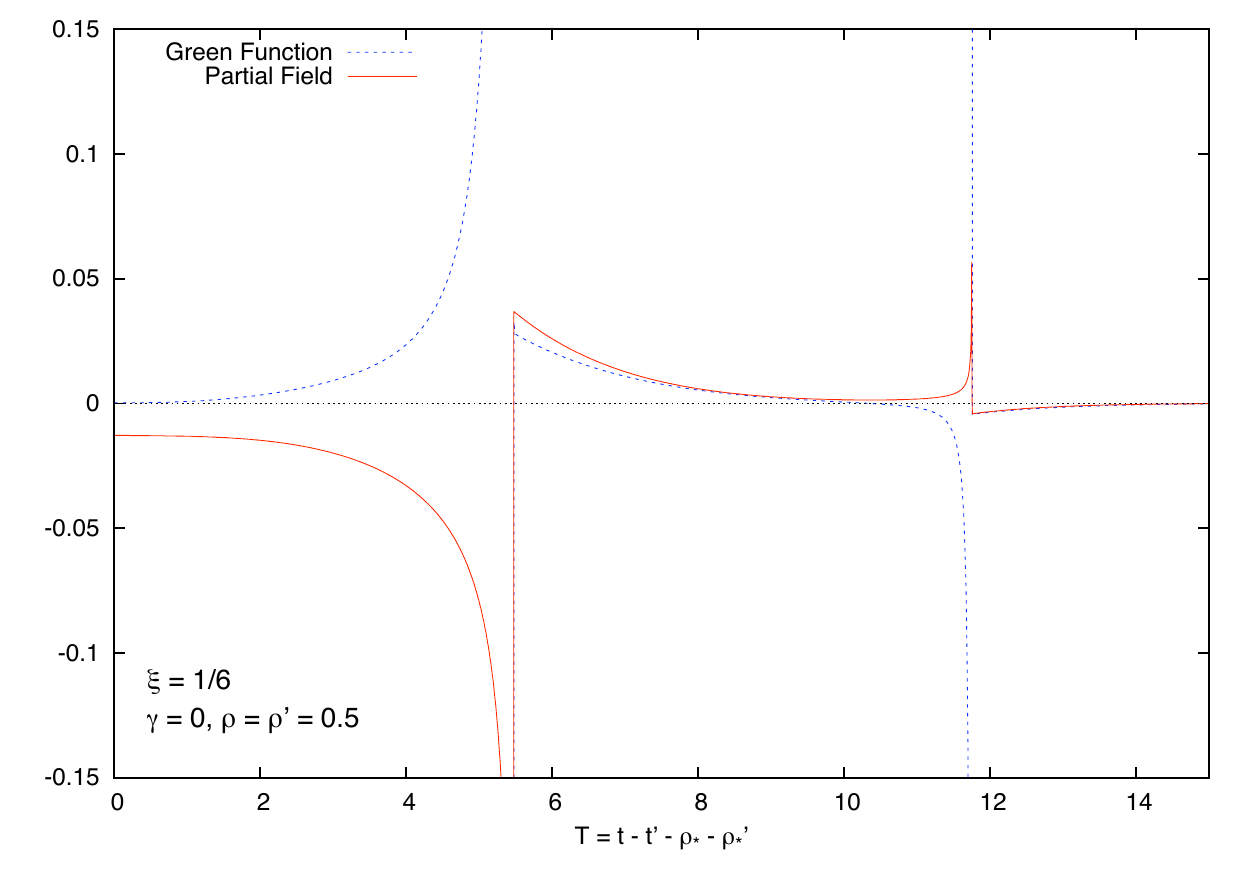}
 \end{center}
 \caption{\emph{`Partial Field' Generated by a Static Particle at $\rho=0.5$}. The dotted line shows the Green function. The solid line shows the partial field $q^{-1} \Phipartial$ defined in Eq.~(\ref{eq:partial-field}). This may be interpreted as the portion of radiative field generated by the segment of the static-particle worldline between $t^\prime = -\infty$ and $t^\prime = t - \Delta t$. Here $\Delta t = T + \Rs + \Rs^\prime $, where $T$ is the time coordinate shown on the horizontal axis.  }
 \label{fig:partial-field}
\end{figure}

The radiative field may be found by integrating the Green function with respect to $\tau^\prime$, where $d\tau^\prime = (1-\R^2)^{1/2} dt^\prime$. Integrating a mode sum like (\ref{G-n0}) with respect to $t^\prime$ is straightforward; we simply multiply each term in the sum by a factor $1/(i \omega_{ln})$. Hence it is straightforward to compute a \emph{partial field} defined by
\beq
\Phipartial(\Delta t) = q \left( 1 - \R^2 \right)^{1/2} \int_{-\infty}^{t - \Delta t} \Gret(t-t', \R=\R', \gam=0) d t^\prime .  \label{eq:partial-field}
\eeq
This may be interpreted as ``the field generated by the segment of the static-particle world line lying between $t^\prime = -\infty$ and $t^\prime = t - \Delta t$''.
In the limit $\Delta t \rightarrow 0$, the partial field $\Phipartial$ will coincide with the radiative field $\Phi_R$. An example of this calculation is shown in Fig.~\ref{fig:partial-field}. Here, $q^{-1} \Phipartial$ is plotted as a function of $T = \Delta t - \Rs - \Rs^\prime$ for a static particle near spatial infinity, $\R \rightarrow 1$. We have used the method of smoothed sums (Sec.~\ref{subsec:nummeth}), with $\lmax = 200$. The `partial field' $\Phipartial$ shares singular points with $\Gret$. Figure \ref{fig:partial-field} shows that a significant amount of the total radiative field arises from the segment of the worldline \emph{after} the first singularity. The Green function tends to zero in the limit $\Delta t \rightarrow 0$ (for $\xi = 1/6$). On the other hand, the partial field tends to a constant non-zero value in this limit. The constant value is the radiative field $q^{-1} \Phi_R$.%

\begin{figure}
 \begin{center}
  \includegraphics[width=7.5cm]{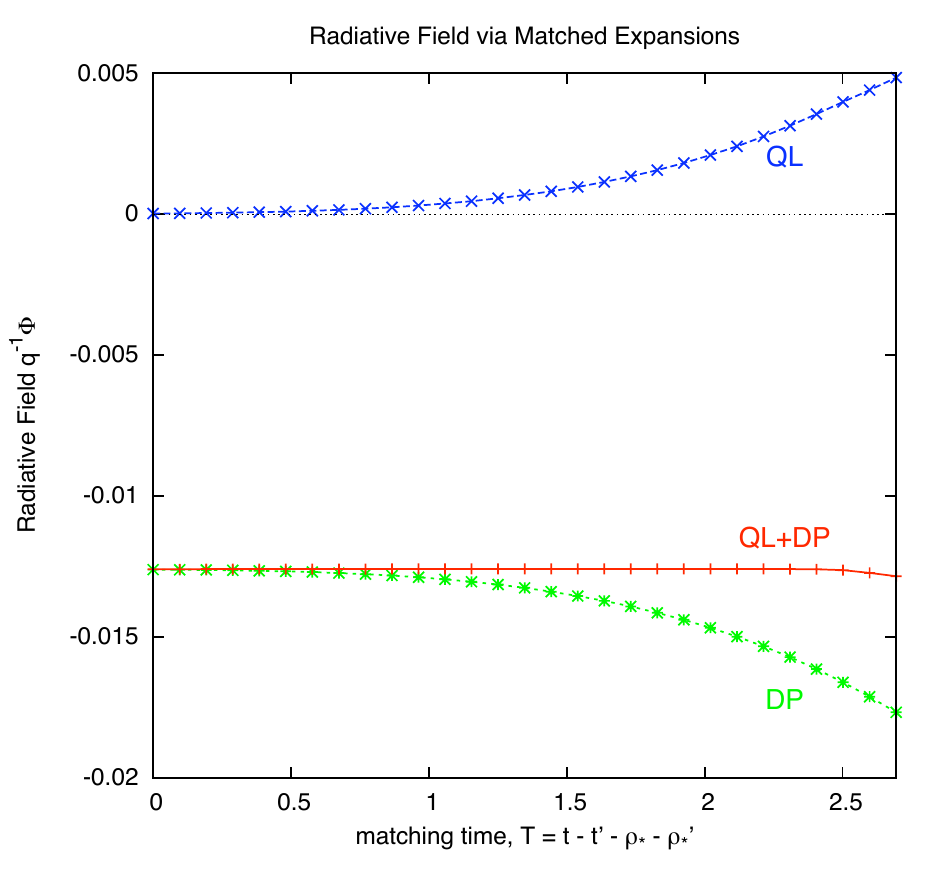}
  \includegraphics[width=7.5cm]{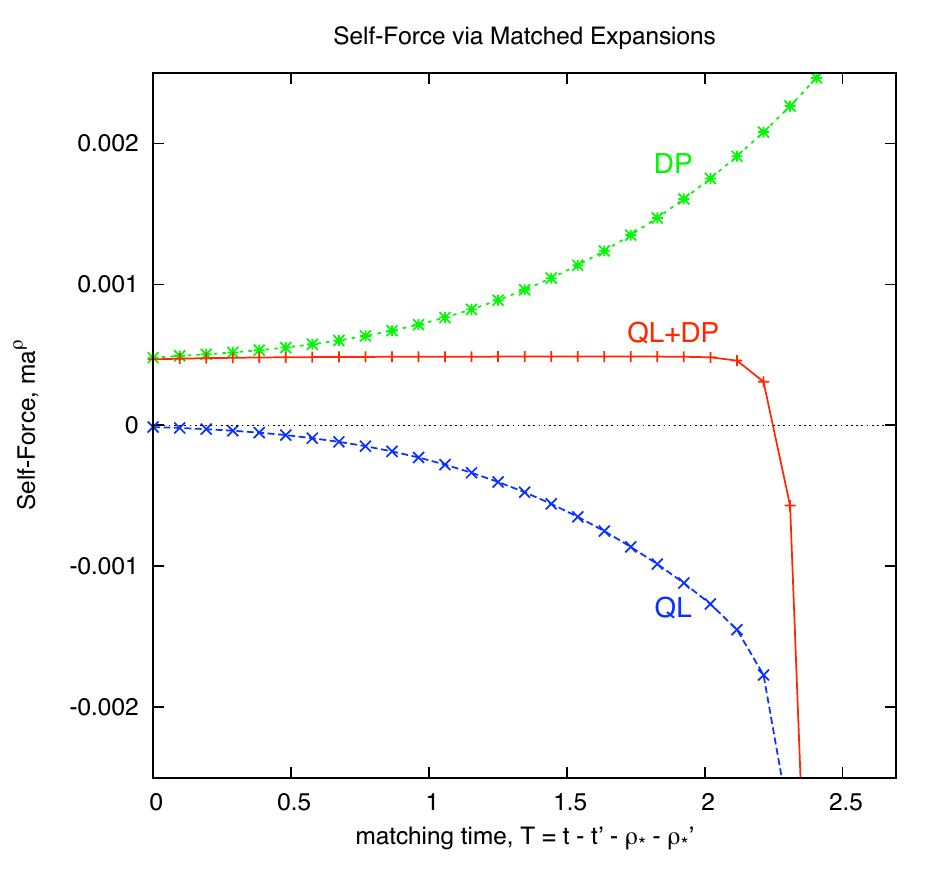}
 \end{center}
 \caption{\emph{Illustration of the matching calculation of the self-force on a static particle at $\R = \R^\prime = 0.5$}. The blue dashed line shows the quasilocal contribution, integrated from $\tau - \Delta \tau = (1-\R^2)^{1/2} [T+ \Rs + \Rs^\prime]$ to coincidence. The green dotted line shows the distant past contribution, integrated from $-\infty$ to $\tau - \Delta \tau$. The red solid line shows the total. In matching region $1 \lesssim T \lesssim 2$ the total approaches a constant, which corresponds to the value of the radiative field (left plot) and radial self-force (right plot). }
 \label{fig:match}
\end{figure} 

An accurate value for the total radiative field is found by using the quasilocal Green function to extend $\Phipartial$ to coincidence, $\Delta t \rightarrow 0^+$. The method is illustrated in Fig.~\ref{fig:match} (left plot). The dashed line shows the quasilocal contribution to the radiative field, and the dotted line shows the distant past contribution to the radiative field, as a function of matching time. The former is the result of integrating from the matching point $\tau - \Delta \tau$ to coincidence, and the latter from integrating from $-\infty$ to the matching point. Here, $\Delta \tau$ varies linearly with the $x$-axis scale $T$ (see caption). The right plot illustrates the same calculation for the radial self-force. Here, the dotted line representing the contribution from the distant past is found from the sum (\ref{eq:sf-dp}). 

\begin{figure}
 \begin{center}
  \includegraphics[width=8cm]{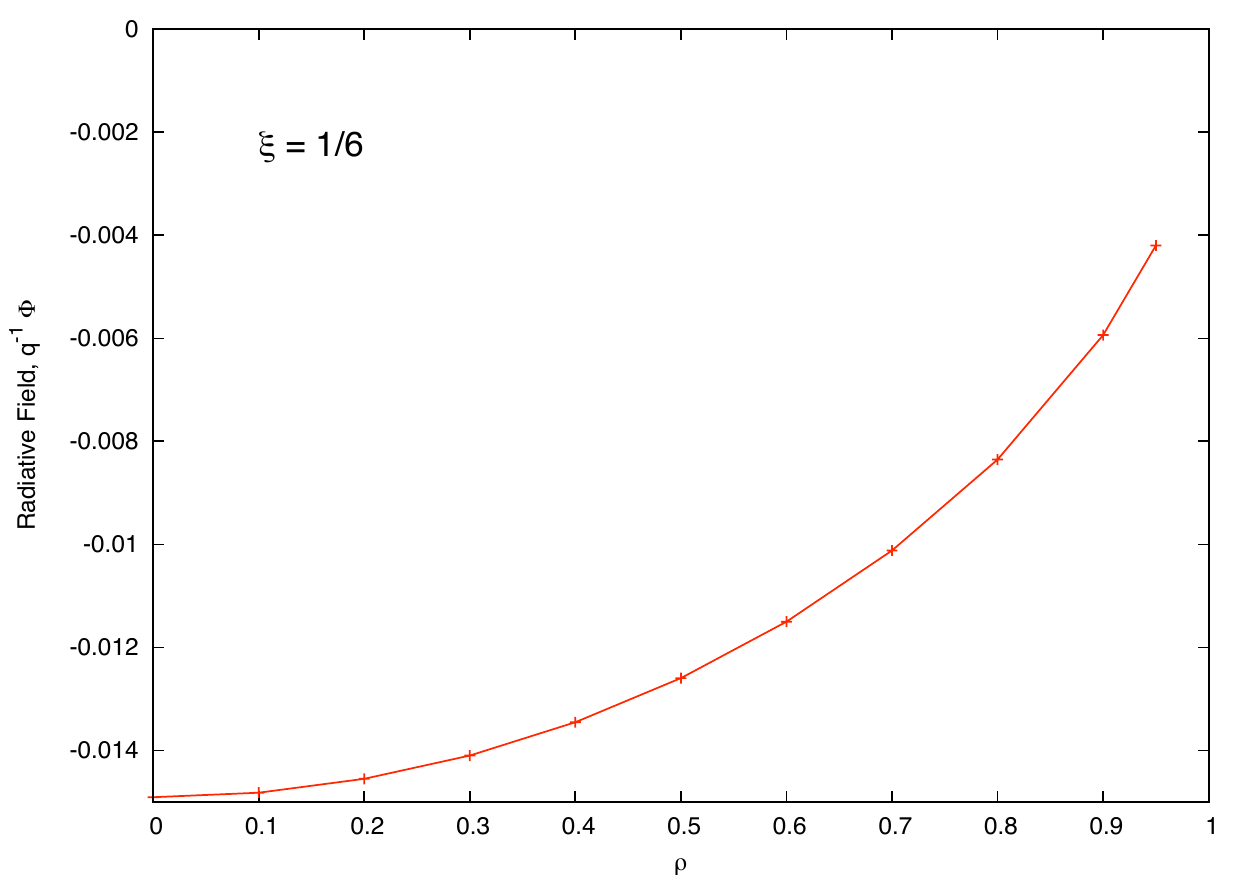}
  \includegraphics[width=8cm]{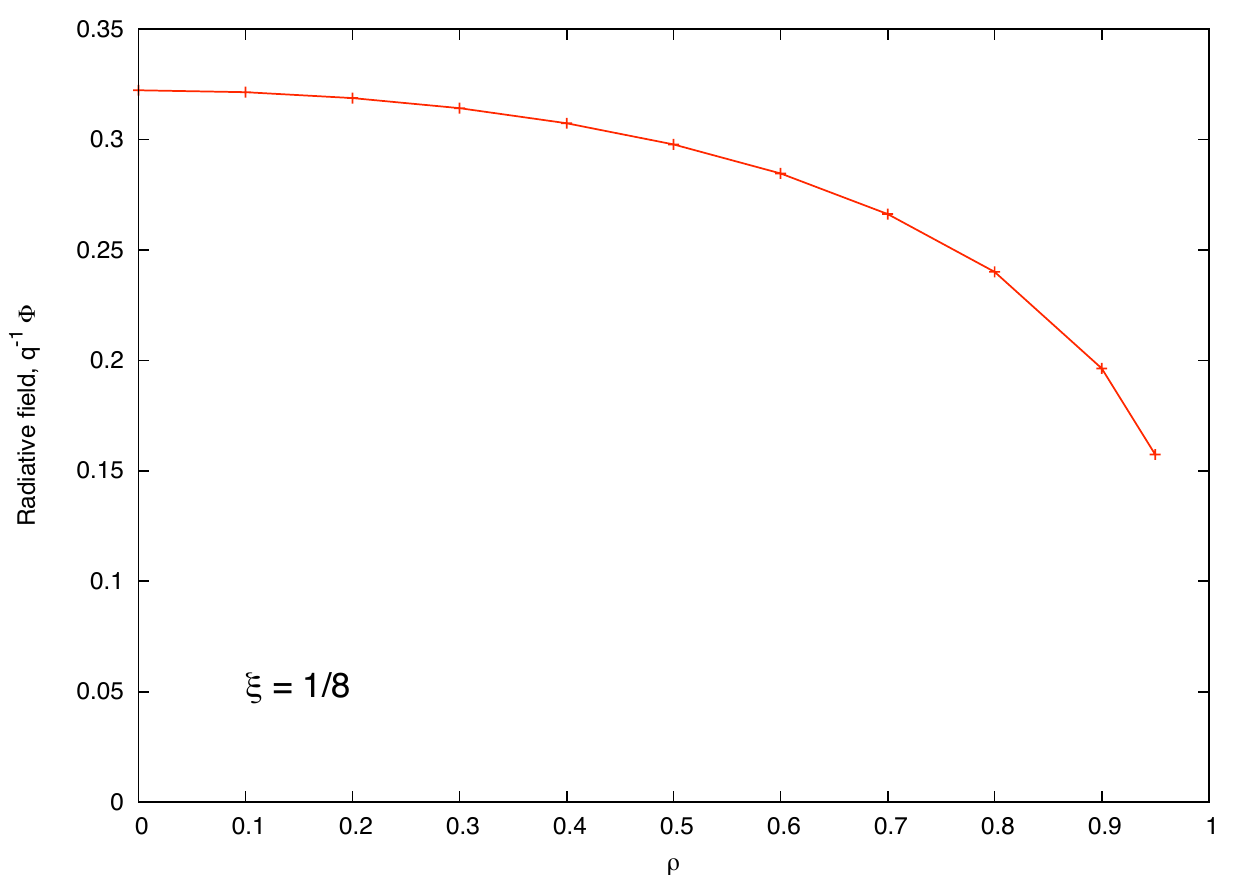}
 \end{center}
 \caption{\emph{The Radiative Field Generated by the Static Particle}. The plot shows the radiative field for the static particle at $\R=\R^\prime$. For the case $\xi = 1/6$ (left) the field is negative whereas for the case $\xi = 1/8$ it is positive (right). Note the differing scales on the vertical axis.}
 \label{fig:field}
\end{figure}

Figure \ref{fig:field} shows the total radiative field $\Phi_R$ generated by a static particle in the Nariai spacetime. The field is plotted as a function of $\R$, for two cases: $\xi = 1/6$ and $\xi = 1/8$. In the former case, the field is negative. In the latter case, the field is positive, and about two orders of magnitude greater in amplitude. In both cases, the amplitude of the field is maximal at $\R = 0$ and tends to zero as $\R \rightarrow 1$ as $\Phi_R \sim (1-\R^2)^{1/2}$. 

\begin{figure}
 \begin{center}
  \includegraphics[width=7.5cm]{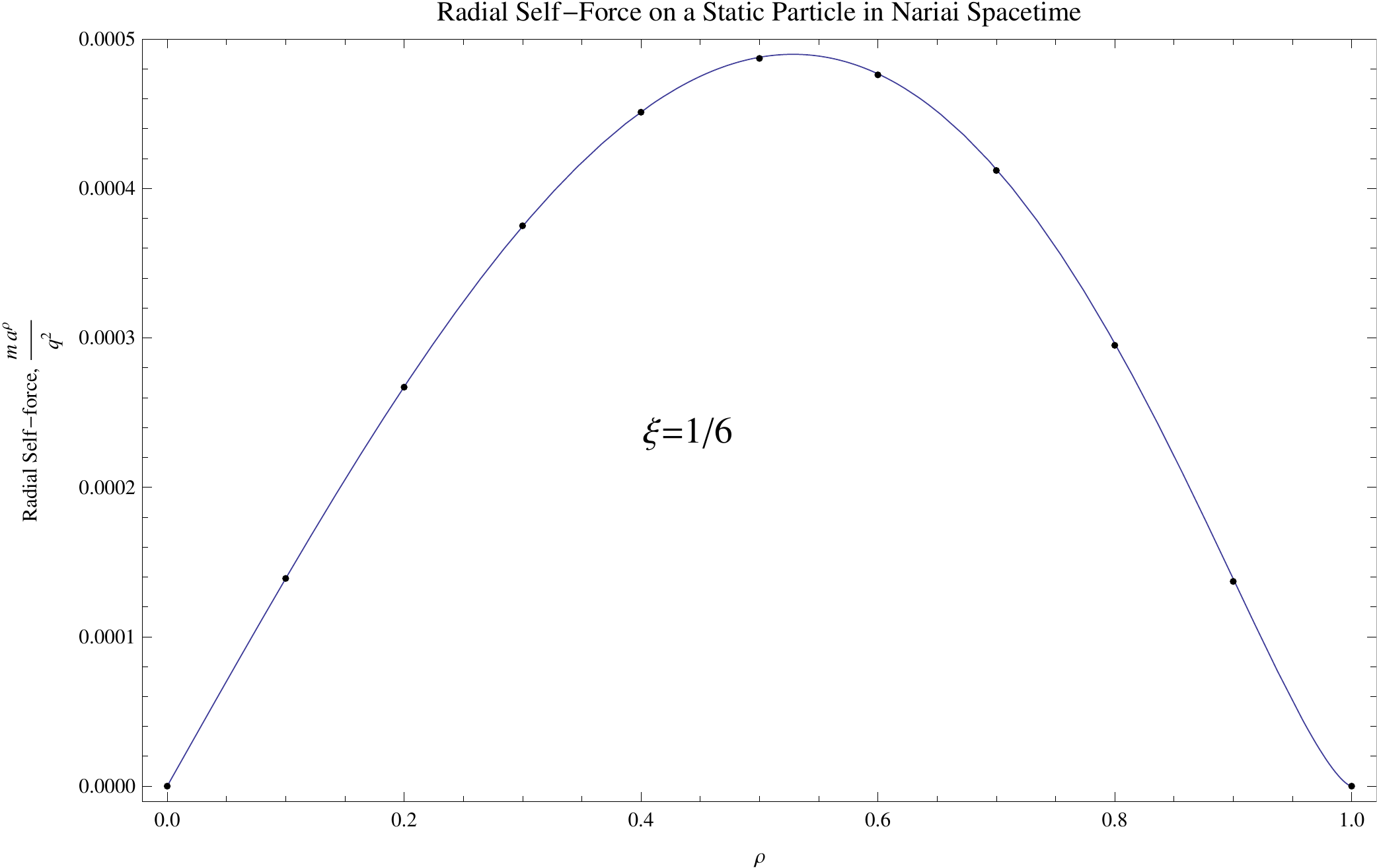}
  \includegraphics[width=7.5cm]{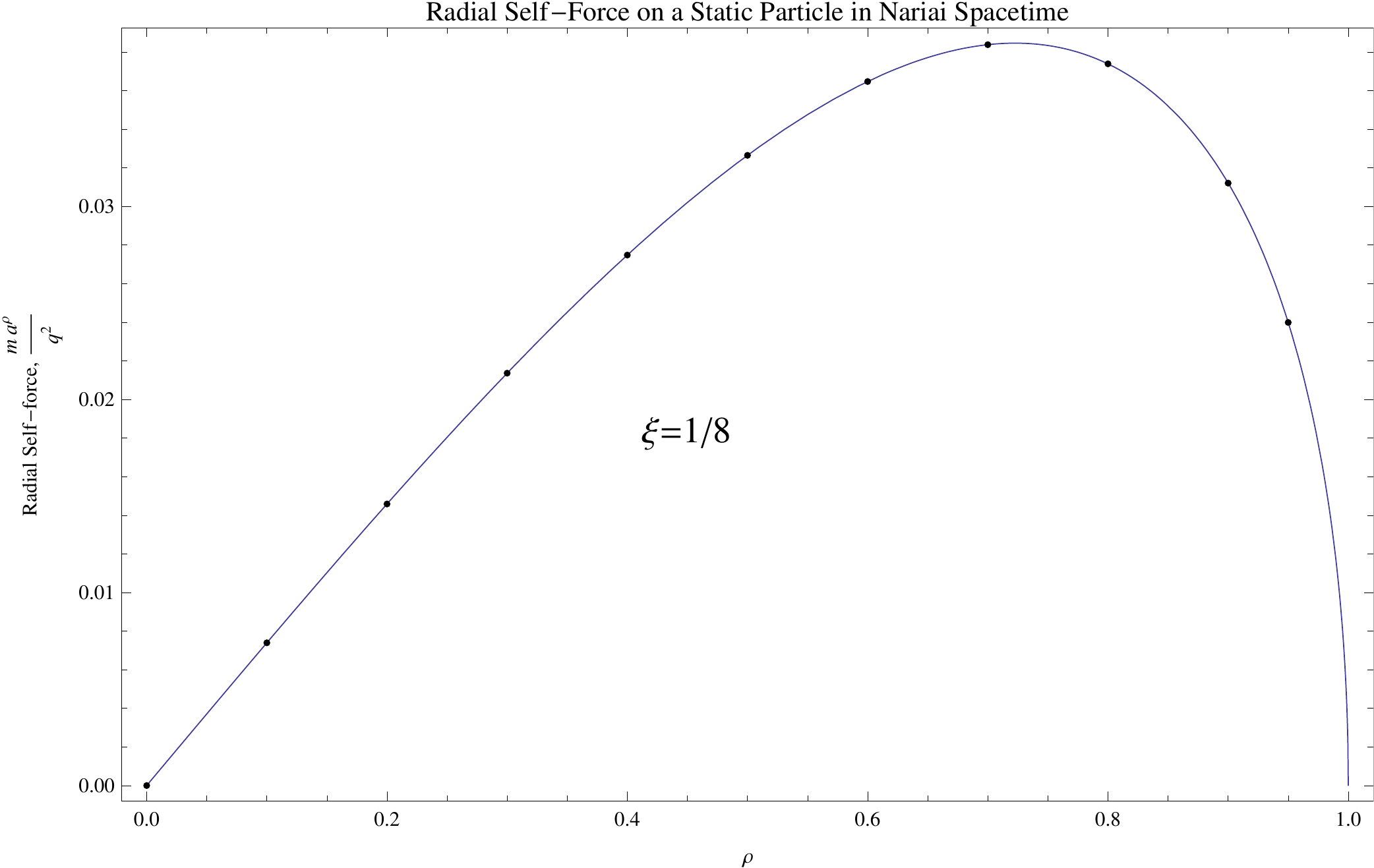}
 \end{center}
 \caption{\emph{The Radial Self-Force on the Static Particle}. The radial component of the self-force, $q^{-2} m a^{\R}$, is plotted as a function of the $\R$ of the static particle. The self-force was calculated by two methods: `matched expansion' (black dots) and `massive field regularization' (blue curve). The left plot shows the curvature coupling $\xi = 1/6$ whereas the right plot shows $\xi = 1/8$. Note the difference in scales for the two cases.}
 \label{radial-sf}
\end{figure}

Figure \ref{radial-sf} shows the self-force $m a^\R$ acting on a static particle. The results of the matched expansion results are shown as points, and the results of the `massive field regularization method' (described in Sec.~\ref{subsec:static full Green}) are shown as the solid line.  The latter method provides an independent check on the accuracy of the former. We find agreement to approximately six decimal places between the two approaches. We find the self-force at $\R = 0$ to be zero, as expected from the symmetry of the spacetime. The self-force also tends to zero as $\R \rightarrow 1$. Between these limits, the self-force rises to a single peak, the magnitude and location of which depends on the curvature coupling $\xi$. We find that the peak of the self-force is approximately $4.9 \times 10^{-4}$ for $\xi = 1/6$ and approximately $3.8 \times 10^{-2}$ for $\xi = 1/8$. 

\section{Discussion and Conclusions}\label{sec:conclusions}
In this paper we have presented the first practical demonstration of a self-force calculation using the method of matched expansions. The matched expansions method was first proposed over a decade ago by Poisson and Wiseman \cite{Poisson:Wiseman:1998}. We have shown that the `quasilocal' expansion in coordinate separation \cite{Anderson:2003, Anderson:Flanagan:Ottewill:2004, Ottewill:Wardell:2008, Ottewill:Wardell:2009}, valid only in the normal neighbourhood, may be accurately matched onto a mode sum expansion, valid in the distant past. Through matching the `quasilocal' and `distant past' expansions, the full retarded Green function may be reconstructed. With full knowledge of the Green function, one may accurately compute the `tail' contribution to the self-force. In this work, we employed the matching method to numerically compute the self-force acting on a static particle, and showed that the resulting self-force is in excellent agreement with the result from an alternative method (Sec.~\ref{subsec:static full Green}), to approximately one part in $10^{6}$.

The key new ingredient in our formulation is the so-called `quasinormal mode sum' expansion for the distant past Green function. Following Leaver's approach \cite{Leaver:1986}, the integral over frequency in the mode sum expansion of the Green function (\ref{eq:Gret:mode-sum}) may be performed by deforming the contour of integration in the complex frequency plane. Poles of the Green function arise at (complex) quasinormal mode frequencies in the lower half-plane. The sum over the residues of the poles gives the quasinormal mode sum -- a key contribution to the Green function (see below).

The QNM sum is only valid at `late times', $t-t^\prime \ge t_c$, where $t_c$ is approximately the time it takes for a geodesic to reflect from the peak of the potential barrier. We have demonstrated that there is a sufficient regime of overlap in $t-t^\prime$ in which both the `quasilocal' and `QNM sum' expansions are valid for the method to be applied successfully.

The QNM spectra of black holes have received much attention in the last three decades. In some studies \cite{Ferrari:Mashhoon:1984, Moss:Norman:2002, Berti:Cardoso:2006}, approximations to QNM frequencies are found by replacing the effective potential $V^{(S)}_l$ (\ref{Veff-Schw}) with the so-called P\"{o}schl-Teller potential (\ref{VPT}). The advantage of this replacement is that the QNMs and radial solutions of the P\"{o}schl-Teller potential are known in closed form. In this paper, we have taken the idea a step further. We have shown that the P\"{o}schl-Teller potential arises naturally if we consider the radial equation resulting from waves on an alternative spacetime: the product spacetime $dS_2 \times  \mathbb{S}^2$, first introduced by Nariai \cite{Nariai:1950, Nariai:1951} in 1950. This is an Einstein spacetime of constant scalar curvature. 

The symmetry of Nariai spacetime undoubtedly makes calculations easier. For example, geodesic motion may be separated into motion on the two submanifolds, $dS_2$ and $ \mathbb{S}^2$; the Van Vleck determinant may be written in closed form (\ref{eq:vanVleckDet}); and the decay rate of the quasinormal modes is independent of $l$. We view the Nariai spacetime as an excellent \emph{testing ground} for our methods. Nevertheless, we should not forget the overall goal of the Poisson-Wiseman-Anderson proposal \cite{Poisson:Wiseman:1998, Anderson:Wiseman:2005}: accurate matched-expansions calculations on physical black hole spacetimes. Below, we review some of the insights provided by  our `experiment' with the Nariai spacetime which will help any future calculations.

At late times $t-t' > \Rs + \Rs^\prime$ it has been previously been established \cite{Beyer:1999} that the quasinormal modes provide a \emph{complete basis} on the Nariai spacetime. Here, we demonstrated that, at late times, the QNM sum (\ref{G-QNM}) fully describes the retarded Green function. This is \emph{not} expected to be the case on the Schwarzschild spacetime \cite{Leaver:1986}. In the latter case, there arises a branch point at $\omega = 0$ and a branch cut in the frequency integral which gives a `power-law tail' contribution \cite{Andersson:1997} to the Green function (Fig. \ref{fig:contours}). The branch point is notably absent for the Nariai spacetime, making the analysis simpler. In this paper we have thoroughly investigated the effect of the quasinormal modes. The contribution of the branch cut integral to the self-force remains to be quantified. We hope to pursue this calculation in a forthcoming work.


This study has provided a number of insights into the properties of the Green function, which are of relevance to any future investigation of the Schwarzschild spacetime. Namely,
\begin{itemize}
 \item The Green function $\Gret(x, x^\prime)$ is singular whenever $x$ and $x^\prime$ are connected by a null geodesic. The nature of the singularity depends on the number of caustics that the wave front has passed through. After an even number of caustics, the singularity is a delta-distribution, with support only on the light cone. After an odd number of caustics, the Green function diverges as $1 / \pi \sigma$, where $\sigma$ is the Synge world function. A four-fold repeating pattern occurs, i.e. $\delta, \, 1/\pi \sig, \, -\delta, \, -1/\pi\sig, \, \delta, $ etc.
 \item The four-fold singular structure can be shown to arise from a Hadamard-like ansatz (\ref{Gdir1}) valid even outside the normal neighborhood, if we allow $U(x,x^\prime)$ to pick up a phase of $-i$ upon passing through a caustic. The accumulation of phase may be deduced by analytically continuing the integral for the Van Vleck determinant through the singularities (due to caustics).
  \item Hadamard's form for the Green function (\ref{eq:Hadamard}) is only strictly valid if $x$ and $x^\prime$ are in a convex normal neighbourhood \cite{Friedlander}. The extension beyond the normal neighbourhood does not seem to be known. However, this work and some previous studies \cite{Ottewill:Grove:Brown:1981} would tentatively suggest that, if $x$ and $x^\prime$ are connected by a countable number of distinct geodesics, then the Green function may be found from the sum of their Hadamard contributions. 
   \item The effect of caustics on wave propagation has been well-studied in a number of other fields, such as optics~\cite{Stavroudis}, acoustics~\cite{Kravtsov:1968}, seismology~\cite{Aki:Richards}, symplectic geometry~\cite{Arnold} and quantum mechanics~\cite{B&M}. It may be that mathematical results developed in other fields may be usefully applied to wave propagation in gravitational physics.
\end{itemize}

A further observation made in this work is confirmation that the `tail' self-force cannot be calculated from the `quasilocal' contribution alone. For instance, Fig.~\ref{fig:partial-field} would appear to show that a significant part of the radiative field is generated by the segment of the world line which lies \emph{outside} of the normal neighbourhood (i.e. beyond the first caustic). Unlike in flat space, the radiated field generated by an accelerated particle may propagate once, twice, etc. around the black hole before later re-intersecting the world-line of the particle (see Fig.~\ref{fig:circ_orbits}). Radiation from near these orbits will give an important contribution to the self-force which cannot be neglected. 

Let us conclude by examining the prospects for a practical `matched expansions' calculation on the Schwarzschild spacetime. Happily, the `quasilocal' expansion is now in excellent shape, as described in \cite{Ottewill:Wardell:2008};  the challenge remains the `distant past' expansion. We have already mentioned that a quasinormal mode sum will not be sufficient; it must be augmented by a branch cut integral. We have reasons to be optimistic that this is a tractable calculation \cite{Andersson:1997}. Perhaps more difficult will be the accurate numerical computation of QNM frequencies and radial functions. 
A further difficulty will be in integrating the mode sum over the worldline; for accuracy we wish to avoid numerical integration if possible. For the static particle, it was straightforward to integrate each term in the mode sum analytically (Sec.~\ref{subsec:matched-static}) with respect to time. For other trajectories (e.g. circular orbits) this may present more of a challenge. Finally, we note that a range of established results are available on Schwarzschild. For the static particle the self-force is zero \cite{Burko:2000a, Wiseman:2000}; for radial trajectories and circular and eccentric orbits, accurate numerical results are available \cite{Burko:2000b,  Anderson:Eftekharzadeh:Hu:2006, Barack:Sago:2007, Haas:2007}. This will surely help the validation of the `matched expansions' method. We hope to undertake such a study in the near future.

\section{Acknowledgments}
The authors would like to thank Leor Barack, Vitor Cardoso, Oscar Dias and Brien Nolan for interesting and helpful discussions. Special thanks are due to Amos Ori for email correspondence about the four-fold singularity structure. MC is also grateful to the Department of Physics and Astronomy of the University of Mississippi for its hospitality during the preparation
of this paper. MC was partially funded by Funda\c c\~ao para a Ci\^encia e Tecnologia (FCT) - Portugal through project PTDC/FIS/64175/2006. MC, BW and SD are supported by the Irish Research Council for Science, Engineering and Technology, funded by the National Development Plan. 

\appendix

\section{Large-$\lam$ Asymptotics of ${}_2F_1(\frac{1}{2} - i \lam,\frac{1}{2} - i \lam;1 - 2 i \lam;-e^{- T})$}
\label{appendix-hypergeom}

\noindent To determine the singularity structure of the Green function in Sec.~\ref{subsec:large-l} we required the large $\lam$ asymptotic behaviour of
$
{}_2F_1(-\beta, -\beta; -2 \beta; z)$ where $\beta = -\frac{1}{2} + i \lam$, $z = -e^{-T}$.
The required asymptotics may be found by applying the WKB method \cite{Bender:Orszag} to the hypergeometric differential equation
\beq
z(1-z)\frac{d^2 u}{dz^2} - \left[2\beta(1 - z)  + z \right] \frac{du}{dz} - \beta^2 u = 0
\eeq
which has solutions $u(z) = {}_2F_1(-\beta,-\beta;-2\beta;z)$. 
Inserting the WKB ansatz
$
u(z) \sim e^{ \lam S_0(z) + S_1(z) + \lam^{-1} S_2(z) \dots}   \label{eq:wkb-def}
$
immediately yields a quadratic equation for $S_0^\prime$,  
\beq
z(1-z) (S_0^\prime)^2  - 2 i (1-z) S_0^\prime + 1 = 0 .
\eeq
In our case, $z = -e^{- T}$, we require  the root which is finite at $z = 0$. We impose $S_0(0) = 0$ to get
\begin{eqnarray}
&& S_0^\prime =  \frac{i}{z} \left(1 - (1-z)^{-1/2}  \right)
 \Rightarrow  S_0(z)= \ln \left( \frac{(-z)}{4} \, \frac{\left[ \sqrt{1-z} + 1 \right]}{\left[ \sqrt{1 - z} - 1 \right]}\right) = -T - 2\ln 2 + \ln \left( \frac{\sqrt{1+e^{-T}} + 1}{\sqrt{1 + e^{-T}} - 1}  \right) \label{eq:wkb-S2}
\end{eqnarray}
At next order, we obtain the equation
\beq
2 \left[  i (1 - z)  -  z (1 - z) S_0^\prime \right]  S_1^\prime = z(1-z) S_0^{\prime \prime} + (1-2z) S_0^\prime + i .
\eeq
It is straightforward to show that this reduces to
\beq
S_1^\prime = \left( 4(1-z) \right)^{-1} \quad \quad \Rightarrow \quad S_1 = -\frac{1}{4} \ln \left(1 + e^{-T}  \right) .   \label{eq:wkb-S1}
\eeq
Inserting (\ref{eq:wkb-S1}) and (\ref{eq:wkb-S2}) into (\ref{eq:wkb-def}) leads to the quoted result, Eq.~(\ref{eq:hypergeom-asymp}). Of course, the asymptotic approximation may be further refined by taking the WKB method to higher orders. 

\section{Poisson Sum Asymptotics}\label{appendix:poisson-sum}
In this appendix we derive asymptotic approximations for the singular structure of the Green function using the Poisson sum formula (\ref{InRN}). Our starting point is expression (\ref{RN-asymp-Hankel}) for the `$n$=0' fundamental modes, in which the Legendre polynomials $P_l(\cos \gam)$ have been replaced by angular waves $\mathcal{Q}_{\nu-1/2}^{(\pm)}(\cos \gam)$ which are, in turn, approximated by Hankel functions $H^{(\mp)}_0(\nu \gam)$ using (\ref{Olver-approx}). 

Let us consider the $\II_1$ and $\II_2$ integrals arising from substituting (\ref{RN-asymp-Hankel}) into (\ref{InRN}). These integrals are singular at $\chi = 2 \pi - \gam$ and $\chi = 2 \pi + \gam$, respectively. 
First, let us consider $\II_1$ (\ref{InRN}) which can be written
\beq
\II_1 \approx -\Agam \text{Re} \int_0^\infty d\nu (-i \nu)^{1/2} e^{i (\chi - 2\pi) \nu} H_{0}^{(+)}(\nu \gam) 
\label{eq:I1-real}
\eeq
with $\Agam$ as defined in Eq.~(\ref{eq:A-gam}) .

For $\chi > 2 \pi - \gam$, the integral may be computed by rotating the contour onto the \emph{positive} imaginary axis ($\nu = i z$) to obtain
\begin{eqnarray}
\II_1 &\approx& -\frac{2 \Agam}{\pi} \int dz z^{1/2} e^{-(\chi - 2\pi) z} K_0 (\gam z) \nonumber \\
       &\approx& -\frac{\Agam \sqrt{\pi}}{2 [\chi - (2\pi - \gam)]^{3/2}} {}_2F_1 \left( 3/2, 1/2; 2; \frac{\chi - 2\pi - \gam}{\chi - 2\pi + \gam} \right)  \label{eq:I1-after}
\end{eqnarray}
Here we have applied the identity $H_0^{(+)}(ix) = 2 K_0(x) / (i \pi)  $, where $K_0$ is the modified Bessel function of the second kind, and the integral is found from Eq.~6.621(3) of Ref.~ \cite{GradRyz}. 


For $\chi < 2 \pi - \gam$, the integral may be computed by rotating the contour onto the \emph{negative} imaginary axis ($\nu = -i z$). First, we make the replacement $H_0^{(+)}(\nu \gam) = 2 J_0(\nu \gam) - H_0^{(-)}(\nu \gam)$ and note $H_0^{(-)}(- i x) = 2 K_0(x) / (-i \pi)$ to obtain
\begin{eqnarray}
  \II_1 & \approx & \frac{2 \Agam}{\pi} \text{Re} \int dz z^{1/2} e^{-(2\pi - \chi) z} \left[ \pi I_0(\gam z) + i K_0 (\gam z) \right] 
\end{eqnarray}
Here $I_0$ is a modified Bessel function of the first kind. Since we are taking the real part, the $K_0$ term is eliminated, and we obtain
\begin{eqnarray}
  \II_1 & \approx & \frac{2 \Agam}{\sqrt{\pi}} \left( 2\pi - \gam - \chi \right)^{-1} \left( 2\pi + \gam - \chi \right)^{-1/2} E\left( \frac{2 \gam}{2 \pi + \gam - \chi} \right)
\end{eqnarray}
where $E$ is the elliptic integral of the second kind defined in, for example, Eq.~8.111(3) of Ref.~\cite{GradRyz}.

The $\II_2$ integral may be calculated in a similar manner. For $\chi < 2\pi + \gam$, we rotate the contour onto the negative imaginary axis,
 \begin{eqnarray}
  \II_2 & \approx & -\Agam \text{Re} \int_0^\infty d \nu (- i \nu)^{1/2} H_0^{(-)}(\nu \gam) e^{i (\chi - 2\pi) \nu} \nonumber \\
   & \approx & \frac{2 \Agam}{\pi} \text{Re} \; i \int_0^\infty dz z^{1/2} e^{-(2\pi - \chi) z} K_0(\gam z)  \quad \quad = 0
\end{eqnarray}
For $\chi > 2\pi + \gam$, we rotate the contour onto the positive imaginary axis after taking the complex conjugate 
 \begin{eqnarray}
  \II_2 & \approx & -\Agam \text{Re} \int_0^\infty d \nu (i \nu)^{1/2} H_0^{(+)}(\nu \gam) e^{i (\chi - 2\pi) \nu} \nonumber \\
   & \approx & \frac{2 \Agam}{\pi} \text{Re} \int_0^\infty dz z^{1/2} e^{-(\chi - \pi) z} \left(i \pi I_0(\gam z) +  K_0(\gam z) \right)  
\end{eqnarray}
The imaginary term does not contribute and hence $\II_2$ is equal and opposite to $\II_1$ defined by Eq.~(\ref{eq:I1-after}) when $\chi > 2 \pi + \gam$.

 \section{Green function on  $T\times \mathbb{S}^2$} \label{appendix-Green on S2}
 
 To the best of our knowledge, the four-fold singularity structure for the Green function of Sec.~\ref{sec:singularities} has not been shown before
 in the literature (with the
 exception of~\cite{Ori1}) within the theory of General Relativity.
 We therefore wish to illustrate its derivation and manifestation in the simplest of spacetimes including $ \mathbb{S}^2$-topology:
 \begin{equation} \label{eq:RxS2}
 ds^2=-dt^2+d\Omega^2_2,
 \end{equation}
where Synge's world function is simply given by
$\sigma=\frac{1}{2}\left(-\Delta t^2+\phif^2\right)$, in the case of a conformally-coupled ($\xi=1/8$) scalar field.

Let $x=(t,\theta,\phi)$ denote any point in this spacetime.
We introduce the Wightman function $G_+(x,x')$ (it satisfies the homogeneous scalar wave equation - see, e.g.,~\cite{Birrell:Davies}), from  
which the `retarded' Green function $G_{ret}(x,x')$ is
easily obtained:
\begin{align} \label{eq:G_+ mode sum}
G_+(x,x')&=
\sum_{l=0}^{+\infty}\sum_{m=-l}^{+l}\Phi_{lm}(x)\Phi^*_{lm}(x')=\frac{1}{4\pi}\sum_{l=0}^{+\infty}e^{- i(l+1/2)\Delta t}P_l(\cos\gamma), \\
G_{ret}(x,x')&=-2\theta(\Delta t)\ \text{Im}\left(G_+(x,x')\right),
\end{align}
where $\Delta t\equiv t-t'$, $\Phi_{lm}(x)=e^{-i(l+1/2)t}Y_{lm}(\theta,\phi)/\sqrt{(2l+1)}$ are the Fourier-decomposed
scalar field modes on (\ref{eq:RxS2}) normalized with respect to the scalar product
\beq 
\left(\Phi_{lm},\Phi_{l'm'}\right)=-i\int_{\Sigma}dVn^{\mu}\left[\Phi_{lm}(x)\partial_{\mu}\Phi^*_{l'm'}(x)-
\Phi^*_{l'm'}(x)\partial_{\mu}\Phi_{lm}(x)\right]=\delta_{ll'}\delta_{mm'},
\eeq
where $\Sigma$ is a Cauchy hypersurface with future-directed unit normal vector $n^{\mu}$ and volume element $dV$.

 We now apply exactly the same tricks as in section \ref{subsec:Poisson} in order to derive the four-fold singularity structure in 
 the Green function
 from the large-$l$ asymptotics of the field modes.
 We use the Poisson sum formula
 \begin{equation}
 \sum_{l=0}^{+\infty}g(l+1/2)=\sum_{s=-\infty}^{+\infty}(-1)^s\int_0^{+\infty}d\nu g(\nu)e^{2\pi is\nu}
 \end{equation}
 to re-write the mode sum in (\ref{eq:G_+ mode sum}) as
 \begin{align}\label{eq:G_+ mode sum,rewritten}
 4\pi G_+(x,x')&=\sum_{s=-\infty}^{+\infty}(-1)^s\int_0^{+\infty}d\nu e^{-i\nu\Delta t}P_{\nu-1/2}(\cos\gamma)e^{2\pi is\nu}
 =
  \sum_{N=1}^{+\infty}G_+^N, \\
 G_+^N(x,x') &\equiv \int_0^{+\infty}d\nu R_N(\cos\gamma)e^{-i\nu\Delta t}.
 \end{align}

 The Legendre functions $P_{\mu}(\cos\gamma)$ and $Q_{\mu}(\cos\gamma)$, as well as $R_N(\cos\gamma)$,
 are standing waves.
 This is in contrast to $Q_{\mu}^{(\pm)}(\cos\gamma)$, which are travelling waves.
 
 We can now use large-order uniform asymptotics (see \cite{Olver:1974,Jones'01}) for the Legendre functions:
 \begin{align} \label{eq:large-l P}
 P_{\nu-1/2}(\cos\gamma)&\sim \left(\frac{\gamma}{\sin\gamma}\right)^{1/2}J_0(\nu\gamma), \quad |\nu|\to \infty,
 \quad \text{``valid in a closed uniform interval containing $\gamma=0$"},
 \\
 Q_{\nu-1/2}(\cos\gamma)&\sim -\frac{\pi}{2}\left(\frac{\gamma}{\sin\gamma}\right)^{1/2}Y_0(\nu\gamma), \quad |\nu|\to \infty,
 \quad \text{``valid with respect to $\gamma\in (0,\pi/2]$"},
 \\ \label{eq:large-l Qpm}
 \mathcal{Q}_{\nu-1/2}^{(\pm)}(\cos\gamma)&\sim \frac{1}{2}\left(\frac{\gamma}{\sin\gamma}\right)^{1/2}H_0^{(\mp)}(\nu\gamma),
 \qquad |\nu|\to\infty,
 \end{align}
 to leading order.
 
 The contribution to the Wightman function from the $N=1$ orbit wave is immediately obtained by using
 the large-order asymptotics of the Legendre function $P_{\nu}(\cos\gamma)$ only, which are ``valid in a closed uniform interval
 containing $\gamma=0$" - this is what we will mean by a result being valid ``near" $\gamma=0$.
 Similarly, we can obtain a result valid ``near" $\gamma=\pi$ by using in (\ref{eq:G_+ mode sum})
 the symmetry $P_l(\cos\gamma)=(-1)^lP_l(\cos(\pi-\gamma))$ for $l\in \mathbb{N}$.
 We then obtain for $N=1$:
 \begin{align} \label{eq:G_+,N=1}
 4\pi G_+^{N=1}(x,x')\sim &\sqrt{\frac{\gamma}{\sin\gamma}} \frac{1}{\sqrt{\gamma^2-\Delta t^2}}, 
 \quad &\text{``near" $\gamma=0$}
 \\
 4\pi G_+^{N=1}(x,x')\sim &\sqrt{\frac{\pi-\gamma}{\sin(\pi-\gamma)}} \frac{-i}{\sqrt{(\pi-\gamma)^2-(\Delta t-\pi)^2}}
 = 
 \sqrt{\frac{\pi-\gamma}{\sin(\pi-\gamma)}} \frac{-i}{\sqrt{-(\Delta t-\gamma)\left[\Delta t-(2\pi-\gamma)\right]}},
 \quad &\text{``near" $\gamma=\pi$}
 \end{align}
 where, for convergence, a small imaginary part was given to $\Delta t$
and/or $\gamma$, in agreement with the Feynman prescription `$\sigma\to \sigma+i\epsilon$'.
 The result for $G_+^{N=1}(x,x')$ valid ``near" $\gamma=0$ is singular at $\Delta t=\pm \gamma$, corresponding
 to $\sigma=0$ before a caustic has been crossed.
It is in accord with the Hadamard form in 3-D~\cite{Decanini:Folacci:2005a} and the 
Van Vleck determinant (\ref{eq:Delta_phi}), before a caustic has been crossed
 (and so without the phase factor).
  The result for $G_+^{N=1}(x,x')$ valid ``near" $\gamma=\pi$ is singular at $\Delta t=\gamma$, corresponding to the
  case where it has not gone through any caustics, and at $\Delta t=2\pi-\gamma$, corresponding to the
  case where it has gone through one caustic; it has thus picked up a factor ``$-i$", as expected.
 Note that these zeros inside the squared root in the denominator are simple zeros along the null geodesic, except
 at the caustic point itself, where the two zeros coincide and so it becomes a double zero. 
  
 Similarly to $N=1$, we can use (\ref{eq:G_+ mode sum,rewritten}) and the asymptotics (\ref{eq:large-l Qpm}) together with~\cite{GradRyz}
 \begin{equation}
 I_{\pm}(T,\gamma)\equiv \int_0^{\infty}d\nu e^{-i\nu(T-i\epsilon)}H_0^{(\pm)}(\nu\gamma)=
 \frac{1}{\sqrt{\gamma^2-(T-i\epsilon)^2}}\left[1\mp \frac{2i}{\pi}\ln\left(iX+\sqrt{1-X^2}\right)\right],
 \quad T\in \mathbb{R},\ \epsilon>|\text{Im}\gamma|,
 \end{equation}
 where $X\equiv (T-i\epsilon)/\gamma$ (again, a small imaginary part needs to be given for convergence, in accordance with
 the Feynman prescription), 
 in order to obtain for $N>1$:
 \begin{align} \label{eq:G_+,N>1}
 &4\pi G_+^N(x,x')
 \sim \sqrt{\frac{\Xi}{\sin\Xi}} \frac{(-1)^{N/2}}{2}
 \begin{cases}
 \left[I_+(\Delta t+N\pi,\Xi)+I_-(\Delta t-N\pi,\Xi)\right],
 \quad  &\text{``near" $\gamma=0$} 
 \\
 -i
 \left[I_+(\Delta t+(N-1)\pi,\Xi)+I_-(\Delta t-(N+1)\pi,\Xi)\right],
 \quad  & \text{``near" $\gamma=\pi$} 
 \end{cases}
 \end{align}
 for $N$ even, where $\Xi=\gamma$ ``near" $\gamma=0$ and $\Xi=\pi-\gamma$ ``near" $\gamma=\pi$. 
 For $N$ odd, merely:  (1) swap $I_{\pm}\to I_{\mp}$, and 
 (2) replace $N$ by $N-1$ if ``near" $\gamma=0$ or  $N$ by $N+1$ if ``near" $\gamma=\pi$ in (\ref{eq:G_+,N>1}).
 We can re-write:
 $\gamma^2-(\Delta t\pm N\pi-i\epsilon)^2=-\left[(\Delta t-i\epsilon)-(\mp N\pi-\gamma)\right]\left[(\Delta t-i\epsilon)-(\gamma\mp N\pi)\right]$.
 Note, however, that $I_{\pm}(T,\gamma)$ is regular at $X=\mp 1$.

 We then have that the singular behaviour goes as
 \begin{equation}
 4\pi G_+(x,x')\sim 
 \frac{1}{2}\sqrt{\frac{\Xi}{\sin\Xi}}
 \begin{cases}
 +I_+(\Delta t,\gamma),  &0<\Delta t<\pi,\quad \gamma\sim 0
 \\
 -iI_+(\Delta t-\pi,\pi-\gamma),  &\pi<\Delta t<2\pi,\quad \gamma\sim \pi
 \\
 -I_+(\Delta t-2\pi,\gamma),  &2\pi<\Delta t<3\pi,\quad \gamma\sim 0
 \\
 +iI_+(\Delta t-3\pi,\pi-\gamma),  &3\pi<\Delta t<4\pi,\quad \gamma\sim \pi
 \end{cases}
 \end{equation}
 The 4-fold singularity structure arises clearly: a phase of $\pi/2$ is picked up everytime the null geodesic
joining $x$ and $x'$ goes through a caustic ($\gamma=0$ or $\pi$).
 
 The expression for $G_+(x,x')$ is simplified by noting that, ``near" $\gamma=0$:
 \begin{align}
 &4\pi \left[G^N_+(x,x')+G^{N+1}_+(x,x')\right]\sim
 \\
 & i^N\sqrt{\frac{\gamma}{\sin\gamma}}
 \left[\frac{1}{\sqrt{\gamma^2-(\Delta t-N\pi-i\epsilon)^2}}+\frac{1}{\sqrt{\gamma^2-(\Delta t+N\pi-i\epsilon)^2}}\right],\quad \text{N even}
 \\
 & i^N\sqrt{\frac{\pi-\gamma}{\sin (\pi-\gamma)}}
 \left[\frac{1}{\sqrt{(\pi-\gamma)^2-(\Delta t-N\pi-i\epsilon)^2}}+\frac{1}{\sqrt{(\pi-\gamma)^2-(\Delta t+N\pi-i\epsilon)^2}}\right],\quad \text{N odd}.
 \end{align}
 Similarly ``near" $\gamma=\pi$.
 The Poisson sum formula has yielded a sum over geodesic paths, labelled by the index $N$, and
 the large-order asymptotics for the Legendre functions have yielded the correct singularity structure near the null geodesics, allowing
 for the correct phase change at each caustic.


\bibliography{Nariai}{}
\bibliographystyle{apsrev}

\end{document}